\documentclass[showkeys,showpacs,aps,amssymb,prd]{revtex4}

\usepackage{appendix}
\usepackage{amsbsy}
\usepackage{amsfonts}
\usepackage{amsmath}
\usepackage{amssymb}
\usepackage{amsthm}
\usepackage{graphicx}
\usepackage{ifthen}
\usepackage{textcomp}
\newcounter{multi} \newcounter{multa}

\newcounter{faki} \newcounter{faka}

    \newtheorem{theorem}{Theorem}

\theoremstyle{definition} 

\theoremstyle{remark} 

\def\fkM{\frak M}

\def\bbf{{\bf f}}
\def\bx{{\bf x}}
\def\bby{{\bf y}}
\def\blm{\boldsymbol \eta}
\def\bs{{\bf s}}

\def\det {\mathop{\rm det}\nolimits}
\def\grad {\mathop{\rm grad}\nolimits}
\def\Jac {\mathop{\rm Jac}\nolimits}
\def\Hess {\mathop{\rm Hess}\nolimits}

\def\RR{\mathbb{R}}

\def\beqan{\begin{eqnarray*}}
\def\eeqan{\end{eqnarray*}}

\newcommand{\beq}{\begin{eqnarray}}
\newcommand{\eeq}{\end{eqnarray}}

\begin{document}

\title{A Universal Magnification Theorem for Higher-Order Caustic Singularities}

\author{A. B. Aazami}\email{aazami@math.duke.edu}\affiliation{Department of
Mathematics, Duke University, Science Drive, Durham, NC 27708}

\author{A. O. Petters}\email{petters@math.duke.edu}\affiliation{Departments of
Mathematics and Physics, and Fuqua School of Business, 
Duke University, Science Drive, Durham, NC 27708}


\begin{abstract}
We prove that, independent of the choice of a lens model, 
the total signed magnification always sums to zero for 
a source anywhere in the four-image region close to
swallowtail, elliptic umbilic, and hyperbolic umbilic caustics. 
This is a more global and higher-order analog of the well-known fold and cusp magnification relations, in which the total signed magnification in the two-image region of the fold, and the three-image region of the cusp, are both always zero.  As an application, we construct
a lensing observable for the hyperbolic umbilic magnification relation and compare it
with the corresponding observables for the cusp and fold relations using a singular isothermal ellipsoid lens.  We demonstrate the 
greater generality of the hyperbolic umbilic magnification relation by showing how it
applies to the fold image doublets and cusp image triplets, and 
extends to image configurations that are neither.  
We show that the results are applicable 
to the study of substructure on galactic scales using observed quadruple images of
lensed quasars.
The magnification relations are also proved for generic 1-parameter families
of mappings between planes, extending their potential range of applicability beyond
lensing.
\end{abstract}


\maketitle

\section{Introduction}

One of the key signatures of gravitational lensing is the occurrence of multiple images
of lensed sources. The magnifications of the images in turn are also known to obey certain relations. 
One of the simplest examples of a magnification relation is that due to a single point-mass lens, where the two images of the source have signed magnifications that sum to unity: 
$\mu_1 + \mu_2 =1$  (e.g., Petters et al. 2001 \cite[p. 191]{Petters}).
 Witt \& Mao 1995 \cite{Witt-Mao} generalized this result to a two point-mass lens.  They showed that when the source lies inside the caustic curve, a region which gives rise to five lensed images, the sum of the signed magnifications of these images is also unity: 
$\sum_{i} \mu_{i} = 1,$ 
where $\mu_{i}$ is the signed magnification of image $i$.  
This result holds independently of the lens's configuration (in this case, the mass of the point-masses and their positions); it is also true for any source position, so long as the
source lies inside the caustic (the region that gives rise to the largest number of images).  
Further examples of magnification relations, involving other families of lens models ($N$ point-masses, elliptical power-law galaxies, etc.), subsequently followed in Rhie 1997 \cite{Rhie}, 
Dalal 1998 \cite{Dalal}, Witt \& Mao 2000 \cite{Witt-Mao2}, Dalal \& Rabin 2001 \cite{Dalal-Rabin}, and 
Hunter \& Evans 2001 \cite{Hunter-Evans}.  More recently,  Werner 2008 \cite{Werner} has shown that the relations for the aforementioned family of lens models are in fact topological invariants.

Although the above relations are ``global'' in that they involve all the images of a given source, they are not {\it universal} because the relations depend on the specific class of lens model used.  
However, it is well-known that for a source near a fold or cusp caustic, the resulting images close to the critical curve are
close doublets and triplets whose signed magnifications always sum to zero 
(e.g, Blandford \& Narayan 1986 \cite{Blan-Nar}, Schneider \& Weiss 1992 \cite{Sch-Weiss92},
Zakharov 1999 \cite{Zakharov}, \cite[Chap. 9]{Petters}):
 \beq
\mu_{1} + \mu_{2}&=&0\ {\rm (fold)}\ , \nonumber \\
\mu_{1} + \mu_{2} + \mu_{3}&=&0\ {\rm (cusp)}\ . \nonumber
\eeq
These magnification relations are ``local'' and universal.  Their locality means that they apply to a subset of the total number of images produced, namely,
a close doublet for the fold and close triplet for the cusp, which requires
the source to be near the fold and cusp caustics, respectively.  Their universality follows from
the fact that the relations hold for a generic family of lens models.  
The higher-order caustics beyond folds and cusps that we consider 
are five generic caustic surfaces or big caustics occurring
in a three-parameter space. Slices of the big caustics give rise  to five generic caustic metamorphoses
(e.g., \cite{Petters}, Chapters 7 and 9).
All five caustic metamorphoses occur in gravitational lensing
(e.g., Blandford 1990 \cite{Blandford90}, Petters 1993 \cite{Petters93}, 
Schneider, Ehlers \& Falco 1992 \cite{Sch-EF}, and \cite{Petters}).  
In addition, the magnification relations for folds and cusps have been shown to provide
powerful diagnostic tools for detecting dark substructure on galactic
scales using quadruple lensed images of quasars (e.g., Mao \& Schneider 1998  \cite{Mao-Sch}; Keeton, Gaudi \& Petters  
2003 and 2005 \cite{KGP-cusps,KGP-folds}).

The aim of this paper is to show that invariants of the following form 
also hold universally for lensing maps
and general mappings with higher-order caustic singularities:
$$\sum_{i} \mu_{i} = 0\ .
$$ 
In particular, we show that such invariants occur not only for folds and cusps, but also for 
lensing maps with {\it elliptic umbilic} and {\it hyperbolic umbilic} caustics, and for
general mappings with {\it swallowtail}, {\it elliptic umbilic}, and {\it hyperbolic umbilic} 
caustics.  Specifically, 
we prove that the total 
signed magnification of a source at any point in the four-image region of these higher-order 
caustic singularities, satisfies: 
$$
\mu_{1} + \mu_{2} + \mu_{3} + \mu_{4}=0\ .
$$  
As an application, we use the hyperbolic umbilic to show how such magnification relations can be used
for substructure studies of four-image lens galaxies.

The outline of the paper is as follows.  Section~\ref{Basic-Concepts} reviews the necessary lensing and singular-theoretic terminologies and results.  Section~\ref{Main-Theorem} states our main theorem, which
is for generic lensing maps and general mappings.  In Section~\ref{Applications}, 
the magnification relations are shown to be relevant to the study of dark substructure in
galaxies. We also employ a singular isothermal ellipsoid lens to compare 
the hyperbolic umbilic relations to the fold and cusp ones.  The proof of the main theorem is quite
long and so is placed in Appendices~\ref{Appendix:ProofLensing}
and \ref{Appendix:ProofGeneric}.

\section{Basic Concepts}
\label{Basic-Concepts}

\subsection{Lensing Theory}
We begin by reviewing the necessary lensing and singular-theory terminologies.  The spacetime geometry for gravitational lensing is treated as a perturbation of a Friedmann universe by a {\lq\lq weak field\rq\rq} spacetime.  To that end, we regard a gravitational lens as being localized in a very small portion of the sky.  Furthermore, we assume that gravity is {\lq\lq weak\rq\rq}, so that near the lens it can be described by a Newtonian potential.  We also suppose that the lens is static.  Respecting these assumptions, the spacetime metric is given by
\beq
{\boldsymbol g}_{GL} = -\left(1+{2\phi \over c^2}\right)c^2d\tau^2+a(\tau)^2\left(1-{2\phi \over c^2}\right)\left({dR^2 \over 1-k R^2}+R^2\left(d\theta^2+{\rm sin}^2\theta\, d\varphi^2\right)\right)\nonumber \ ,
\eeq
\noindent where $\tau$ is cosmic time, $\phi$ the time-independent Newtonian potential of the perturbation caused by the lens, $k$ is the curvature constant, and $(R,\theta, \varphi)$ are the coordinates in space. Here
terms of order greater than $1/c^2$ are ignored in any calculation involving $\phi$.

The above metric is used to derive the time delay function $T_{\boldsymbol {\rm y}} : L \longrightarrow \mathbb{R}$, which for a single lens plane is given by
$$T_{\boldsymbol {\rm y}}({\boldsymbol {\rm x}}) = {1 \over 2}|{\boldsymbol {\rm x}}-{\boldsymbol {\rm y}}|^{2}-\psi({\boldsymbol {\rm x}})\ ,$$
\noindent where ${\boldsymbol {\rm y}}=(s_{1},s_{2}) \in S$ is the position of the source on the light source plane $S = \mathbb{R}^2$, ${\boldsymbol {\rm x}} = (u,v) \in L$ is the impact position of a light ray on the lens plane $L \subseteq \mathbb{R}^2$, and $\psi : L \longrightarrow \mathbb{R}$ is the gravitational lens potential.  As its name suggests, the time delay function gives the time delay of a lensed light ray emitted from a source in $S$, relative to the arrival time of a light ray emitted from the same source in the absence of lensing. Fermat's principle yields that light rays emitted from a source that reach an observer are realized as critical points of the time delay function.  In other words, a {\it lensed image} of a light source at $\bf y$
is a solution ${\boldsymbol {\rm x}} \in L$ of the equation $({\rm grad}\,T_{\boldsymbol {\rm y}})({\boldsymbol {\rm x}})  = {\boldsymbol 0}$, where the gradient is taken with respect to ${\boldsymbol {\rm x}}$. 
When there is no confusion with the mathematical image 
of a point, we shall follow common practice and sometimes call a lensed image simply an {\it image}.

The time delay function also induces a {\it lensing map}  ${\boldsymbol \eta} : L \longrightarrow S$, which 
is defined by
$$
{\boldsymbol {\rm x}} \longmapsto {\boldsymbol \eta}({\boldsymbol {\rm x}}) = {\boldsymbol {\rm x}} - ({\rm grad}\,\psi)({\boldsymbol {\rm x}})\ .
$$
We call 
$
{\boldsymbol \eta} ({\bf x}) = {\bf y}
$
the {\it lens equation}.  Note that  ${\bf x}\in L$ is a solution of the lens equation if and only
if it is a lensed image because 
$({\rm grad}\,T_{\boldsymbol {\rm y}})({\boldsymbol {\rm x}})  = {\boldsymbol \eta} ({\bf x}) - 
{\bf y}.$
Critical points of the lensing map $\boldsymbol \eta$
are those ${\boldsymbol {\rm x}} \in L$ for which ${\rm det(Jac}\,{\boldsymbol \eta})({\boldsymbol {\rm x}}) = 0$.  Generically, the locus of critical points of the lensing map form curves called
{\it critical curves}.
The value 
${\boldsymbol \eta}({\boldsymbol {\rm x}})$ of a critical
point $\bf x$ under ${\boldsymbol \eta}$ is called a {\it caustic point}.  These typically form curves,
but could be isolated points.   Examples of
caustics are shown in the third column of Figure~\ref{Figure0}.
  For a generic lensing scenario,
the number of lensed images of a given source can change (by $\pm 2$ for generic crossings) if and only if the source crosses a caustic.  The {\it signed magnification} of a lensed image ${\boldsymbol {\rm x}} \in L$ of a light source at ${\bf y} =  {\boldsymbol \eta}({\boldsymbol {\rm x}})  \in S$ is given by 
\beq
\label{lensmap}
\mu({\boldsymbol {\rm x}}) = {1 \over {\rm det(Jac}\,{\boldsymbol \eta})({\boldsymbol {\rm x}})}\ ,
\eeq  
where we used the fact that ${\rm det(Jac}\,{\boldsymbol \eta}) = {\rm det(Hess\,T_{\boldsymbol {\rm y}})}$
for single plane lensing.  Considering the graph of the time delay function, its principal curvatures coincide with the eigenvalues of ${\rm Hess}\,T_{\boldsymbol {\rm y}}({\boldsymbol {\rm x}})$.
In addition, its Gaussian curvature at $({\boldsymbol {\rm x}},T_{\boldsymbol {\rm y}}({\boldsymbol {\rm x}}))$ 
equals ${\rm det}({\rm Hess}\,T_{\boldsymbol {\rm y}})({\boldsymbol {\rm x}})$. In other words, the magnification
of an image $\bf x$ can be expressed as
\beq
\label{Gauss-lensmap}
\mu({\bf x}) = \frac{1}{{\rm Gauss}({\bf x}, \tiny{T_{\bf y} ({\bf x})})}\ ,
\eeq
where ${\bf y} = {\boldsymbol \eta} ({\bf x})$ and
${\rm Gauss}({\bf x}, T_{\bf y} ({\bf x}))$ is the Gaussian curvature of the
graph of $T_{\bf y}$ at the point $({\bf x}, T_{\bf y} ({\bf x}))$.
Therefore, the magnification relations are also geometric invariants involving the Gaussian curvature of the graph of $T_{\boldsymbol {\rm y}}$ at its critical points.
Readers are referred to \cite[Chap. 6]{Petters}  for a full treatment of these
aspects of lensing.

\subsection{Higher-Order Caustic Singularities}

This section briefly reviews those aspects of the theory of
singularities that will be needed for our main theorem.
The central theorem we shall employ is actually summarized in Table~\ref{table1} below. 
It is also worth noting  that the terms ``universal'' and
``generic'' will be used often.   Formally, a property is called {\it generic}
or {\it universal}  if it holds for an open, dense subset of mappings in the
given space of mappings.  Elements of the open, dense subset are then
referred to as being {\it generic} (or {\it universal}).
See
\cite[Chap. 8]{Petters} for a discussion of genericity.

\begin{figure}[ht]
\label{cuspcross}
\begin{center}
\begin{tabular}{| c | c | c |}
\hline
& & \\
{\bf Type} & {\bf Big Caustic} & {\bf Caustic Metamorphosis}\\
& & \\
\hline\hline
& & \\
Swallowtail &
$~~~~~$\includegraphics[scale=.23]{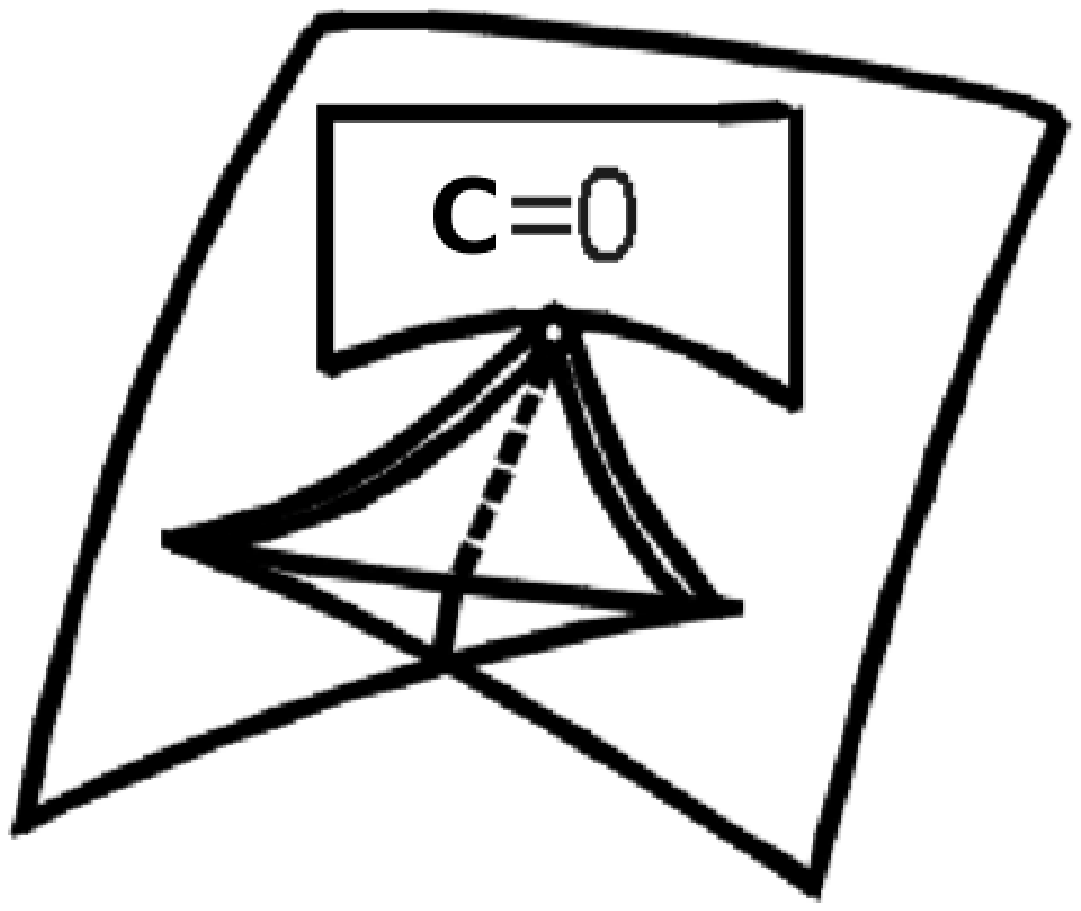}$~~~~~$ &
$~~~~$\includegraphics[scale=.25]{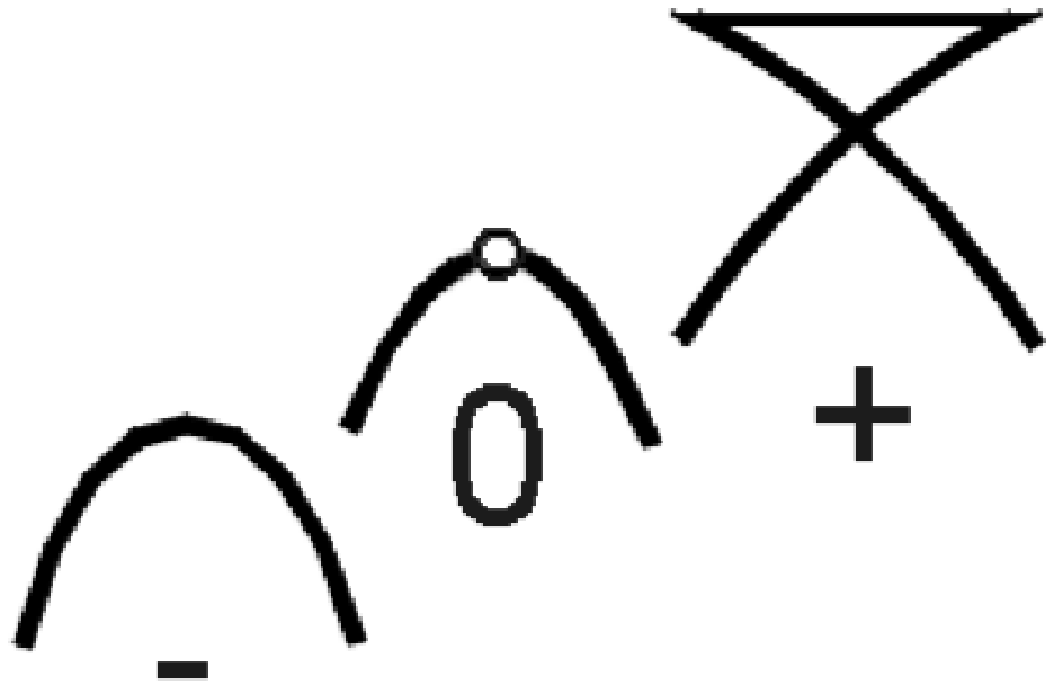}$~~~~~$\\
& & \\
\hline
& & \\
Elliptic Umbilic &
$~~~~~$\includegraphics[scale=.24]{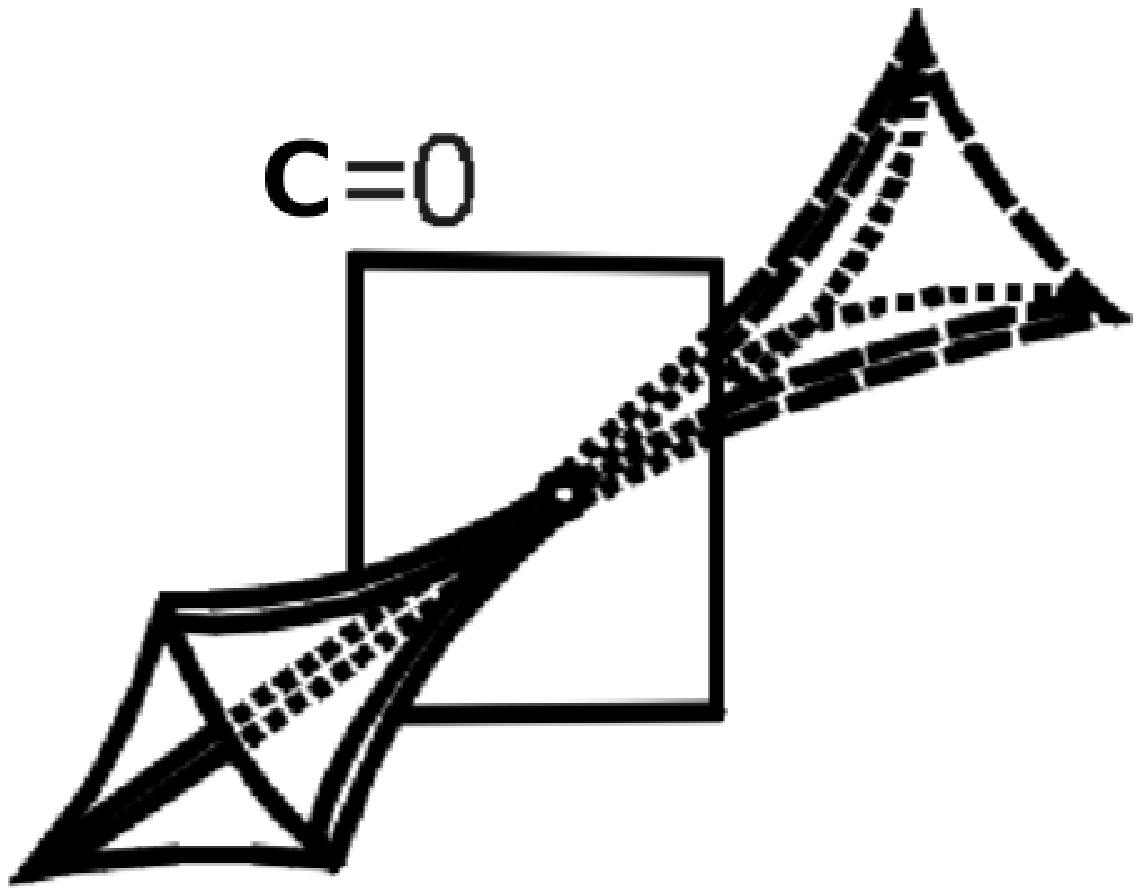}$~~~~~$ &
\,\,\,\, \includegraphics[scale=.25]{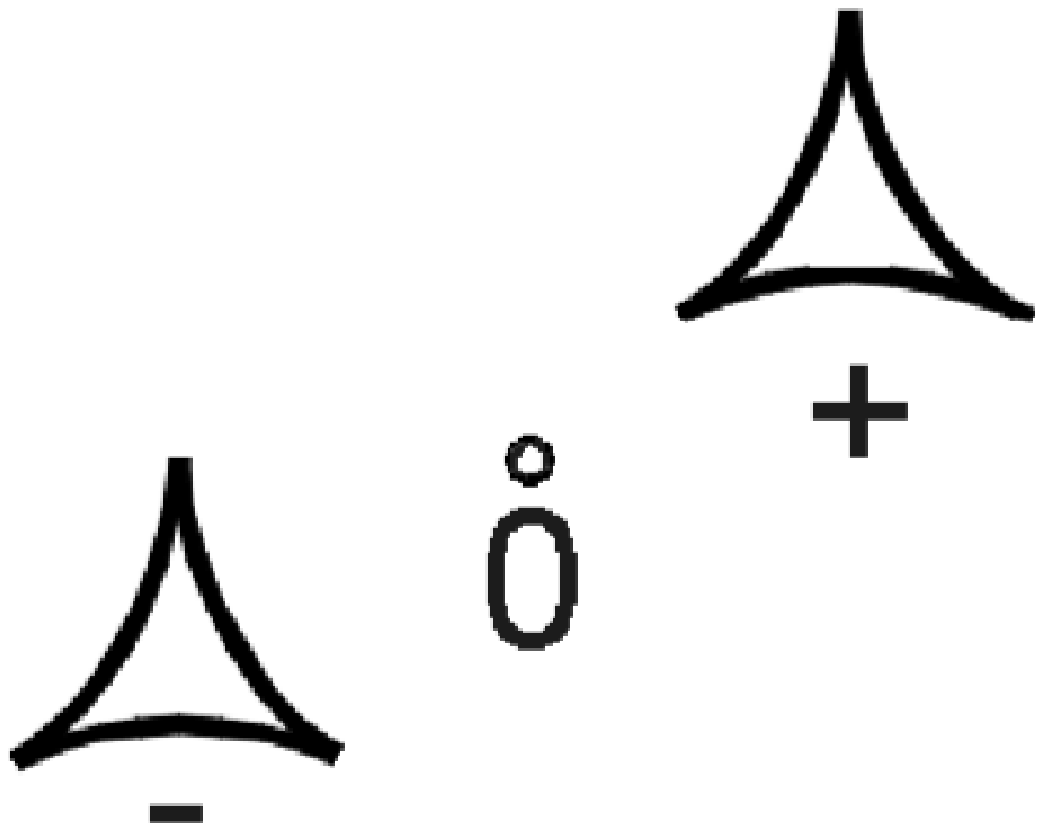}$~~~~~$\\
& & \\
\hline
& & \\
\,\,\, Hyperbolic Umbilic\,\,\, &
$~~~~~$\includegraphics[scale=.23]{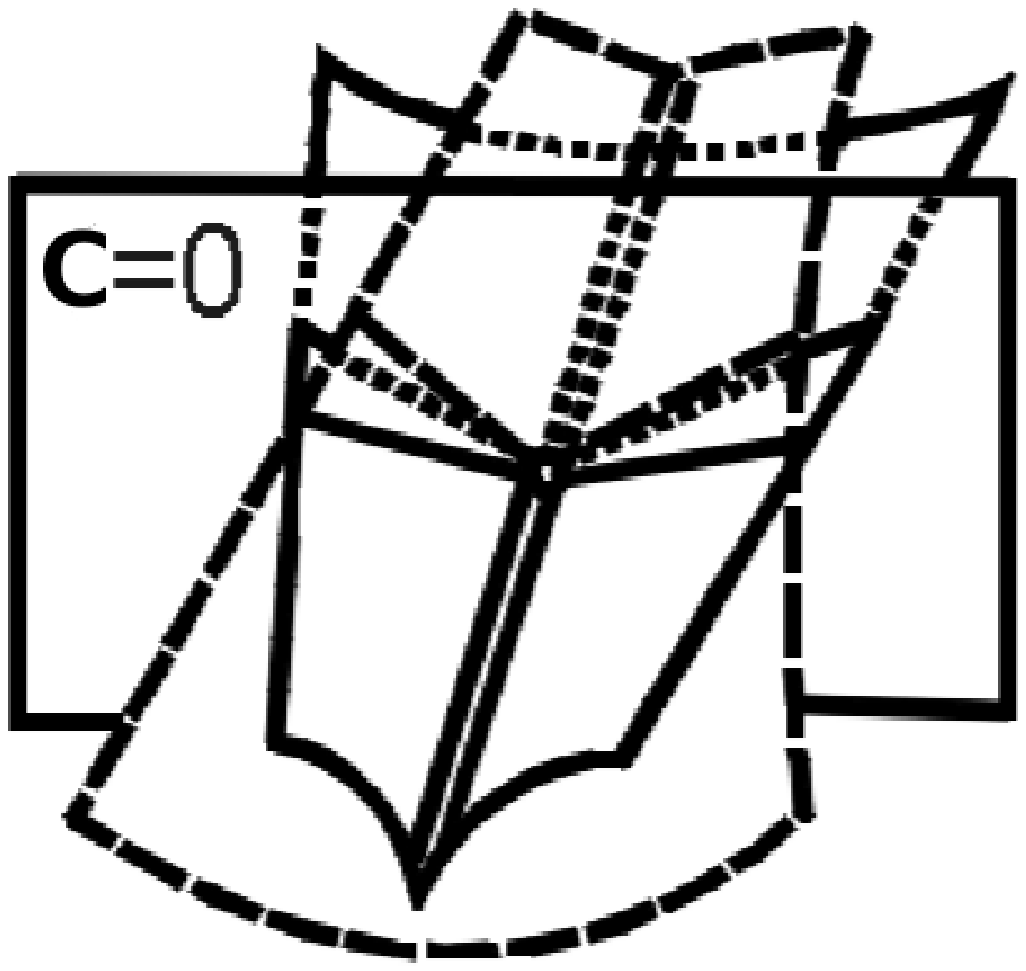}$~~~~~$ &
\,\,\,\, \includegraphics[scale=.25]{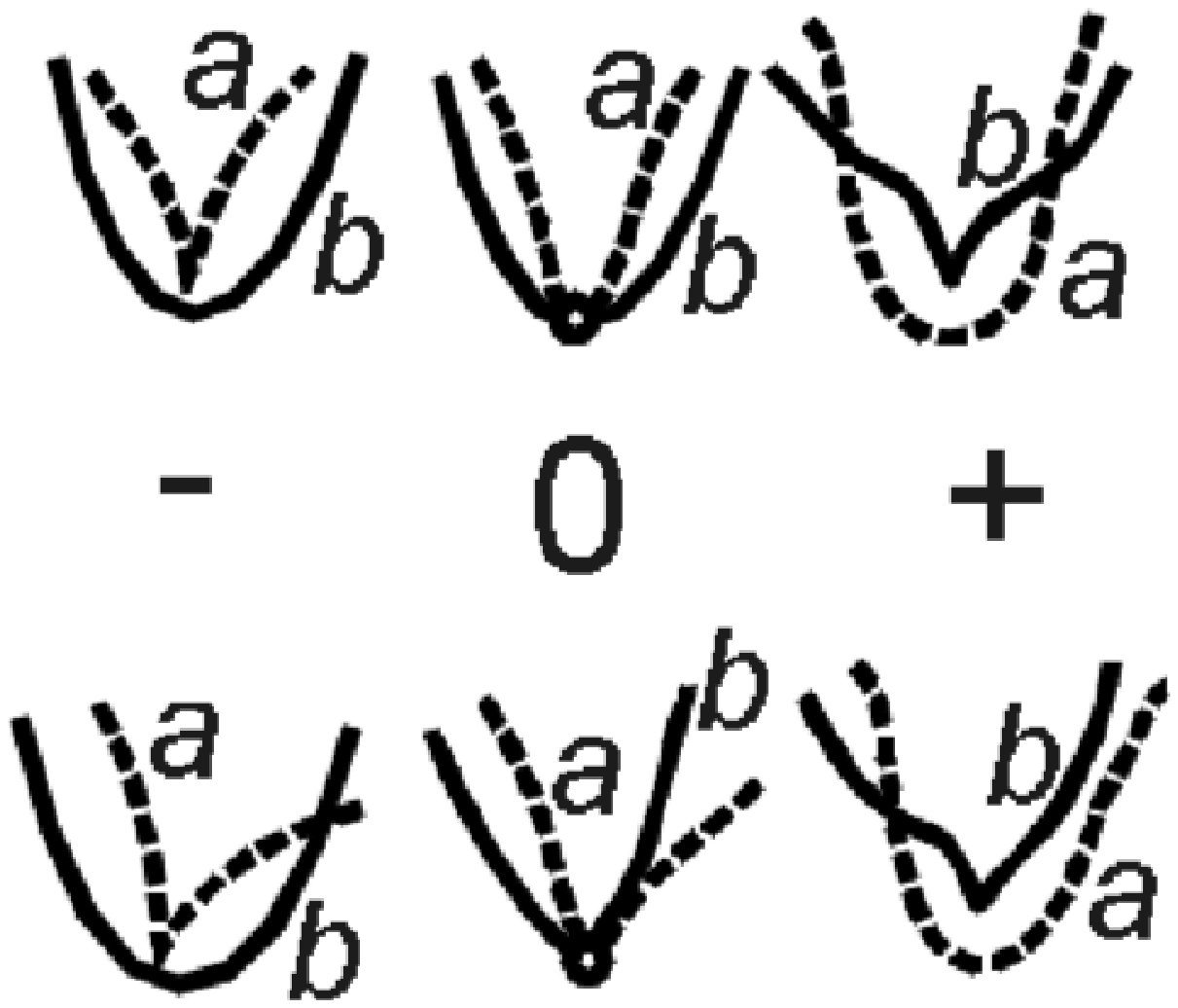}$~~~~~$\\
& & \\
\hline
\end{tabular}
\end{center}
\caption{The swallowtail, elliptic umbilic, and hyperbolic umbilic are higher-order caustics
shown as surfaces or big caustics in the three-parameter space $\{(c,{\bf y})\}$
(middle column). Each $c$-slice of a big caustic yields  caustic curves, which
for generic slices evolve according to the metamorphoses in the rightmost column. 
The point $\circ$ occurs for the slice $c=0$.  In the case of the hyperbolic umbilic, note that 
the $a$ and $b$ caustic curves are exchanged when $c$ varies through $c =0$.
This classification is due to Arnold 1986 \cite{Arnold86}.}
\label{Figure0}
\end{figure}

We saw in the previous section that the time delay function
$T_{{\bf y}} ({\bf x})$, which can be viewed as a two-parameter
family of functions with parameter $\bf y$, gives rise to the lensing map
${\boldsymbol \eta} : L \longrightarrow \RR^2$.  The set of critical points of ${\boldsymbol \eta}$ consists
of all $\bx \in L$ such that  ${\rm det(Jac}\,{\boldsymbol \eta})({\boldsymbol {\rm x}}) = 0$. 
In this two-dimensional setting, a generic lensing map will have
only two types of generic critical points: folds and cusps (see \cite[Chap. 8]{Petters}).
The fold critical points map over to caustic arcs that abut isolated cusp caustic points;
e.g, see the astroid caustic in Figure~\ref{Figure0}.

Now, let  $T_{c,\bby}({\boldsymbol {\rm x}})$ denote a family of time delay functions
parametrized by the source position $\bf y$ and 
 $c \in \mathbb{R}$.  In the context of gravitational lensing, the parameter $c$ may denote external shear, core radius, redshift, or some other physical input.  
The three-parameter family $T_{c,\bby} ({\bf x})$ gives rise to a one parameter family of lensing maps ${\boldsymbol \eta_{c}}$.  Varying $c$ causes the caustic curves in the light source plane $S$ to evolve with $c$.  This traces out a caustic surface, called a {\it big caustic}, in the three-dimensional space 
$\RR \times \RR^2 = \{(c,\bby)\}$; see  Figure~\ref{Figure0}.
Beyond folds and cusps, these surfaces form higher-order caustics that are
classified into three {\it universal} or {\it generic} types for  locally stable
families ${\boldsymbol \eta_{c}}$, 
namely, {\it swallowtails, elliptic umbilics}, and {\it hyperbolic umbilics}
(e.g., Arnold 1986 \cite{Arnold86} and \cite[Chap. 9]{Petters}).
Generic $c$-slices of these big caustics also yield {\it caustic metamorphoses;}
see Figure~\ref{Figure0}. Note that the point $\circ$ is a
degenerate point of the lensing map ${\boldsymbol \eta_{c}}$ on the slice $c =0$.

For the three-parameter family  $T_{c,\bby}({\boldsymbol {\rm x}})$ of time delay functions,  
the universal {\it quantitative} form of the lensing map can be derived locally 
using {\it rigid} coordinate transformations and
Taylor expansions, along appropriate constraint equations for
the caustics (see \cite[Chap. 6]{Sch-EF} for details).  
Table~\ref{table1} summarizes the quantitative forms of ${\boldsymbol \eta}_{c}$ for the elliptic umbilic and hyperbolic umbilic critical points.  The quantitative form for the swallowtail will be dealt with in future work.
Observe that the
elliptic and hyperbolic umbilics for $T_{c,\bby}$ (or $\blm_c$) do not depend on the lens potential,
apart perhaps from $c$ in the event that $c$ is a lens parameter.

One can also consider a general, smooth three-parameter family 
$F_{c,\bs}({\boldsymbol {\rm x}})$
of functions on an open subset of $\RR^2$ that
induces a one-parameter family of mappings
$\bbf_c$ between planes, which are analogs of  the lensing map. 
The {\it universal} form  (also known as the {\it generic} or {\it qualitative} form)
of the one-parameter family $\bbf_c$
is obtained basically by 
using differentiable equivalence classes of $F_{c,\bs}$
that distinguish $c$ from the coordinates of
$\bs$, to construct catastrophe manifolds that are projected into the
space  $\{ (c,\bs)\} = \RR \times \RR^2$ to obtain local coordinate expressions 
for $\bbf_{c}$
(e.g.,  Majthay 1985 \cite{Majthay}, Castrigiano \& Hayes 1993 \cite{C-Hayes},
Golubitsky \& Guillemin  1973 \cite{Gol-G}).
These projections of the catastrophe manifolds are called
{\it catastrophe maps} or {\it Lagrangian maps}, and they are differentiably equivalent
to  $\bbf_c$ (see \cite[pp. 273-275]{Petters}).
Similar to the case for $T_{c,\bby}$ and its induced lensing map
$\blm_c$, a generic family $F_{c,\bs}$ and its associated map
$\bbf_c$ has three types of caustic singularities beyond folds and cusps:
{\it swallowtails, elliptic umbilics}, and {\it hyperbolic umbilics}
(e.g., \cite{Arnold86}, \cite[Chap. 9]{Petters}).
The generic forms of $\bbf_{c}$ about fold, cusp,
elliptic umbilic, hyperbolic umbilic, and swallowtail singularities
are shown in Table~\ref{table1}. 
A detailed treatment of these issues is given in \cite{Petters,Sch-EF,Majthay}.  

In summary, the central result about caustic singularities that we shall use can be
stated as follows: 
\begin{itemize}
\item A {\it generic, smooth} three-parameter family of time delay functions $T_{c,\bby}$
can be transformed in a neighborhood of a caustic into one of the forms in 
the second column of Table~\ref{table1} using {\it rigid}
coordinate transformations that distinguish $c$ from the component parameters of $\bby$
\cite{Sch-EF,Petters}.
\item A {\it generic, smooth} three-parameter family of
general functions $F_{c,\bs}({\boldsymbol {\rm x}})$, which need not be a time delay family,
can be transformed in a neighborhood of a caustic into one of the forms in the third column of Table~\ref{table1}
using coordinate transformations distinguishing $c$ from the parameters of $\bs$
\cite{Majthay,Petters}.
\end{itemize}

\begin{table}[tbh]
\centering
\vskip 6pt
\begin{tabular}{| c | c | c |}
\hline
 & & \\
$~~~${\bf Caustic}$~~~$ & $~~~$ ${\bf Quantitative\ Lensing\ Map}$ $~~~$ &
$~~~$
${\bf Generic\ Map}$ $~~~$\\ [0.5ex]
 & & \\
\hline\hline
& & \\
& $T_{{\boldsymbol {\rm y}}}(u,v) =  {1 \over 2}{\boldsymbol {\rm
y}}^{2}-{\boldsymbol {\rm x \cdot y}}+{1 \over
2}a_{11}u^{2}+{1\over6}a_{111}u^{3}$\,\,\,\, &\\ 
&
\hskip42pt$+{1\over2}a_{112}u^{2}v+{1\over2}a_{122}uv^{2}+{1\over6}a_{222}v^{3}$
&
$F_{\boldsymbol {\rm s}} (u,v) =
s_{1}u+s_{2}v-{1\over2}u^{2}-{1\over3}v^{3}$ \\
 Fold (2D) & & \\
& ${\boldsymbol \eta}(u,v)=\left(a_{11}u+{1 \over
2}a_{122}v^{2}+a_{112}uv\right.,$\,\,\,\, & ${\bf f}(u,v)=\left(u\ ,\
v^{2}\right)$ \\
& \hskip42pt$\left.{1 \over 2}a_{112}u^{2}+a_{122}uv+{1 \over
2}a_{222}v^{2}\right)$ & \\
& & \\
\hline
& & \\
& \,\,\,\,$T_{{\boldsymbol {\rm y}}}(u,v) = {1 \over 2}{\boldsymbol {\rm
y}}^{2}-{\boldsymbol {\rm x \cdot y}}+{1 \over
2}a_{11}u^{2}+{1\over6}a_{111}u^{3}$\,\,\,\,& \\
Cusp (2D) &
$\hskip55pt+{1\over2}a_{112}u^{2}v+{1\over2}a_{122}uv^{2}+{1\over24}a_{2222}v^{4}$
& $F_{\boldsymbol {\rm s}}(u,v) =
s_{1}u+s_{2}v-{1\over2}u^{2}-{1\over2}s_{1}v^{2}-{1\over4}v^{4}$
\\
& & \\
& \,\,\,${\boldsymbol
\eta}(u,v)=\left(a_{11}u+{1 \over 2}a_{122}v^{2}\ ,\ a_{122}uv+{1 \over
6}a_{2222}v^{3}\right)$\,\,\, & ${\bf f}(u,v)=\left(u\ ,\ uv+v^{3}\right)$
\\
& & \\
\hline
& & \\
& $T_{c,{\boldsymbol {\rm y}}}(u,v) = {1 \over 2}{\boldsymbol {\rm
y}}^{2}-{\boldsymbol {\rm x \cdot
y}}+{1 \over 3}u^{3}-uv^{2}+2cv^{2}$ & \,\,\,$F_{c,{\boldsymbol {\rm s}}}
(u,v) =
s_{1}u+s_{2}v+c(u^{2}+v^{2})$\,\,\, \\
Elliptic Umbilic (3D) & & \ $+\,u^{3}-3uv^{2}$\\ 
& & \\
& \,\,\,\,${\boldsymbol \eta}_{c}(u,v)=\left(u^{2}-v^{2}\ ,\
-2uv+4cv\right)$\,\,\,\, &
\,\,\,\,${\bf f}_{c}(u,v)=\left(3v^{2}-3u^{2}-2cu\ ,\
6uv-2cv\right)$\,\,\,\,\\
& & \\
\hline
& & \\
& $T_{c,{\boldsymbol {\rm y}}} (u,v) = {1 \over 2}{\boldsymbol {\rm
y}}^{2}-{\boldsymbol {\rm x \cdot y}}+{1 \over 3}(u^{3}+v^{3})+2cuv$ &
\,\,\,$F_{c,{\boldsymbol {\rm s}}} (u,v) =
s_{1}u+s_{2}v+cuv+u^{3}+v^{3}$\,\,\\\
\,\,\,\,Hyperbolic Umbilic (3D)\,\,\,\, & & \\
& ${\boldsymbol \eta}_{c}(u,v)=\left(u^{2}+2cv\ ,\ v^{2}+2cu\right)$ &
\,\,\,\,${\bf f}_{c}(u,v)=\left(-3u^{2}-cv\ ,\
-3v^{2}-cu\right)$\,\,\,\,\\
& & \\
\hline
& & \\
& & \,\,\,$F_{c,{\boldsymbol {\rm s}}}(u,v) = s_{1}u + s_{2}v -
{1\over2}s_{2}u^{2}-{1\over2}v^{2}$\,\,\, \\
Swallowtail (3D) & & $\,-{1\over3}cu^{3}-{1\over5}u^{5}$ \\
& & \\ 
& & \,\,\,\,${\bf f}_{c}(u,v)=\left(uv+cu^{2}+u^{4}\ ,\
v\right)$\,\,\,\, \\
& & \\
\hline
\end{tabular}
\caption{For each type of caustic singularity, the second and third columns show the 
respective universal local forms of the smooth three-parameter family of
time delay functions $T_{c,\bby}$ and family of general functions $F_{c,\bs}$,
along with their one-parameter family of lensing maps $\blm_c$
and induced general maps
$\bbf_c$. For the two-parameter case of the fold and cusp, 
the constants $a_{ijk}$
denote partial derivatives of $T_{c,\bby}(\bx)$ with respect to $\bx = (u,v) \equiv (x_1,x_2)$, 
evaluated at the
origin: 
$a_{ijk} = (\partial^3 T_{c,\bby}/\partial x_i \partial x_j \partial x_k) ({\bf 0})$.  
These constants do not
appear in the quantitative forms of the elliptic and hyperbolic umbilic
in the second column. 
This implies that the local behavior of $T_{c,\bby}$ and ${\boldsymbol \eta}_{c}$ 
about an elliptic or hyperbolic
umbilic does not depend on the lens potential, except possibly through $c$ when $c$ is
a lens parameter.  We have omitted the quantitative 
form of the swallowtail for $T_{c,\bby}$ and $\blm_c$
because the proof of its magnification relation will appear in forthcoming work.}
\label{table1}
\end{table}

\newpage

\section{Main Theorem}
\label{Main-Theorem}

Consider the universal one-parameter family of lensing maps $\blm_{c}$
in Table~\ref{table1}. Let $\bx_i$ denote a lensed image of
a source at $\bby$, that is, $\bby = \blm_{c}(\bx_i)$,
and let $\mu_i$ be the magnification of 
$\bx_i$, which by (\ref{lensmap}) is
$
\mu_i = 1/\det(\Jac\,\blm_c)(\bx_i).
$
For the generic mappings
$\bbf_c$ in Table~\ref{table1}, we define the analog of magnification as follows:
$$
\fkM_i = \frac{1}{\det(\Jac \bbf_c)(\bx_i)}\ ,
$$
where $\bbf_c (\bx_i) = \bs$ or, equivalently, the point
$(\bx_i, F_{c,\bs} (\bx_i))$ is a critical point in the
graph of $F_{c,\bs}$.

\begin{theorem}
\label{theorem-main}
For any of the smooth generic three-parameter family of 
time delay functions $T_{c,\bby}$ {\rm(}or
lensing maps ${\boldsymbol \eta}_c${\rm)}and family of
general functions $F_{c,\bs}$
{\rm(}or general mappings $\bbf_c${\rm)} in
Table~\ref{table1},
and for
any source position  $\bf y$ and point $\bf s$
in the indicated region, the following results hold:
\begin{enumerate}
\item $A_{2}$ {\rm (Fold)} Magnification relations in two-image region: 
$$
\mu_{1} + \mu_{2} = 0\ , \qquad
\fkM_{1} + \fkM_{2} = 0\ .
$$

\vskip 12pt
\item $A_{3}$ {\rm (Cusp)} Magnification relations in three-image region:
$$\mu_{1} + \mu_{2} + \mu_{3}=0\ , \qquad
\fkM_{1} + \fkM_{2} + \fkM_3 =0\ .
$$
\vskip 12pt
\item $A_{4}$ {\rm (Swallowtail)}
 Magnification relation in four-image region: 
$$
\fkM_{1} + \fkM_{2} + \fkM_3 + \fkM_4 = 0\ .
$$
\vskip12pt
\item $D_{4}^{-}$ {\rm (Elliptic Umbilic)}
 Magnification relations in four-image region: 
$$\mu_{1} + \mu_{2} + \mu_{3} + \mu_{4}=0\ , \qquad
\fkM_{1} + \fkM_{2} + \fkM_3 + \fkM_4 = 0\ .
$$
\vskip 12pt
\item $D_{4}^{+}$ {\rm (Hyperbolic Umbilic)}
Magnification relations in four-image region: 
$$
\mu_{1} + \mu_{2} + \mu_{3} + \mu_{4}= 0\ , \qquad
\fkM_{1} + \fkM_{2} + \fkM_3 + \fkM_4 = 0\ .
$$
\end{enumerate}
\end{theorem}
\vskip 12pt

In the theorem, the $\mu$-magnification (resp., $\fkM$-magnification) relations are {\it universal}
or {\it generic}  in the sense that
they hold for an open, dense set of three-parameter families $T_{c,\bby}$
(resp., general families $F_{c,\bs}$) in the space of such families;
see \cite{Arnold86} and \cite[Chaps. 7,8]{Petters}. Readers are referred to
\cite[Chap. 8]{Petters} for a discussion of universality/genericity.

The magnification relations in Theorem~\ref{theorem-main} are also geometric invariants.
In fact, we saw in equation (\ref{Gauss-lensmap}) that each $\mu_i$ is a 
reciprocal of the Gaussian curvature. This is also true of the quantities
$\fkM_i$. To see this, recall that the Gaussian curvature at the point
$(\bx_i,F_{c,\bs}(\bx_i))$ in the graph of $F_{c,\bs}$ is given
by
$${\rm Gauss}(\bx_i,F_{c,\bs}(\bx_i))
= \frac{\det(\Hess F_{c,\bs})(\bx_i)}{1 + |\grad F_{c,\bs}(\bx_i)|^2}\ \cdot
$$
But $(\bx_i,F_{c,\bs}(\bx_i))$ is a critical point of the graph, so
$\grad F_{c,\bs}(\bx_i) = {\bf 0}$.
A computation also shows that 
$$
\det(\Jac \bbf_c) = \det(\Hess F_{c,\bs})\ .
$$
Hence
$$
\fkM_i
= \frac{1}{{\rm Gauss}(\bx_i,F_{c,\bs}(\bx_i))}\ \cdot
$$

We use the A, D classification notation of Arnold 1973 \cite{Arnold73} in the theorem.
This notation highlights a deep link between the above singularities and Coxeter-Dynkin
diagrams appearing in the theory of simple Lie algebras.  Theorem~\ref{theorem-main} 
is also apparently related to a deep result in singularity theory, namely, the inverse Jacobian 
Theorem and its corollary, the Euler-Jacobi formula (see Arnold, Gusein-Zade, \& Varchenko 
1985 \cite{AGV1}).  We are thankful to the referee for pointing out this link, which is 
currently being pursued by the authors.

As mentioned in the introduction, the fold and cusp magnification relations are known
\cite{Blan-Nar,Sch-Weiss92,Zakharov,Petters}, but we restate
them in the theorem for completeness. In addition, note that the magnification relation 
for the swallowtail
is established only for the generic form; the quantitative lensing case will be taken up
in future work. 

The proof of Theorem~\ref{theorem-main} is very long.
Appendix~\ref{Appendix:ProofLensing} gives a detailed proof of
the $\mu$-magnification relations, while
Appendix~\ref{Appendix:ProofGeneric} provides 
a proof of the $\fkM$-magnification relations.


\section{Applications}
\label{Applications}

Before discussing the applications, we recall that 
the magnification $\mu_i$ of a lensed image is the flux $F_{i}$ of the image
divided by the flux $F_{S}$ of the unlensed source (e.g., \cite[pp. 82-85]{Petters}):
$$
\mu_i = \pm \frac{F_i}{F_S}\ ,
$$
where the ``$+$'' choice is for even index images (minima and maxima) and
the ``$-$'' choice is for odd index images (saddles). Though $F_i$ is an observable, the
source's flux  $F_S$ is generally unknown. Consequently, 
the magnification $\mu_i$ is not directly
observable and so magnification sums
$
\sum_i \mu_i 
$
are also not observable.  However, we can construct an observable by introducing 
the following quantity:
\beq
\label{R-quantity}
R \equiv \frac{\sum_i \mu_i}{\sum_i |\mu_i|} = \frac{\sum_i (\pm) F_i}{\sum_i F_i}\  ,
\eeq
where the $\pm$ choice is the same as above. This quantity is in
terms of the observable image fluxes $F_i$ and image signs, which can be determined
for real systems \cite{KGP-folds,KGP-cusps}.

Now, aside from their natural theoretical interest, the importance of
magnification relations in gravitational lensing arises in their applications
to detecting dark substructure in galaxies using ``anomalous'' flux
ratios of multiply imaged quasars. 
The setting consists typically of four images of a quasar lensed by a foreground
galaxy.   The {\it smooth} mass density models used for the galaxy lens
usually
accurately reproduce the number and relative positions of the images, but
fail to reproduce the image flux ratios.  For the case of a cusp, where
a close image  triplet appears,
Mao \& Schneider
1998 \cite{Mao-Sch} showed that the cusp $\mu$-magnification relation
fails (i.e., deviates from zero) and argued that it does so since the {\it smoothness
assumption} about the galaxy lens breaks down on the scale of the fold image doublet.
In other words, a violation of the cusp magnification relation
in a real lens system implies a violation of smoothness in the lens, which in turn
invokes the presence  of substructure or graininess 
in the galaxy lens on 
the scale of the image separation.
Soon thereafter Metcalf \& Madau
2001 \cite{M-M} and Chiba 2002 \cite{Chiba} showed that dark matter was a plausible candidate
for this substructure.  

\begin{figure}[tbh]
\label{cuspcross}
\begin{center}
\begin{tabular}{| c | c |}
\hline
& \\
{\bf SIE for  $\bf e = 0.35$ ,
$\bf \boldsymbol \gamma  = 0.05$} & {\bf Hyperbolic Umbilic $\blm_c$ for $\boldsymbol c = 0.2$} \\
& \\
\hline\hline
{\scriptsize {\rm {\bf (a)}}}\hskip 10pt{\bf fold} & {\scriptsize {\rm
{\bf (d)}}}\hskip 10pt{\bf fold}\\
& \\
$~~~~~$\includegraphics[scale=.25]{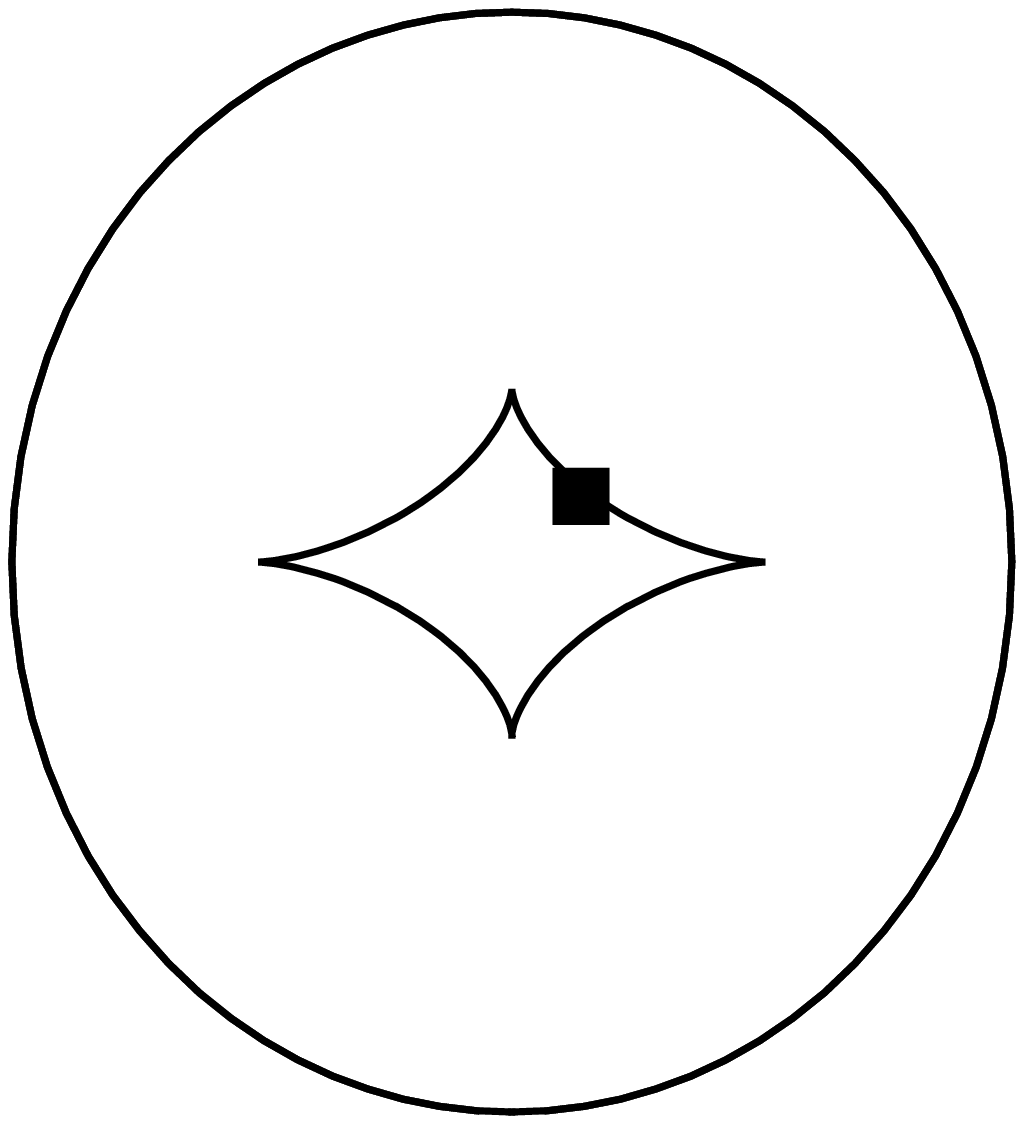}$~~~~~~~~~~$
\includegraphics[scale=.25]{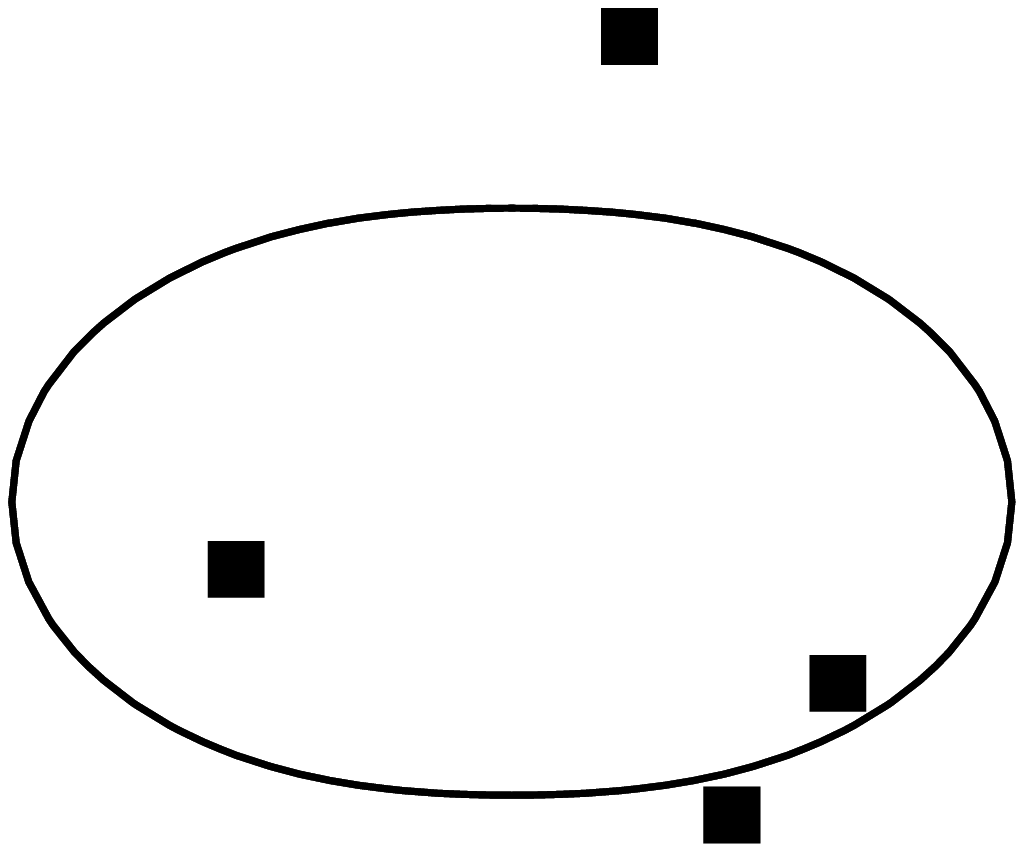}$~~~~~$
&
$~~~~~$\includegraphics[scale=.25]{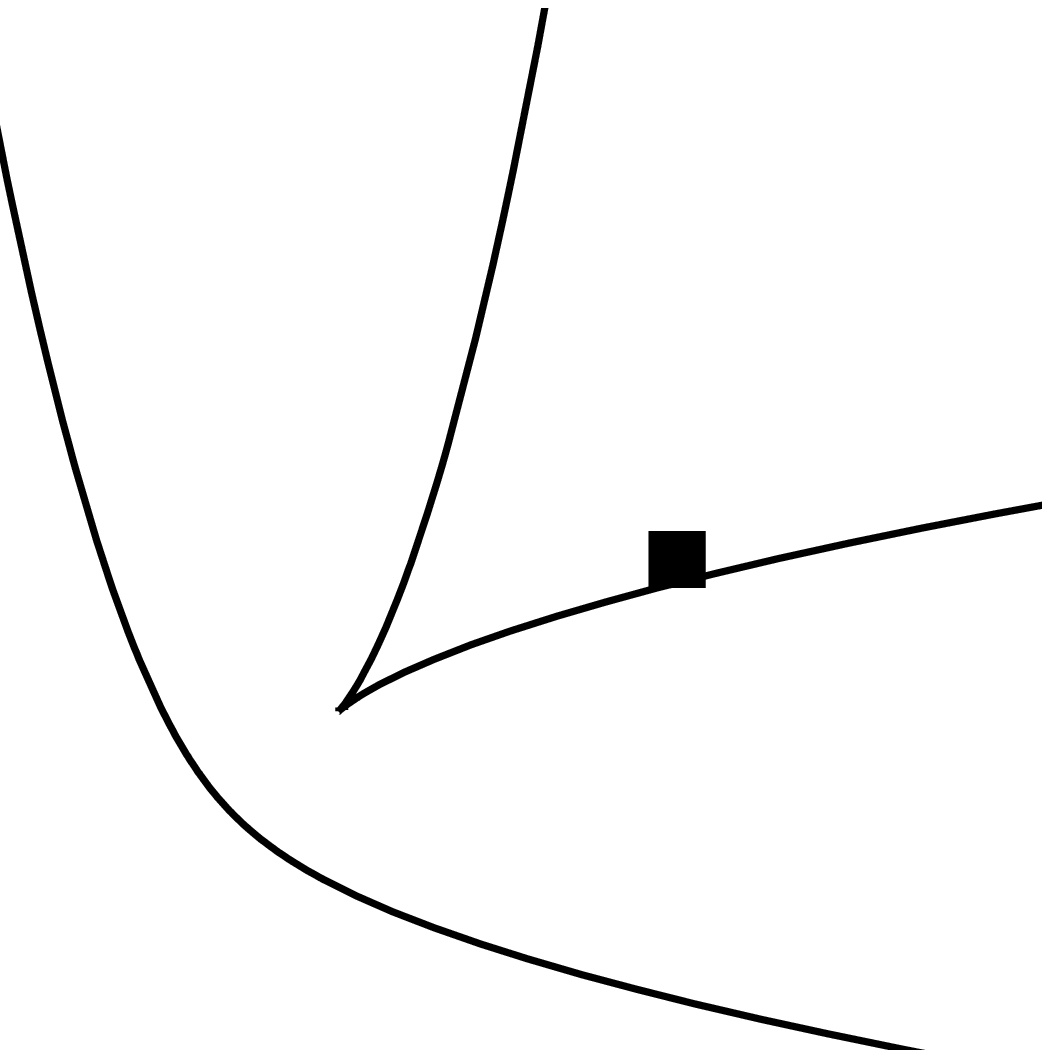}$~~~~~~~~~~$
\includegraphics[scale=.25]{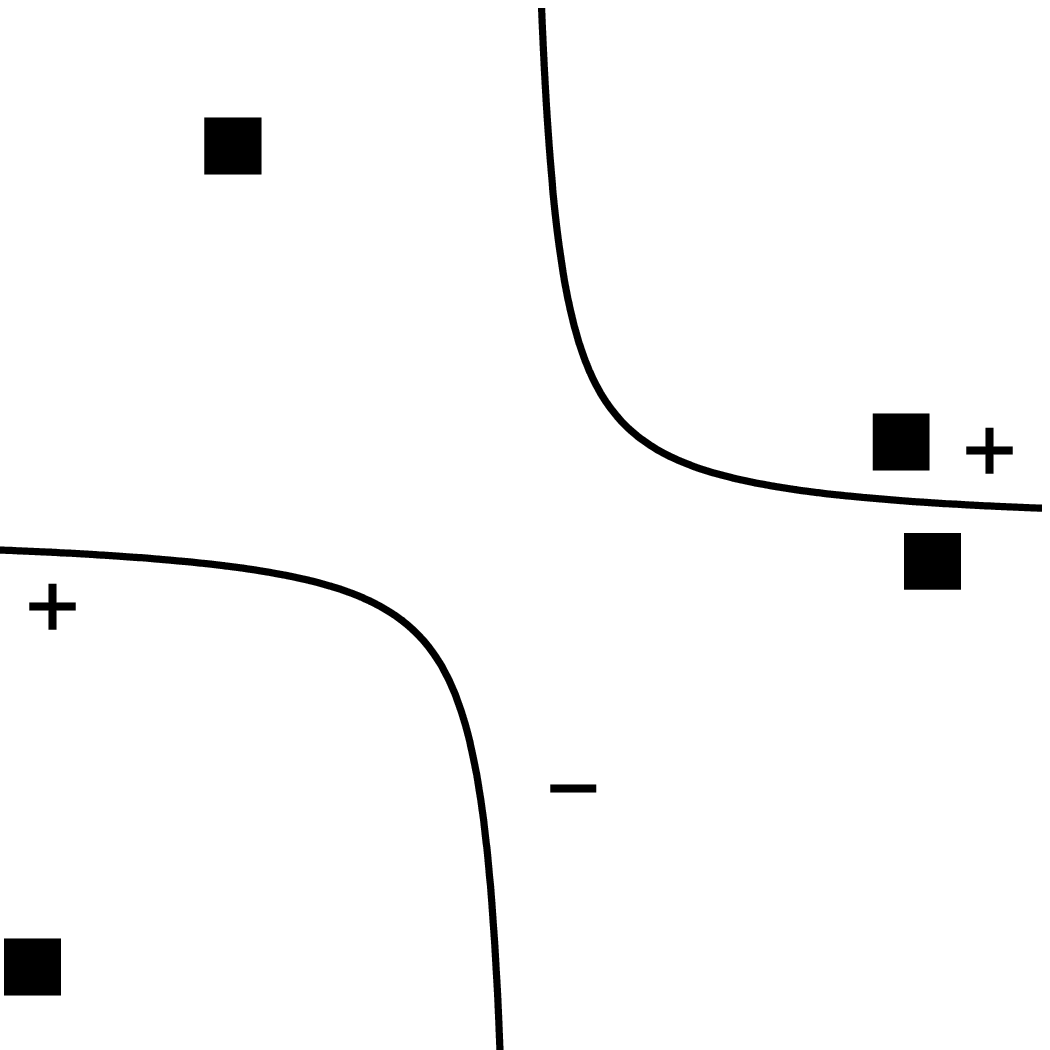}$~~~~~$\\
& \\
\hline
{\scriptsize {\rm {\bf (b)}}}\hskip 10pt{\bf cusp} & {\scriptsize {\rm
{\bf (e)}}}\hskip 10pt{\bf cusp}\\
& \\
$~~~~~$\includegraphics[scale=.25]{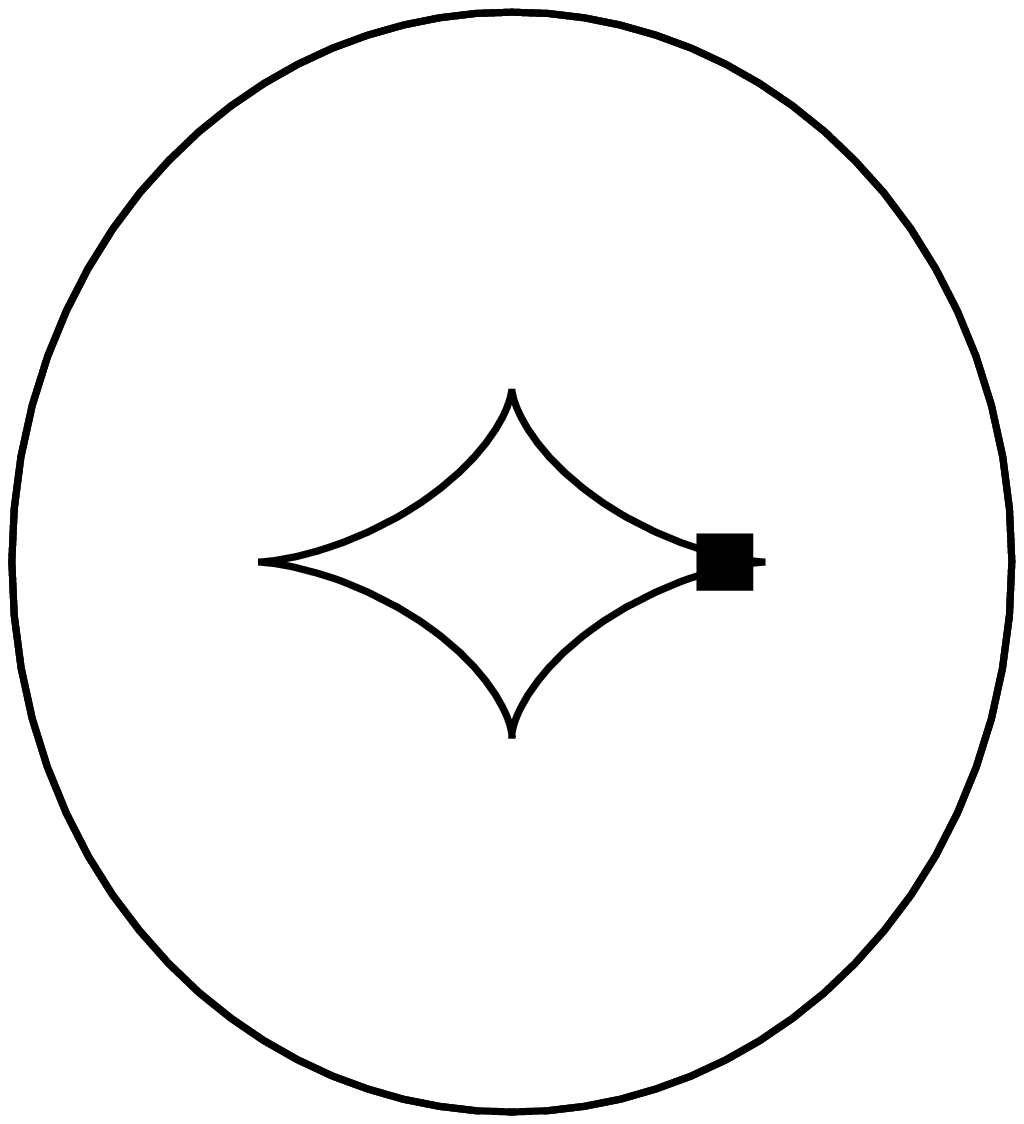}$~~~~~~~~~~$
\includegraphics[scale=.25]{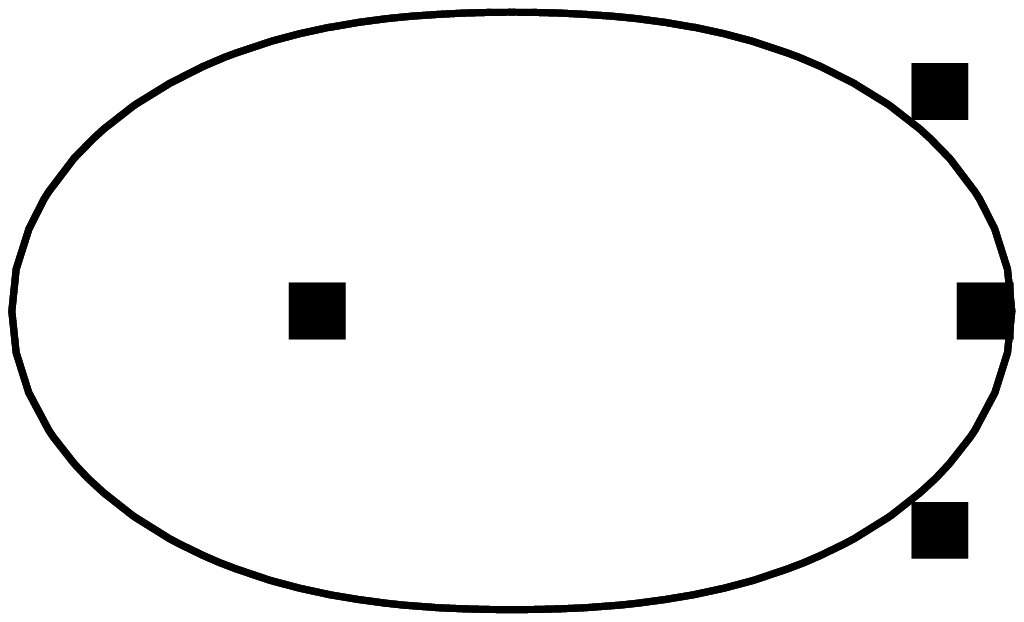}$~~~~~$
&
$~~~~~$\includegraphics[scale=.25]{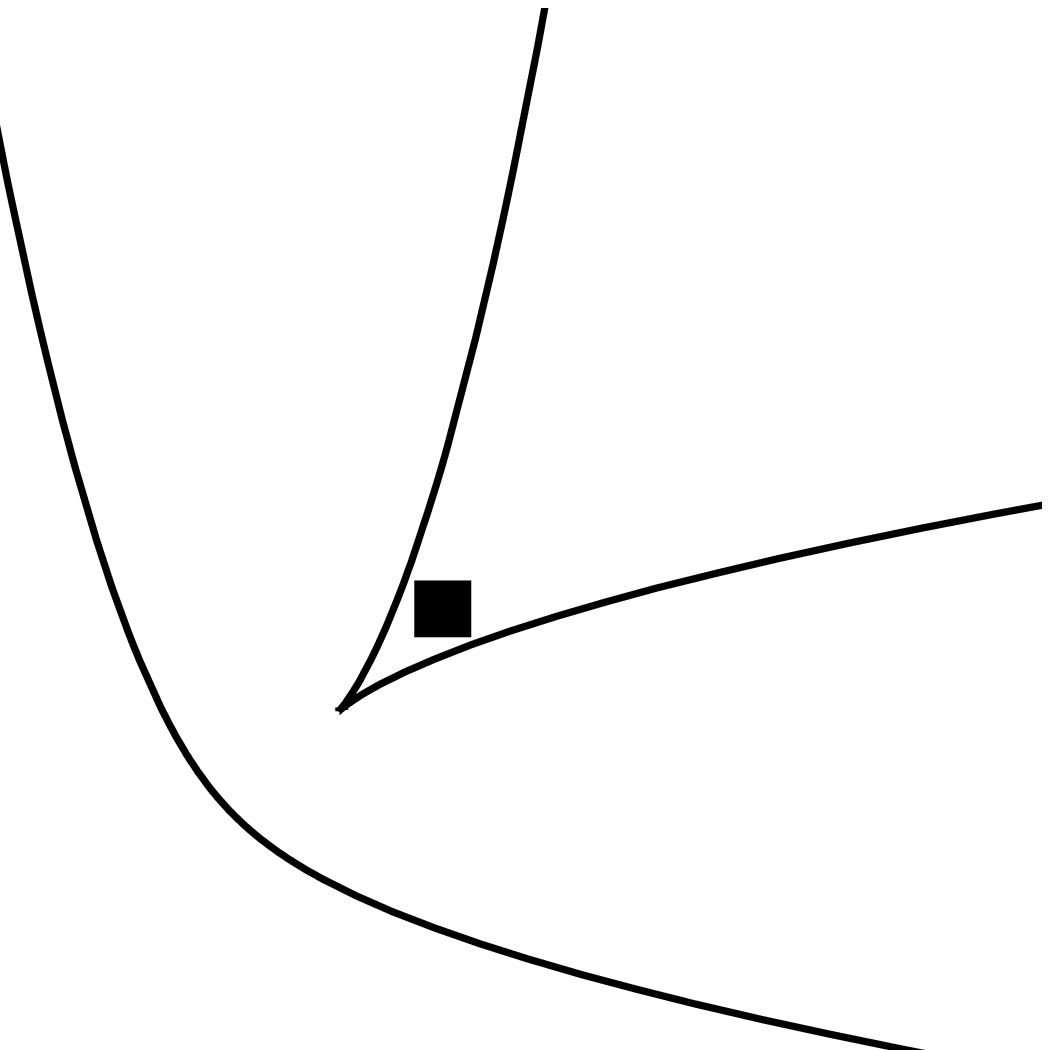}$~~~~~~~~~~$
\includegraphics[scale=.25]{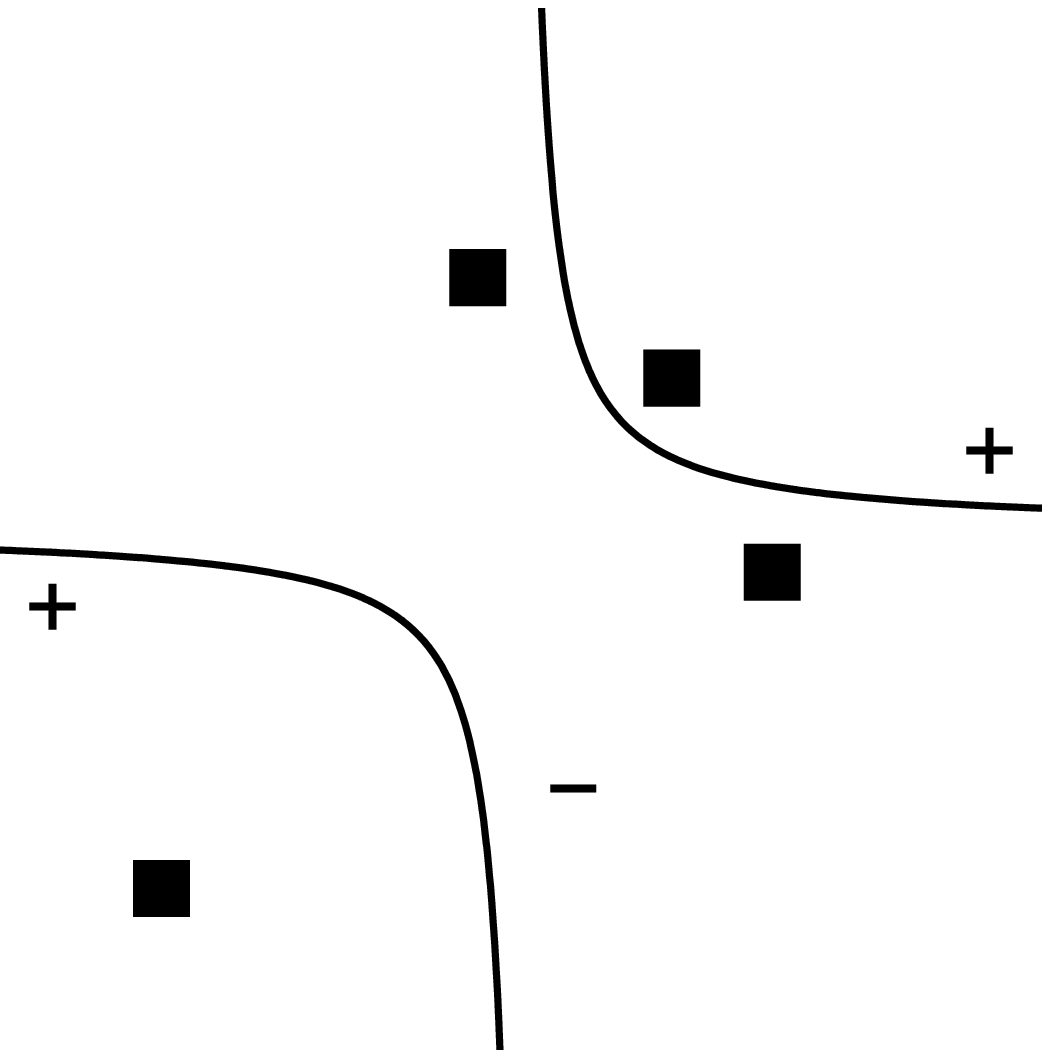}$~~~~~$\\
& \\
\hline
{\scriptsize {\rm {\bf (c)}}}\hskip 10pt{\bf cross} & {\scriptsize {\rm
{\bf (f)}}}\hskip 10pt{\bf cross}\\
& \\
$~~~~~$\includegraphics[scale=.25]{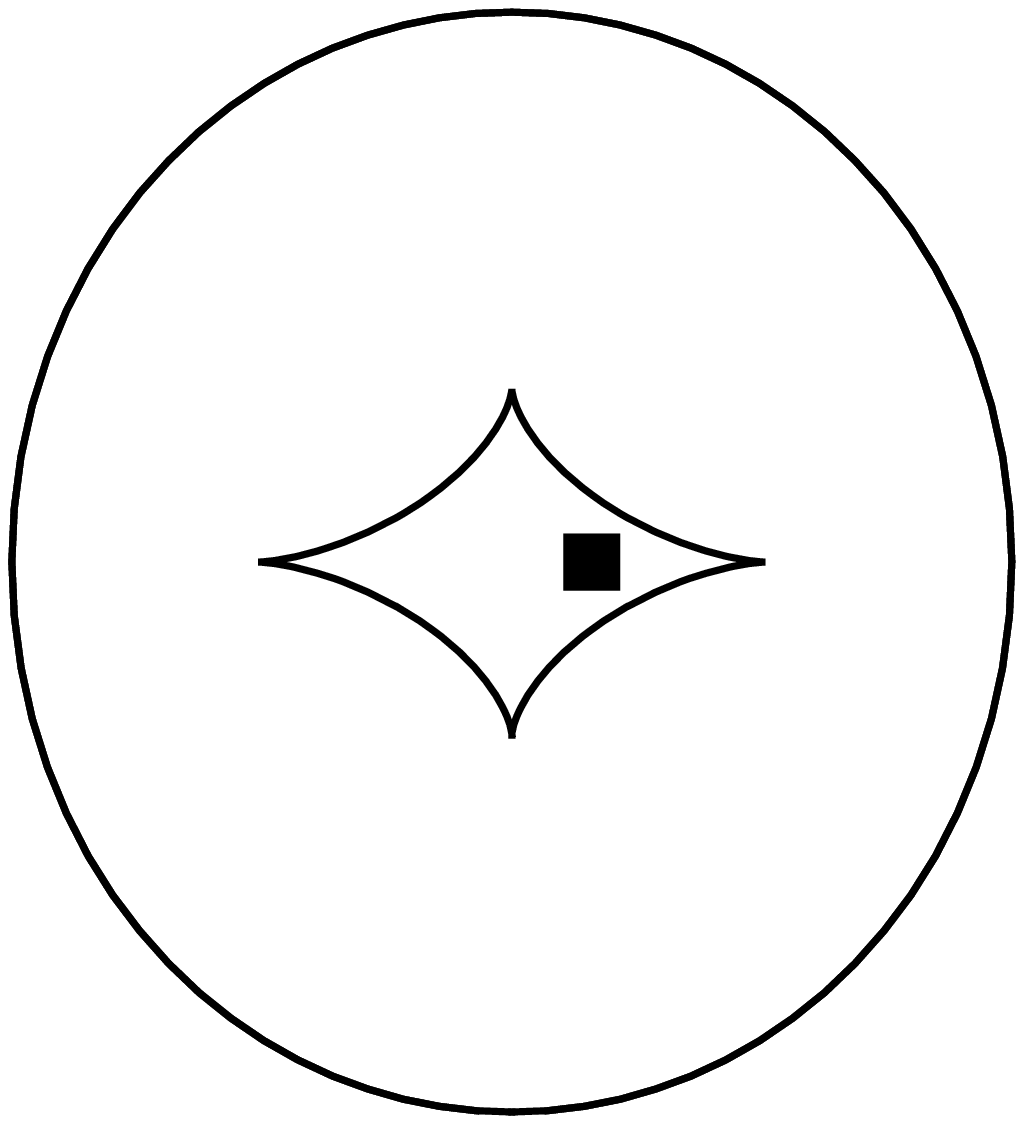}$~~~~~~~~~~$
\includegraphics[scale=.25]{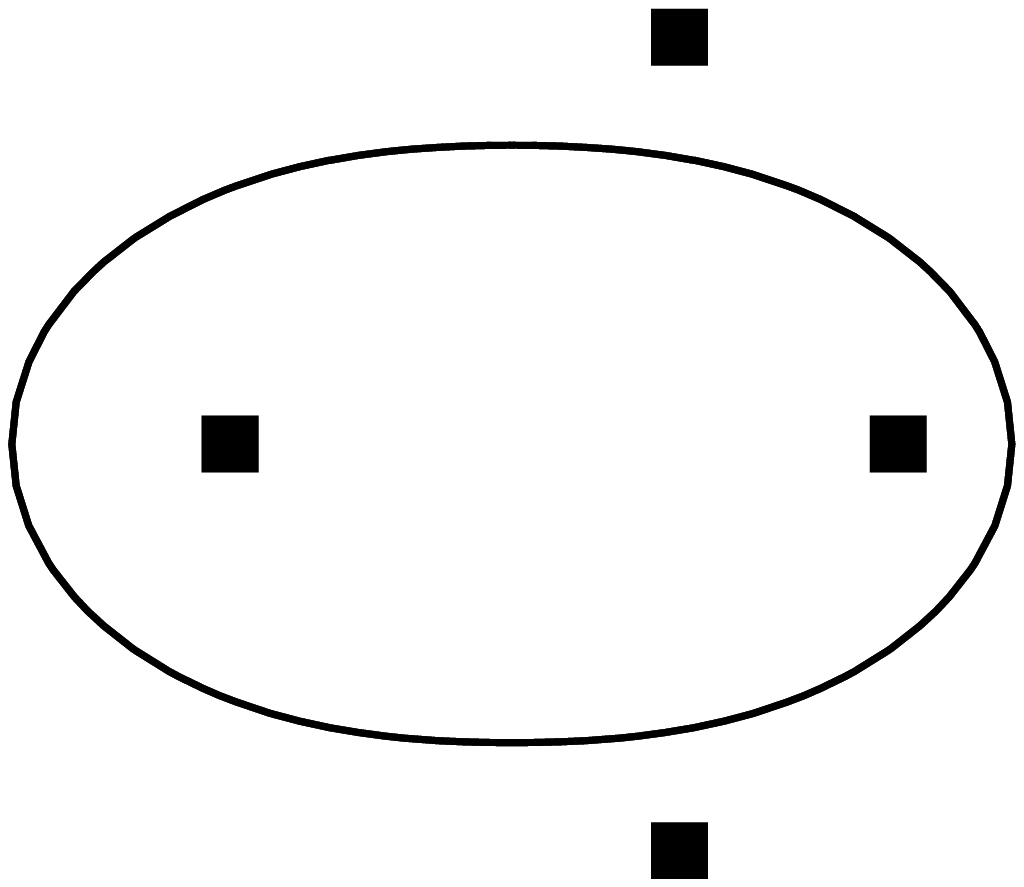}$~~~~~$
&
$~~~~~$\includegraphics[scale=.25]{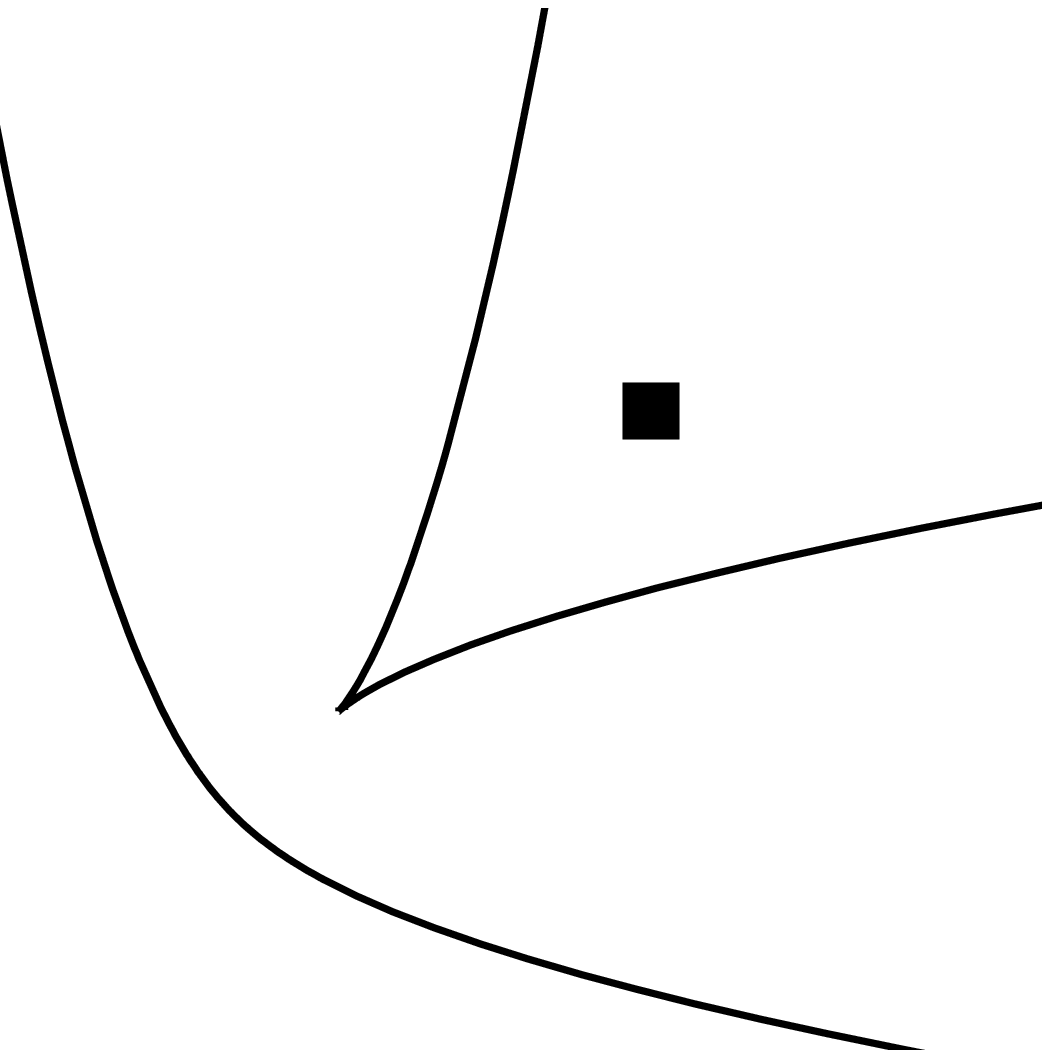}$~~~~~~~~~~$
\includegraphics[scale=.25]{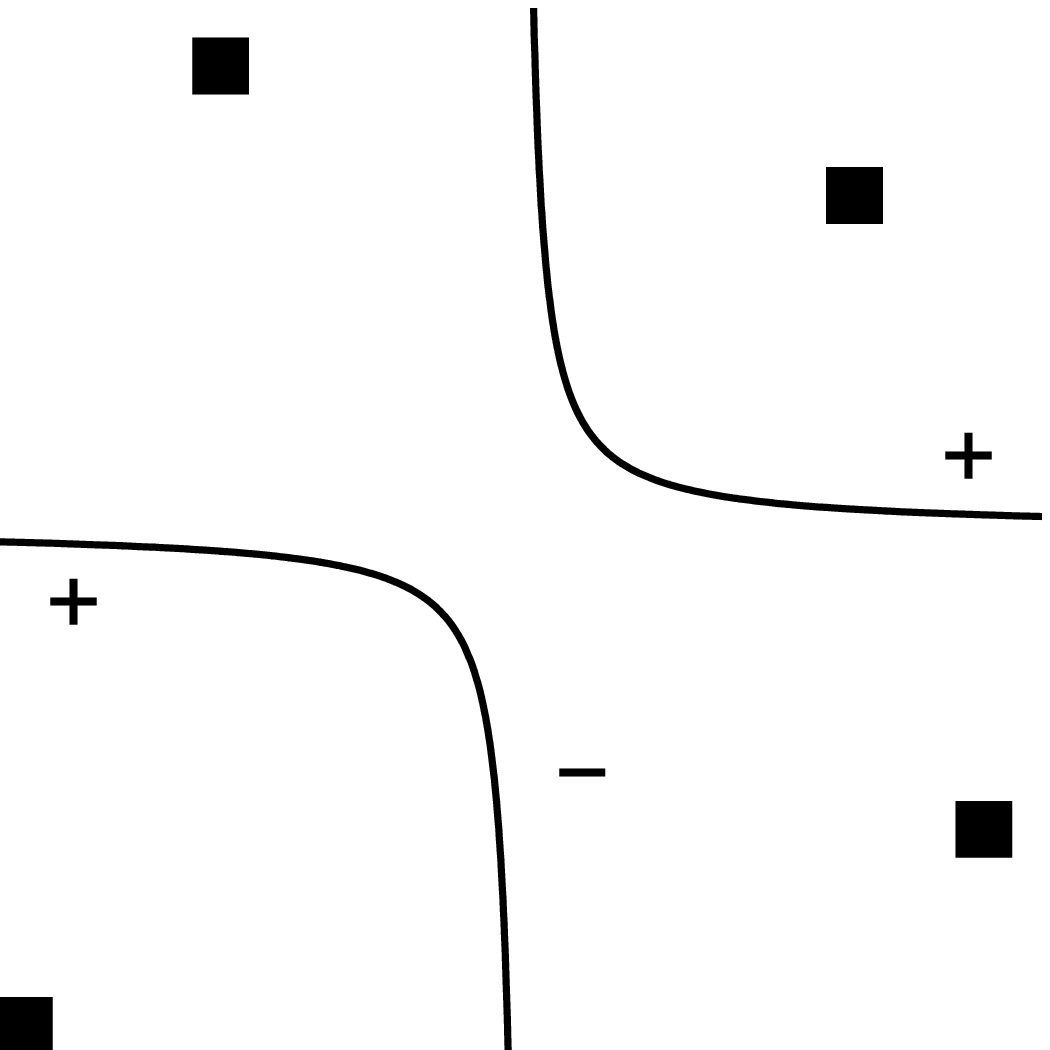}$~~~~~$\\
& \\
\hline
\end{tabular}
\end{center}
\caption{The first column shows fold, cusp, and cross-like configurations due to an SIE with
ellipticity $e = 0.35$ and shear $\gamma = 0.05$ oriented along the horizontal axis (Panels a,b,c).  The second column shows the same configurations due to
the hyperbolic umbilic lensing map $\blm_c$ in Table~\ref{table1} 
with parameter value $c =
0.2$ (Panels d,e,f).  In each panel, the sub-figure on the left depicts 
the caustic
curves with source position (solid box) in the light source plane, while the sub-figure on the
right shows the critical curves with image positions (solid boxes) in the lens plane. 
For the hyperbolic umbilic, image parities have been indicated through
$\pm$ in the given regions.  Note
that the cross-like configuration shown for the SIE is not a perfect cross,
which would be the case if the source were centered inside the
astroid-shaped inner caustic.  Also, for the SIE fold and cusp
configurations, the source is actually located inside (rather than over) the cusped curve of
the astroid.}
\label{figure0b}
\end{figure}

In 2003 and 2005,  Keeton, Gaudi \& Petters \cite{KGP-cusps,KGP-folds} developed a rigorous
theoretical framework showing how the fold and cusp $\mu$-magnification relations
provide a diagnostic for detecting substructure on galactic scales. 
Their analysis employs the $R$-quantity (\ref{R-quantity}) for folds and cusps:
$$
R_{\rm{fold}} \equiv {\mu_{\rm{1}} + \mu_{\rm{2}} \over |\mu_{\rm{1}}| +
|\mu_{\rm{2}}|} = {F_{\rm{1}} - F_{\rm{2}} \over F_{\rm{1}} +
F_{\rm{2}}}\ , \qquad \qquad
R_{\rm{cusp}} \equiv {\mu_{1} + \mu_{2} + \mu_{3} \over |\mu_{1}| +
|\mu_{2}| + |\mu_{3}|} = {F_{1} - F_{2} + F_{3} \over F_{1} + F_{2} +
F_{3}}\ ,
$$
where $F_{i}$ is the observable flux of image $i$ and 
image $2$ has negative parity.  
For a source
sufficiently close to a fold (resp., cusp) caustic, the images will
have a close image pair (resp., close image triplet); see the close
doublets and triplets in Figure~\ref{figure0b}(a,b,d,e).
Theoretically, these images should have vanishing $R_{\rm{fold}}$ and $R_{\rm{cusp}}$
due to the fold and cusp
magnification relations and so nontrivial deviations from zero
would signal the presence of substructure.  In \cite{KGP-cusps,KGP-folds}, it was shown 
that $5$ of the  $12$ fold-image systems and
$3$ of the $4$ cusp-image ones showed evidence for substructure.

The study above would look at a multiple-image system and consider subsets of two and three
images to analyze  $R_{\rm{fold}}$ and $R_{\rm{cusp}}$, respectively.  Such analyses are then  
``local'' when more than three images occur since only two
or three images are studied at a time.
Theorem~\ref{theorem-main} generalizes the above $R$-quantities from folds and cusps
to generic smooth lens systems that exhibit swallowtail, elliptic umbilic, 
and hyperbolic umbilic singularities.  The $R$-quantities resulting from these higher-order singularities allow one to
consider {\it four} images at a time and so are more global than the fold and cusp relations
in terms of how many images are incorporated.  The singularity that is most applicable
to observed quadruple-images produced by the lensing of quasars is 
the hyperbolic umbilic (cf. Figure~\ref{figure0b}).  The associated
$R$-quantity is
$$
R_{\rm{h.u.}} \equiv {\mu_{1} + \mu_{2} + \mu_{3} + \mu_{4} \over
|\mu_{1}| + |\mu_{2}| + |\mu_{3}| + |\mu_{4}|} = {F_{1} - F_{2} + F_{3} -
F_{4} \over F_{1} + F_{2} + F_{3} + F_{4}}\ , \nonumber
$$
where images $2$ and $4$ have negative parity.

We now illustrate the hyperbolic umbilic quantity
$R_{\rm{h.u.}}$ using a well-known model for a galaxy
lens, namely,  a singular isothermal ellipsoid (SIE) lens.  The SIE
lens potential and surface mass density are given respectively as follows:
$$
\psi(r,\varphi)=rF(\varphi) - {\gamma \over 2}r^{2}{\rm cos}\,2\varphi, \qquad
\kappa(r,\varphi)={G(\varphi) \over 2r},
$$
where $F(\varphi)$ and $G(\varphi)$ satisfy $G(\varphi) = F(\varphi) +
F''(\varphi)$ by Poisson's equation, and are given explicitly by
\beq
G(\varphi)&=&{R_{\rm ein} \over \sqrt{1-\varepsilon\,{\rm cos}\,2\varphi}}\
,\nonumber \\
F(\varphi)&=&{R_{\rm ein} \over \sqrt{2\varepsilon}}\left[{\rm cos}\,\varphi
\,{\rm tan}^{-1}\left({\sqrt{2\varepsilon} \,{\rm cos}\,\varphi \over
\sqrt{1-\varepsilon\,{\rm cos}\,2\varphi}}\right)+{\rm sin}\,\varphi \,{\rm
tanh}^{-1}\left({\sqrt{2\varepsilon} \,{\rm sin}\,\varphi \over
\sqrt{1-\varepsilon\,{\rm cos}\,2\varphi}}\right)\right], \nonumber
\eeq
where $R_{{\rm ein}}$ is the angular Einstein ring radius.
The parameter $\varepsilon$ is related to the axis ratio $q$ by
$\varepsilon = (1-q^2)/(1+q^2)$, and should not be confused with the
ellipticity $e = 1- q$.  The cusp at $\varphi = 0$ is given by
\beq
\label{SIE:cusp-gen}
\bby_{\rm cusp} = \left({2\gamma F(0)+(1+\gamma) F''(0) \over 1-\gamma}\,
,\, 0\right)\ \cdot
\eeq

Using the
{\it Gravlens} software by Keeton 2001 \cite{Keeton}, we now solve the 
SIE lens equation
for sources on the positive horizontal axis in the four-image
region of the light source plane, and compute $R_{{\rm h.u.}}$.
Let the SIE have ellipticity $e = 0.35$ and shear $\gamma = 0.05$ oriented along the horizontal axis; both
of these values are observationally motivated \cite{KGP-folds,KGP-cusps}.  
 Figure~\ref{figure0b}(a,b,c) shows three important image
configurations for the SIE:  the fold, when the source lies close to a
fold arc and produces a close pair of images about a critical curve; 
the cusp, when the source lies close to a cusp caustic
and produces a close triplet of images about a critical curve; the
cross-like configuration of four images, when the source sits nearer to the center of the astroid-shaped
inner caustic curve.  Figure~\ref{figure0b}(d,e,f) illustrates how the SIE image configurations
are similar to those of the hyperbolic umbilic lensing map
$\blm_c$ given in Table~\ref{table1}.  See Appendix~\ref{Appendix:hu}
for more on the hyperbolic umbilic $\blm_c$.

We now look at the behavior of $R_{{\rm fold}}$, $R_{{\rm cusp}}$, and
 $R_{{\rm h.u.}}$ for an SIE.
Table~\ref{table2} compares $R_{\rm fold}$ and $R_{\rm h.u.}$ for a source
approaching a fold arc diagonally from the center of the astroid-shaped
inner caustic; see Figure~\ref{figure0b}(a).  The fold point where the
diagonal intersects the
fold arc is at
$$
\bby_{\rm fold} \approx (0.14055 R_{\rm ein},\ 0.14055 R_{\rm ein})\ .
$$
As the source at $\bby$ approaches $\bby_{\rm fold}$ along the diagonal, 
the values in Table~\ref{table2} show that $R_{\rm fold}$ and $R_{\rm h.u.}$
each approach the ideal value of $0$, and that $R_{\rm h.u.}$ approaches
$R_{\rm fold}$ from {\it above}.  
The reason for this is as 
follows:  From Figure~\ref{figure0b}(a) we see that  there are two pairs of images in
a hyperbolic umbilic configuration: the fold image doublet  straddling the critical
curve, and whose two images we denote by $d_{1}, d_{2}$, and the pair
consisting of the outer two images, which we denote by $o_{1}, o_{2}$.  
The quantity $R_{\rm h.u.}$ then becomes
$$
R_{\rm{h.u.}} = {|\mu_{d_{1}}| - |\mu_{d_{2}}| + |\mu_{o_{1}}| -
|\mu_{o_{2}}| \over |\mu_{d_{1}}| + |\mu_{d_{2}}| + |\mu_{o_{1}}| +
|\mu_{o_{2}}|}\ \cdot
$$
As the source approaches $\bby_{\rm fold}$ along the diagonal,
Table~\ref{table2} shows that the quantities $|\mu_{d_{1}}| - |\mu_{d_{2}}|$ and $|\mu_{o_{1}}| -
|\mu_{o_{2}}|$ stay roughly constant, though the individual magnifications
vary.  In addition, near the fold, we see that $|\mu_{d_{1}}| + |\mu_{d_{2}}|$
dominates $|\mu_{o_{1}}| + |\mu_{o_{2}}|$, causing the
denominator of $R_{\rm h.u.}$ to approach $|\mu_{d_{1}}| +
|\mu_{d_{2}}|$, which is the denominator of $R_{\rm fold}$.  
This leads to
$$
R_{\rm{h.u.}} \approx 
 {|\mu_{d_{1}}| - |\mu_{d_{2}}| \over |\mu_{d_{1}}| + |\mu_{d_{2}}|} +
 {|\mu_{o_{3}}| -
|\mu_{o_{4}}| \over |\mu_{d_{1}}| + |\mu_{d_{2}}|}
\ge {|\mu_{d_{1}}| - |\mu_{d_{2}}| \over |\mu_{d_{1}}| + |\mu_{d_{2}}|}
= R_{\rm fold}\ .
$$
The net
effect is  that $R_{\rm{h.u.}}$
approaches 
$R_{\rm fold}$ from above (at least for the path along the diagonal).
Furthermore, since the quantity $|\mu_{d_{1}}| + |\mu_{d_{2}}|$ diverges, we see that both $R_{\rm{h.u.}}$ and
$R_{\rm fold}$ approach 
the magnification relation value of $0$.

\begin{table}[tbh]
\centering
\vskip 6pt
\begin{tabular}{| c | c | c | c | c | c | c |}
\hline
$~~~$Source$~~~$ & $~~~~~$$R_{{\rm fold}}$$~~~~~$ & $~~~~~$$R_{{\rm
h.u.}}$$~~~~~$ & $~~$$|\mu_{d_{1}}| - |\mu_{d_{2}}|$$~~$ &
$~~$$|\mu_{o_{1}}| - |\mu_{o_{2}}|$$~~$ & $~~$$|\mu_{d_{1}}| +
|\mu_{d_{2}}|$$~~$ & $~~$$|\mu_{o_{1}}| + |\mu_{o_{2}}|$$~~$
\\ [0.5ex]
\hline
\,\,\,(0.10$R_{\rm ein}$\,,\,0.10$R_{\rm ein}$)\,\,\, & 0.14 & 0.19 & 1.22
& 1.21 & 8.51 & 4.35\\
\,\,\,(0.11$R_{\rm ein}$\,,\,0.11$R_{\rm ein}$)\,\,\, & 0.13 & 0.18 & 1.22
& 1.22 & 9.64 & 4.28\\
\,\,\,(0.12$R_{\rm ein}$\,,\,0.12$R_{\rm ein}$)\,\,\, & 0.11 & 0.15 & 1.22
& 1.22 & 11.55 & 4.21\\
\,\,\,(0.13$R_{\rm ein}$\,,\,0.13$R_{\rm ein}$)\,\,\, & 0.08 & 0.12 & 1.22
& 1.22 & 15.83 & 4.15\\
\,\,\,(0.14$R_{\rm ein}$\,,\,0.14$R_{\rm ein}$)\,\,\, & 0.02 & 0.04 & 1.21
& 1.23 & 65.17 & 4.081\\
\,\,\,(0.1405$R_{\rm ein}$\,,\,0.1405$R_{\rm ein}$)\,\,\, & 0.008 & 0.015
& 1.21 & 1.23 & 156.80 & 4.078\\
\hline
\end{tabular}
\caption{The quantities $R_{{\rm h.u.}}$ and $R_{{\rm fold}}$ for an SIE
with $e = 0.35$ and $\gamma = 0.05$ oriented along the horizontal axis.  The source approaches the fold point
$
\bby_{\rm fold} \approx (0.14055 R_{\rm ein},\ 0.14055 R_{\rm ein})
$
diagonally from the center of the astroid-shaped inner caustic.  
The quantity  $|\mu_{d_{1}}| - |\mu_{d_{2}}|$ is the difference in the magnifications of the
images in the close doublet, while $|\mu_{o_{1}}| - |\mu_{o_{2}}|$ is the difference
for the remaining two outer images; cf. Figure~\ref{figure0b}(a).
}
\label{table2}
\end{table}

Table~\ref{table3} compares $R_{{\rm h.u.}}$ with $R_{{\rm cusp}}$ for 
a source approaching a cusp along the horizontal axis from the center 
of the astroid-shaped caustic curve; see Figure~\ref{figure0b}(b,c).  
For these values of the
ellipticity and shear, we see from (\ref{SIE:cusp-gen}) that the two cusps on
the horizontal axis are located at
\beq
\label{SIE:cusp}
\bby^\pm_{\rm cusp}\approx (\pm
0.48R_{\rm ein},0)\ .
\eeq 
The table shows that as the source approaches $\bby^{+}_{\rm cusp}$ along 
the horizontal axis, the quantity $R_{{\rm h.u.}}$
approaches $R_{{\rm fold}}$ from {\it below}. In other words, 
$R_{{\rm h.u.}}$ is smaller than $R_{{\rm fold}}$.  To see why this
happens,
consider the triplet of sub-images in
 Figure~\ref{figure0b}(b), which we denote by $t_{1}, t_{2}, t_{3}$,
and the extra outer image, denote by $o$. With this notation,
$$
R_{\rm{h.u.}} = {|\mu_{t_{1}}| - |\mu_{t_{2}}| + |\mu_{t_{3}}| -
|\mu_{o}| \over |\mu_{t_{1}}| + |\mu_{t_{2}}| + |\mu_{t_{3}}| +
|\mu_{o}|}\ \cdot
$$
As the source approaches $\bby^{+}_{\rm cusp}$ along the horizontal axis,
the values in Table~\ref{table3} of the cusp relation 
$|\mu_{t_{1}}| -
|\mu_{t_{2}}| + |\mu_{t_{3}}|$ 
are {\it positive}.  The inclusion of the outer, negative parity
magnification $\mu_{o}$ then subtracts from that positive value, yielding
$$
\left(|\mu_{t_{1}}| - |\mu_{t_{2}}| + |\mu_{t_{3}}|\right) - |\mu_{o}| \leq
|\mu_{t_{1}}| 
- |\mu_{t_{2}}| + |\mu_{t_{3}}|\ ,
$$
which implies that
$$
R_{\rm h.u.} \leq R_{\rm cusp}\ .
$$
Furthermore,
Table~\ref{table3} shows that $|\mu_{o}|$  grows fainter faster than the value of the
signed magnification of the triplet, which yields
$$
|\mu_{t_{1}}| + |\mu_{t_{2}}| + |\mu_{t_{3}}|\, \ge \,
|\mu_{t_{1}}| - |\mu_{t_{2}}| + |\mu_{t_{3}}| \gg |\mu_{o}|\ .
$$
In other words, as the source approaches $\bby^{+}_{\rm cusp}$ along the horizontal axis,
the contribution of the outer image $|\mu_{o}|$ to the numerator and denominator
of $R_{\rm h.u.}$ becomes
negligible.  The net effect, at least for the given horizontal axis approach, is that
$R_{\rm h.u.}$ and $R_{\rm cusp}$ converge, with $R_{\rm h.u.}$ approaching $R_{\rm cusp}$ from below
as they both approach the magnification relation value of $0$.

\begin{table}
\centering
\vskip 6pt
\begin{tabular}{| c | c | c | c | c | c |}
\hline
$~~~$Source$~~~$ & $~~~$$R_{{\rm cusp}}$$~~~$ & $~~~$$R_{{\rm h.u.}}$$~~~$
& \,\,\,$|\mu_{t_{1}}| + |\mu_{t_{2}}| + |\mu_{t_{3}}|$\,\,\, &
\,\,\,$|\mu_{t_{1}}| - |\mu_{t_{2}}| + |\mu_{t_{3}}|$\,\,\,&
$~~~$$|\mu_{o_{1}}|$$~~~$
\\ [0.5ex]
\hline
\,\,\,(0\,,\,0) $({\rm center})$ & 0.52 & 0.23 & 8.49 & 4.46 & 2.02\\
\,\,\,(0.10$R_{\rm ein}$\,,\,0)\,\,\, & 0.41 & 0.22 & 9.58 & 3.94 &
1.49\\
\,\,\,(0.15$R_{\rm ein}$\,,\,0)\,\,\, & 0.36 & 0.21 & 10.57 & 3.76 &
1.29\\
\,\,\,(0.20$R_{\rm ein}$\,,\,0)\,\,\, & 0.30 & 0.19 & 12.02 & 3.61 &
1.12\\
\,\,\,(0.25$R_{\rm ein}$\,,\,0)\,\,\, & 0.25 & 0.17 & 14.20 & 3.48 &
0.98\\
\,\,\,(0.30$R_{\rm ein}$\,,\,0)\,\,\, & 0.19 & 0.14 & 17.71 & 3.38 &
0.85\\
\,\,\,(0.35$R_{\rm ein}$\,,\,0)\,\,\, & 0.14 & 0.10 & 24.10 & 3.30 &
0.74\\
\,\,\,(0.40$R_{\rm ein}$\,,\,0)\,\,\, & 0.08 & 0.07 & 39.02 & 3.23 &
0.64\\
\,\,\,(0.45$R_{\rm ein}$\,,\,0)\,\,\, & 0.03 & 0.02 & 111.5 & 3.18 &
0.55\\
\hline
\end{tabular}
\caption{The quantities $R_{{\rm h.u.}}$ and $R_{{\rm cusp}}$ for an SIE
with $e = 0.35$ and $\gamma = 0.05$ oriented along the horizontal axis.  The source approaches the cusp point
$
\bby_{\rm cusp}^{+} \approx (0.48 R_{\rm ein},0)
$
along the horizontal axis from the center of the astroid-shaped inner
caustic.  
The quantity  $|\mu_{t_{1}}| - |\mu_{t_{2}}| + |\mu_{t_{3}}|$ is the
signed magnification 
sum of the cusp triplet, while $|\mu_{o}|$ 
is the magnification of the outer image; see Figure~\ref{figure0b}(b).
}
\label{table3}
\end{table}

Finally, though
$R_{{\rm h.u.}}$ can approximate 
$R_{\rm fold}$ and $R_{\rm cusp}$ for fold image doublets and cusp image triplets, resp.,
{\it the hyperbolic umbilic magnification relation has a more global reach} in terms of the number of images included.  This is because $R_{{\rm h.u.}}$ also
applies directly to image configurations that are neither
close doublets nor triplets; e.g., to cross-like configurations as in Figure~\ref{figure0b}(c).
For instance, it was determined in \cite{KGP-cusps}
that to
satisfy the relation $|R_{\rm cusp}| < 0.1$ at $99\%$ confidence, the
opening angle must be $\theta \stackrel{{\textstyle <}}{_{\textstyle
\sim}} 30^{\circ}$.  By opening angle we mean the angle of the polygon
spanned by the three images in the cusp triplet, measured from the
position of the lens galaxy, which in our case, is centered at the origin
in the lens plane.  For the SIE cross-like configuration shown in
Figure~\ref{figure0b}(c), the opening angle is $\theta \approx
140^{\circ}$; a
perfect cross, which would be the case if the source were centered inside
the astroid-shaped inner caustic, has $\theta = 180^{\circ}$.  In other words,
to satisfy the cusp relation reasonably well, the cusp triplet
must be quite tight as, for example, in the SIE cusp triplet shown in
Figure~\ref{figure0b}(b).  By contrast, the quantity $R_{\rm h.u.}$
applies even for values $\theta \gg 30^{\circ}$. 
(In Table~\ref{table3} note how $R_{\rm h.u.}$ is smaller than $R_{\rm cusp}$
for source positions closer to the center $(0,0)$, which yield more cross-like 
image configurations.)

A more detailed study of the properties of $R_{{\rm h.u.}}$ would
require a separate paper and involve a Monte Carlo analysis similar to that employed in
\cite{KGP-folds,KGP-cusps} 
to study $R_{{\rm cusp}}$ and $R_{{\rm fold}}$.  
Such an analysis of $R_{\rm h.u.}$ would be applicable to the currently
known $26$ four-image
lens systems (courtesy of the CASTLES lens sample \footnote{http://www.cfa.harvard.edu/castles.})

\section{Conclusion}
\label{Conclusion}
We showed that magnification invariants
hold universally not only for folds and cusps, but also for 
swallowtails, elliptic umbilics, and hyperbolic umbilics.  Specifically, for a source anywhere in the
four-image region close to each of these caustic singularities, the total signed
magnification is identically zero.  This result is universal in that it
does not depend on the class of lens models used, and is thus an extension
of the familiar fold and cusp magnification sum relations.  We proved
that these relations hold for generic one-parameter families of lensing maps with the elliptic umbilic and
hyperbolic umbilic singularities. We also established the relations for 
generic one-parameter families of general mappings, which need not relate
to lensing,  for the swallowtail, elliptic umbilic,
and hyperbolic umbilic singularities.   We emphasized that these
universal sum relations are {\it geometric} invariants, because they
are sums of reciprocals of Gaussian curvatures at critical points.

The relevance of these higher order magnification invariants 
to the study of dark substructure in galaxies was shown.
Using a singular isothermal ellipsoidal model of a galaxy lens, 
we constructed a lensing observable for the
hyperbolic umbilic, denoted $R_{\rm h.u.}$, and compared it to
the well-known fold and cusp analogues, $R_{\rm fold}$ and $R_{\rm cusp}$.
These three observables approach their magnification relation value of
$0$ the closer a source gets to a caustic. 
Significant deviations from this value indicate that the lens in question
is not smooth, but has some kind of substructure on the scale of the image
separations. 
We showed that $R_{\rm h.u.}$ is a more global quantity than $R_{\rm fold}$ and $R_{\rm cusp}$
because $R_{\rm h.u.}$ considers four lensed images simultaneously,
while $R_{\rm fold}$ considers two and $R_{\rm cusp}$ three.
At the same time, we showed that, as the source approached a fold arc
or cusp point, the
quantity $R_{\rm h.u.}$ approaches $R_{\rm fold}$ or $R_{\rm cusp}$, respectively.  
More
stringent conclusions about the properties of $R_{\rm h.u.}$ await a full
Monte Carlo analysis, akin to the one employed recently to examine $R_{\rm
fold}$ and $R_{\rm cusp}$.

\section{Acknowledgments}
\noindent The authors thank the referee for a careful reading of the manuscript, and especially for pointing out the inverse Jacobian Theorem.  ABA acknowledges the hospitality of the Petters Research Institute, where this work was conceived.  He also thanks A. Teguia for helpful discussions.  AOP acknowledges the support of NSF Grant DMS-0707003.


\appendix

\section{Proof of the Main Theorem for Lensing Maps}
\label{Appendix:ProofLensing}
\subsection{Elliptic Umbilic}
\label{Appendix:eu}

\begin{figure}[ht]
\begin{center}
\includegraphics[scale=.5]{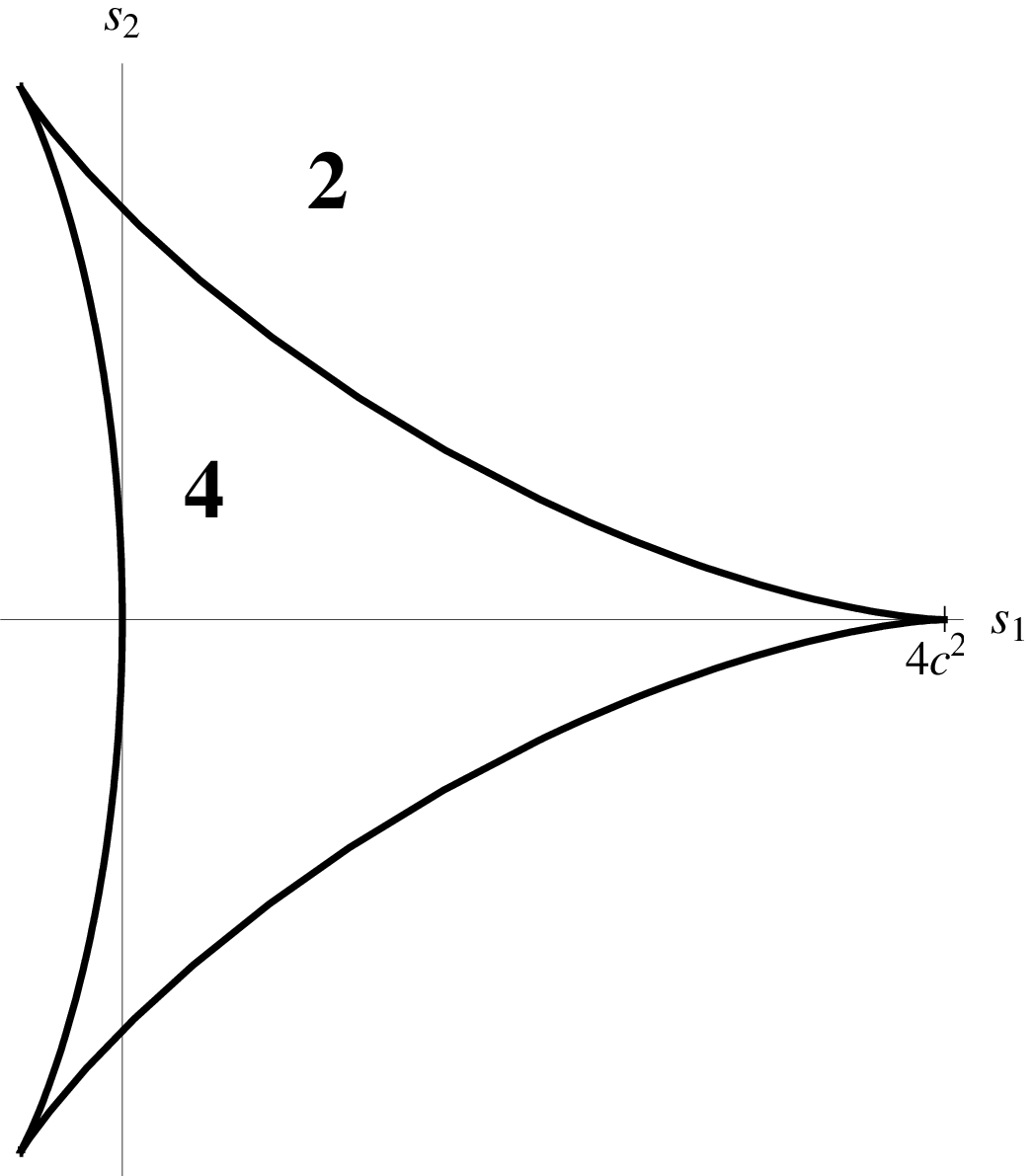}
\end{center}
\caption{Caustic curve for an elliptic umbilic lensing
map $\blm_c$ in equation (\ref{lenseqn}) or Table~\ref{table1}.  The number of lensed
images for sources in their respective regions is indicated.}
\label{Figure3}
\end{figure}

The derivation of the quantitative form of the lensing map in the neighborhood of an elliptic umbilic critical point, can be found in \cite[Chap. 6]{Sch-EF}.  The resulting map is
\beq
\label{lenseqn}
s_{1}&=&u^{2}-v^{2}\ ,\nonumber \\
s_{2}&=&-2uv+4cv\ .
\eeq  

\noindent Here ${\bf y} = (s_{1},s_{2})$ is the location of a source on the light source plane $S$, $(u,v)$ the location of a corresponding lensed image on the lens plane $L$, and $c$ is a constant which signifies that the lens mapping under consideration is one in a one-parameter family of lens mappings.  The magnification of an image $(u,v)$ is
\beq
\label{mag}
\left({\rm det}({\rm Jac}\ {\bf s})\right)^{-1}(u,v)&=&{1 \over 8cu - 4(u^{2}+v^{2})}\ \cdot
\eeq

\noindent A parameter representation of the critical curve is 
\beq
u&=&c(1+\cos{\phi})\ ,\nonumber \\
v&=&c\sin{\phi}\ .\nonumber
\eeq

\noindent Inserting these into the lens equation (\ref{lenseqn}) gives the caustic curve:
\beq
s_{1}&=&2c^{2}\cos{\phi}\,(1+\cos{\phi})\ ,\nonumber \\
s_{2}&=&2c^{2}\sin{\phi}\,(1-\cos{\phi})\ .\nonumber
\eeq

\noindent (Note that our notation differs from that of \cite{Sch-EF}.)  The caustic curve is shown in Figure~\ref{Figure3}.  The region inside the closed caustic curve constitutes the four-image region.  We now show that for all sources inside this region, the total signed magnification is identically zero.

We begin by considering a special case: sources in the four-image region lying on the horizontal axis; that is, with $s_{2}=0$.  In this case the lens equation (\ref{lenseqn}) is solvable.  The lensed images are $\left(\pm \sqrt{s_{1}}, 0\right), \left(2 c,\pm \sqrt{4 c^{2} - s_{1}}\,\right),$ where we note that all four of these images are real because $0 < s_{1} < 4c^{2}$ inside the caustic curve (see Figure~\ref{Figure3}).  The total signed magnification, obtained by inserting each of these four images into (\ref{mag}) and summing over, will be zero.  For the remainder of this section, therefore, we can restrict ourselves to sources $(s_{1},s_{2})$ inside the caustic curve with $s_{2} \not = 0$.  Note from the second lens equation (\ref{lenseqn}) that $s_{2} \not = 0$ forces the $v$-coordinate of each lensed image to be nonzero: $v_{i} \not = 0$.  This fact will prove useful below.

\vskip 12pt

Let $(s_{1},s_{2} \not = 0)$ denote the position of an arbitrary source lying off the $s_{1}$-axis inside the caustic curve.  Let $(u_{i},v_{i})$ denote the corresponding lensed images.
\noindent The total signed magnification $\mu$ at $(s_{1},s_{2})$ is
\beq
\label{totalmag}
\mu(u_{i},v_{i})={1 \over 8cu_{1} - 4(u_{1}^{2} + v_{1}^{2})} + {1 \over 8cu_{2} - 4(u_{2}^{2}+v_{2}^{2})} + {1 \over 8cu_{3} - 4(u_{3}^{2}+v_{3}^{2})} + {1 \over 8cu_{4} - 4(u_{4}^{2}+v_{4}^{2})}\ \cdot
\eeq 

\noindent Our goal is to show that this sum is in fact identically zero.  Let us begin by eliminating $u$ from the lens equation (\ref{lenseqn}) to obtain a (depressed) quartic in $v$:
\beq
\label{vquartic}
v^{4} + (s_{1} - 4c^{2})v^{2} + 2cs_{2}v - {s_{2}^{2}\over4}=0\ .
\eeq

\noindent Knowing that this quartic must factor as
\beq
\label{roots}
(v - v_{1})(v - v_{2})(v - v_{3})(v - v_{4})=0\ ,
\eeq 

\noindent we expand (\ref{roots}) and equate its coefficients to those of (\ref{vquartic}).  As a result we obtain four equations involving the $v_{i}$:
\beq 
\label{eqn1}
{\bf v_{1} + v_{2} + v_{3} + v_{4}}&=&{\bf 0}\ ,\\
\label{eqn2}
{\bf v_{1}v_{2}+v_{1} v_{3} + v_{1} v_{4} + v_{2} v_{3} + v_{2} v_{4} + v_{3} v_{4}}&=&{\bf s_{1} - 4 c^{2}}\ ,\\
\label{eqn3}
{\bf v_{1} v_{2} v_{3} + v_{1} v_{2} v_{4} + v_{1} v_{3} v_{4} + v_{2} v_{3} v_{4}}&=&{\bf -2 c s_{2}}\ ,\\
\label{eqn4}
{\bf v_{1} v_{2} v_{3} v_{4}}&=&{\bf -{s_{2}^{2}\over 4}}\ \cdot
\eeq

\noindent Next, we use (\ref{lenseqn}) to express each $u_{i}$ in terms of $v_{i}$, bearing in mind that all $v_{i} \not = 0$:
\beq
\label{uintermsofv}
u_{i}(v_{i}) = {-s_{2} + 4 c v_{i} \over 2 v_{i}}\ \cdot
\eeq

\noindent Our procedure is to insert (\ref{uintermsofv}) into the total magnification (\ref{totalmag}), thereby obtaining an expression involving only the $v_{i}$, $\mu = \mu(v_{i})$, and to then simplify this expression to zero using (\ref{eqn1})--(\ref{eqn4}).

\vskip 12pt

Unfortunately, the equation $\mu = \mu(v_{i})$, when written over a common denominator, is quite unwieldy.  To simplify proceedings, we factor the numerator in terms of powers of $s_{2}$:
\beq
\label{s_{2}^6term}
s_{2}^{6}\,\left(v_{1}^{2} + v_{2}^{2} + v_{3}^{2} + v_{4}^{2}\right)\ ,
\eeq
\beq
\label{s_{2}^5term}
-4c\,s_{2}^{5}\,\left(v_{1}^{2} v_{2} + v_{1} v_{2}^{2} + v_{1}^{2} v_{3} + v_{2}^{2} v_{3} + v_{1} v_{3}^{2} + v_{2} v_{3}^{2} + v_{1}^{2} v_{4} + v_{2}^{2} v_{4} + v_{3}^{2} v_{4} + v_{1} v_{4}^{2} + v_{2} v_{4}^{2} + v_{3} v_{4}^{2}\right)\ ,
\eeq
\beq
\label{s_{2}^4term}
4\,s_{2}^{4}\,\left(\left(v_{1}^{4} v_{2}^{2}\right.\right.&+&\left.\left.v_{1}^{2} v_{2}^{4} + v_{1}^{4} v_{3}^{2} + v_{2}^{4} v_{3}^{2} + v_{1}^{2} v_{3}^{4} + v_{2}^{2} v_{3}^{4} + v_{1}^{4} v_{4}^{2} + v_{2}^{4} v_{4}^{2} + v_{3}^{4} v_{4}^{2} + v_{1}^{2} v_{4}^{4} + v_{2}^{2} v_{4}^{4} + v_{3}^{2} v_{4}^{4}\right.\right)\nonumber \\
&+&\left.4 c^{2}(v_{1}^{2} v_{2} v_{3} + v_{1} v_{2}^{2} v_{3} + v_{1} v_{2} v_{3}^{2} + v_{1}^{2} v_{2} v_{4} + v_{1} v_{2}^{2} v_{4} + v_{1}^{2} v_{3} v_{4} + v_{2}^{2} v_{3} v_{4}\right.\nonumber \\
&+&\left.\left.v_{1} v_{3}^{2} v_{4} +v_{2} v_{3}^{2} v_{4} + v_{1} v_{2} v_{4}^{2} + v_{1} v_{3} v_{4}^{2} + v_{2} v_{3} v_{4}^{2}\right)\right) \ ,
\eeq
\beq
\label{s_{2}^3term}
-16 c\,s_{2}^{3}\,\left(\left(v_{1}^{4} v_{2}^{2} v_{3}\right.\right.&+&\left.\left.v_{1}^{2} v_{2}^{4} v_{3} + v_{1}^{4} v_{2} v_{3}^{2} + v_{1} v_{2}^{4} v_{3}^{2} + v_{1}^{2} v_{2} v_{3}^{4} + v_{1} v_{2}^{2} v_{3}^{4} + v_{1}^{4} v_{2}^{2} v_{4} + v_{1}^{2} v_{2}^{4} v_{4} + v_{1}^{4} v_{3}^{2} v_{4} + v_{2}^{4} v_{3}^{2} v_{4}\right.\right.\nonumber \\
&+&\left.\left.v_{1}^{2} v_{3}^{4} v_{4} + v_{2}^{2} v_{3}^{4} v_{4} + v_{1}^{4} v_{2} v_{4}^{2} + v_{1} v_{2}^{4} v_{4}^{2} + v_{1}^{4} v_{3} v_{4}^{2} + v_{2}^{4} v_{3} v_{4}^{2} + v_{1} v_{3}^{4} v_{4}^{2} + v_{2} v_{3}^{4} v_{4}^{2}\right.\right.\nonumber \\
&+&\left.\left.v_{1}^{2} v_{2} v_{4}^{4} + v_{1} v_{2}^{2} v_{4}^{4} + v_{1}^{2} v_{3} v_{4}^{4} + v_{2}^{2} v_{3} v_{4}^{4} + v_{1} v_{3}^{2} v_{4}^{4} + v_{2} v_{3}^{2} v_{4}^{4}\right.\right)\nonumber \\
&+&4c^{2}\left(\left.v_{1}^{2} v_{2} v_{3} v_{4} + v_{1} v_{2}^{2} v_{3} v_{4} + v_{1} v_{2} v_{3}^{2} v_{4} + v_{1} v_{2} v_{3} v_{4}^{2}\right)\right)\ ,
\eeq
\beq
\label{s_{2}^2term}
16\,s_{2}^{2}\,\left(\left(v_{1}^{4} v_{2}^{4} v_{3}^{2}\right.\right.&+&\left.\left.v_{1}^{4} v_{2}^{2} v_{3}^{4} + v_{1}^{2} v_{2}^{4} v_{3}^{4} + v_{1}^{4} v_{2}^{4} v_{4}^{2} + v_{1}^{4} v_{3}^{4} v_{4}^{2} + v_{2}^{4} v_{3}^{4} v_{4}^{2} + v_{1}^{4} v_{2}^{2} v_{4}^{4} + v_{1}^{2} v_{2}^{4} v_{4}^{4} + v_{1}^{4} v_{3}^{2} v_{4}^{4} + v_{2}^{4} v_{3}^{2} v_{4}^{4}\right.\right.\nonumber \\
&+&\left.\left.v_{1}^{2} v_{3}^{4} v_{4}^{4} + v_{2}^{2} v_{3}^{4} v_{4}^{4}\right) + 4c^{2}\left(v_{1}^{4} v_{2}^{2} v_{3} v_{4} + v_{1}^{2} v_{2}^{4} v_{3} v_{4} + v_{1}^{4} v_{2} v_{3}^{2} v_{4} + v_{1} v_{2}^{4} v_{3}^{2} v_{4} + v_{1}^{2} v_{2} v_{3}^{4} v_{4}\right.\right.\nonumber \\
&+&\left.\left.v_{1} v_{2}^{2} v_{3}^{4} v_{4} + v_{1}^{4} v_{2} v_{3} v_{4}^{2} + v_{1} v_{2}^{4} v_{3} v_{4}^{2} + v_{1} v_{2} v_{3}^{4} v_{4}^{2} + v_{1}^{2} v_{2} v_{3} v_{4}^{4} + v_{1} v_{2}^{2} v_{3} v_{4}^{4} + v_{1} v_{2} v_{3}^{2} v_{4}^{4}\right)\right) \ ,
\eeq
\beq
\label{s_{2}^1term}
-64 c\,s_{2}\,\left(v_{1}^{4} v_{2}^{4} v_{3}^{2} v_{4}\right.&+&\left.v_{1}^{4} v_{2}^{2} v_{3}^{4} v_{4} + v_{1}^{2} v_{2}^{4} v_{3}^{4} v_{4} + v_{1}^{4} v_{2}^{4} v_{3} v_{4}^{2} + v_{1}^{4} v_{2} v_{3}^{4} v_{4}^{2} + v_{1} v_{2}^{4} v_{3}^{4} v_{4}^{2} + v_{1}^{4} v_{2}^{2} v_{3} v_{4}^{4}\right.\nonumber \\
&+&\left.v_{1}^{2} v_{2}^{4} v_{3} v_{4}^{4} + v_{1}^{4} v_{2} v_{3}^{2} v_{4}^{4} + v_{1} v_{2}^{4} v_{3}^{2} v_{4}^{4} + v_{1}^{2} v_{2} v_{3}^{4} v_{4}^{4} + v_{1} v_{2}^{2} v_{3}^{4} v_{4}^{4}\right)\ ,
\eeq
\beq
\label{s_{2}^0term}
64 \left(v_{1}^{4} v_{2}^{4} v_{3}^{4} v_{4}^{2} + v_{1}^{4} v_{2}^{4} v_{3}^{2} v_{4}^{4} + v_{1}^{4} v_{2}^{2} v_{3}^{4} v_{4}^{4} + v_{1}^{2} v_{2}^{4} v_{3}^{4} v_{4}^{4}\right)\ .
\eeq

\noindent When written over a common denominator, the numerator of $\mu = \mu(v_{i})$ is therefore $(\ref{s_{2}^6term}) + (\ref{s_{2}^5term}) + (\ref{s_{2}^4term}) + (\ref{s_{2}^3term}) + (\ref{s_{2}^2term}) + (\ref{s_{2}^1term}) + (\ref{s_{2}^0term})$.  We now proceed to use (\ref{eqn1})--(\ref{eqn4}) to simplify each of these terms, beginning with (\ref{s_{2}^6term}), the $s_{2}^{6}$-term.

\vskip 24pt
\noindent {\bf The $s_{2}^{6}$-term.} We use (\ref{eqn1}) and (\ref{eqn2}):
\beq
\label{eqn5}
(v_{1}+v_{2} + v_{3} + v_{4})^{2}-2(v_{1}v_{2}+v_{1} v_{3} + v_{1} v_{4} + v_{2} v_{3} + v_{2} v_{4} + v_{3} v_{4})&=&v_{1}^{2} + v_{2}^{2} + v_{3}^{2} + v_{4}^{2}\nonumber \\
&=&0-2(s_{1} - 4 c^{2})=8c^{2}-2s_{1}\ .
\eeq

\noindent The $s_{2}^{6}$-term thus simplifies to 
\beq
\label{finals_{2}^6term}
s_{2}^{6}(8c^{2}-2s_{1})={\bf 8c^2s_{2}^{6}-2s_{1}s_{2}^{6}}\ .
\eeq
\noindent Because now there is no $v_{i}$-dependence, this term has been fully simplified.  

\vskip 24pt
\noindent {\bf The $s_{2}^{5}$-term.} We use (\ref{eqn1})--(\ref{eqn3}):
\beq
\label{eqn6}
(v_{1}&+&v_{2} + v_{3} + v_{4})\left(v_{1}v_{2}+v_{1} v_{3} + v_{1} v_{4} + v_{2} v_{3} + v_{2} v_{4} + v_{3} v_{4}\right)-3\left(v_{1} v_{2} v_{3} + v_{1} v_{2} v_{4} + v_{1} v_{3} v_{4} + v_{2} v_{3} v_{4}\right)\nonumber \\
&=&v_{1}^{2} v_{2} + v_{1} v_{2}^{2} + v_{1}^{2} v_{3} + v_{2}^{2} v_{3} + v_{1} v_{3}^{2} + v_{2} v_{3}^{2} + v_{1}^{2} v_{4} + v_{2}^{2} v_{4} + v_{3}^{2} v_{4} + v_{1} v_{4}^{2} + v_{2} v_{4}^{2} + v_{3} v_{4}^{2}\nonumber \\
&=&0\ (s_{1} - 4 c^{2})-3(-2 c s_{2})=6cs_{2}\ .
\eeq

\noindent The $s_{2}^{5}$-term thus simplifies to
\beq
\label{finals_{2}^5term}
-4cs_{2}^{5}(6cs_{2})={\bf -24c^{2}s_{2}^{6}}\ .
\eeq

\vskip 24pt
\noindent {\bf The $s_{2}^{4}$-term.} We proceed in steps.  First,
\beq
\label{step1}
(v_{1}&+&v_{2} + v_{3} + v_{4})\left(v_{1} v_{2} v_{3} + v_{1} v_{2} v_{4} + v_{1} v_{3} v_{4} + v_{2} v_{3} v_{4}\right)-4v_{1} v_{2} v_{3} v_{4}\nonumber \\
&=&v_{1}^{2} v_{2} v_{3} + v_{1} v_{2}^{2} v_{3} + v_{1} v_{2} v_{3}^{2} + v_{1}^{2} v_{2} v_{4} + v_{1} v_{2}^{2} v_{4} + v_{1}^{2} v_{3} v_{4} + v_{2}^{2} v_{3} v_{4} + v_{1} v_{3}^{2} v_{4}\nonumber \\
&&~~~~~+v_{2} v_{3}^{2} v_{4} + v_{1} v_{2} v_{4}^{2} + v_{1} v_{3} v_{4}^{2} + v_{2} v_{3} v_{4}^{2}\nonumber \\
&=&0\ (-2 c s_{2})-4\left(-{s_{2}^{2}\over 4}\right)=s_{2}^{2}\ .
\eeq

\noindent Second,
\beq
\label{step2}
(v_{1}v_{2}&+&v_{1} v_{3} + v_{1} v_{4} + v_{2} v_{3} + v_{2} v_{4} + v_{3} v_{4})^{2}-2\left(v_{1}^{2} v_{2} v_{3} + v_{1} v_{2}^{2} v_{3} + v_{1} v_{2} v_{3}^{2} + v_{1}^{2} v_{2} v_{4} + v_{1} v_{2}^{2} v_{4}\right.\nonumber \\
&+&\left.v_{1}^{2} v_{3} v_{4} + v_{2}^{2} v_{3} v_{4} + v_{1} v_{3}^{2} v_{4}+v_{2} v_{3}^{2} v_{4} + v_{1} v_{2} v_{4}^{2} + v_{1} v_{3} v_{4}^{2} + v_{2} v_{3} v_{4}^{2}\right)-6v_{1} v_{2} v_{3} v_{4}\nonumber \\
&=&v_{1}^{2} v_{2}^{2} + v_{1}^{2} v_{3}^{2} + v_{2}^{2} v_{3}^{2} + v_{1}^{2} v_{4}^{2} + v_{2}^{2} v_{4}^{2} + v_{3}^{2} v_{4}^{2}\nonumber \\
&=&(s_{1} - 4 c^{2})^{2}-2(s_{2}^{2})-6\left(-{s_{2}^{2}\over 4}\right)=(s_{1} - 4 c^{2})^{2}-{s_{2}^{2}\over 2}\ \cdot
\eeq

\noindent Third,
\beq
\label{step3}
(v_{1} v_{2} v_{3}&+&v_{1} v_{2} v_{4} + v_{1} v_{3} v_{4} + v_{2} v_{3} v_{4})^{2}-2\left(v_{1}v_{2}+v_{1} v_{3} + v_{1} v_{4} + v_{2} v_{3} + v_{2} v_{4} + v_{3} v_{4})(v_{1} v_{2} v_{3} v_{4}\right)\nonumber \\
&=&v_{1}^{2} v_{2}^{2} v_{3}^{2} + v_{1}^{2} v_{2}^{2} v_{4}^{2} + v_{1}^{2} v_{3}^{2} v_{4}^{2} + v_{2}^{2} v_{3}^{2} v_{4}^{2}\nonumber \\
&=&(-2 c s_{2})^{2}-2(s_{1} - 4 c^{2})\left(-{s_{2}^{2}\over 4}\right)=2c^{2}s_{2}^2+{s_{1}s_{2}^2 \over 2}\ \cdot
\eeq

\noindent Fourth, we combine ({\ref{eqn5}), (\ref{step2}), and (\ref{step3}) as follows:
\beq
\label{step4}
(v_{1}^{2}&+&v_{2}^{2} + v_{3}^{2} + v_{4}^{2})\left(v_{1}^{2} v_{2}^{2} + v_{1}^{2} v_{3}^{2} + v_{2}^{2} v_{3}^{2} + v_{1}^{2} v_{4}^{2} + v_{2}^{2} v_{4}^{2} + v_{3}^{2} v_{4}^{2}\right)-3\left(v_{1}^{2} v_{2}^{2} v_{3}^{2} + v_{1}^{2} v_{2}^{2} v_{4}^{2} + v_{1}^{2} v_{3}^{2} v_{4}^{2} + v_{2}^{2} v_{3}^{2} v_{4}^{2}\right)\nonumber \\
&=&v_{1}^{4} v_{2}^{2} + v_{1}^{2} v_{2}^{4} + v_{1}^{4} v_{3}^{2} + v_{2}^{4} v_{3}^{2} + v_{1}^{2} v_{3}^{4} + v_{2}^{2} v_{3}^{4} + v_{1}^{4} v_{4}^{2} + v_{2}^{4} v_{4}^{2} + v_{3}^{4} v_{4}^{2} + v_{1}^{2} v_{4}^{4} + 
 v_{2}^{2} v_{4}^{4} + v_{3}^{2} v_{4}^{4}\nonumber \\
 &=&\left(8c^{2}-2s_{1}\right)\left((s_{1} - 4 c^{2})^{2}-{s_{2}^{2}\over 2}\right)-3\left(2c^{2}s_{2}^2+{s_{1}s_{2}^2 \over 2}\right)\nonumber \\
 &=&128 c^{6} - 96 c^{4} s_{1} + 24 c^{2} s_{1}^{2} - 2 s_{1}^{3} - 10 c^{2} s_{2}^{2} - {s_{1} s_{2}^{2} \over 2}\ \cdot
\eeq

\noindent Finally, using (\ref{step4}) and (\ref{step1}), the $s_{2}^{4}$-term simplifies to
\beq
\label{finals_{2}^4term}
4s_{2}^{4}\left(\left(128 c^{6} - 96 c^{4} s_{1} + 24 c^{2} s_{1}^{2} - 2 s_{1}^{3} - 10 c^{2} s_{2}^{2} - {s_{1} s_{2}^{2} \over 2}\right)+4c^{2}(s_{2}^{2})\right)\nonumber \\
={\bf 512 c^{6} s_{2}^{4} - 384 c^{4} s_{1} s_{2}^{4} + 96 c^{2} s_{1}^{2} s_{2}^{4} - 8 s_{1}^{3} s_{2}^{4} - 
 24 c^{2} s_{2}^{6} - 2 s_{1} s_{2}^{6}}\ .
\eeq

\vskip 24pt
\noindent {\bf The $s_{2}^{3}$-term.} Once again we proceed in steps.  First, we note that
\beq
4c^{2}\left( v_{1}^{2} v_{2} v_{3} v_{4} + v_{1} v_{2}^{2} v_{3} v_{4} + v_{1} v_{2} v_{3}^{2} v_{4} + v_{1} v_{2} v_{3} v_{4}^{2}\right)=4c^{2}v_{1} v_{2} v_{3} v_{4}(v_{1} + v_{2} + v_{3} + v_{4})=0\nonumber \ ,
\eeq

\noindent so that the $s_{2}^{3}$-term reduces to 
\beq
-16 c\,s_{2}^{3}\,\left(v_{1}^{4} v_{2}^{2} v_{3}\right.&+&\left.v_{1}^{2} v_{2}^{4} v_{3} + v_{1}^{4} v_{2} v_{3}^{2} + v_{1} v_{2}^{4} v_{3}^{2} + v_{1}^{2} v_{2} v_{3}^{4} + v_{1} v_{2}^{2} v_{3}^{4} + v_{1}^{4} v_{2}^{2} v_{4} + v_{1}^{2} v_{2}^{4} v_{4} + v_{1}^{4} v_{3}^{2} v_{4} + v_{2}^{4} v_{3}^{2} v_{4}\right.\nonumber \\
&+&\left.v_{1}^{2} v_{3}^{4} v_{4} + v_{2}^{2} v_{3}^{4} v_{4} + v_{1}^{4} v_{2} v_{4}^{2} + v_{1} v_{2}^{4} v_{4}^{2} + v_{1}^{4} v_{3} v_{4}^{2} + v_{2}^{4} v_{3} v_{4}^{2} + v_{1} v_{3}^{4} v_{4}^{2} + v_{2} v_{3}^{4} v_{4}^{2}\right.\nonumber \\
&+&\left.v_{1}^{2} v_{2} v_{4}^{4} + v_{1} v_{2}^{2} v_{4}^{4} + v_{1}^{2} v_{3} v_{4}^{4} + v_{2}^{2} v_{3} v_{4}^{4} + v_{1} v_{3}^{2} v_{4}^{4} + v_{2} v_{3}^{2} v_{4}^{4}\right)\nonumber \ .
\eeq

\noindent Second, we multiply our equation by $1 = {\left({-s_{2}^{2} / 4}\right) \over v_{1} v_{2} v_{3} v_{4}}$, which is a valid operation since each $v_{i} \not = 0$, and group together terms with a common denominator to obtain
\beq
\label{s_{2}^3termintermediate}
4 c\,s_{2}^{5}\,\left({v_{3} v_{4}^{3} + v_{2}^{3} v_{3} + v_{2} v_{3}^{3} + v_{2}^{3} v_{4} + v_{3}^{3} v_{4} + v_{2} v_{4}^{3} \over v_{1}} + {v_{1}^{3} v_{3} + v_{1} v_{3}^{3} + v_{1}^{3} v_{4} + v_{3}^{3} v_{4} + v_{1} v_{4}^{3} + v_{3} v_{4}^{3} \over v_{2}}\right.\nonumber \\
\left.+ {v_{1} v_{4}^{3} + v_{2} v_{4}^{3} + v_{1}^{3} v_{2} + v_{1} v_{2}^{3} + v_{1}^{3} v_{4} + v_{2}^{3} v_{4} \over v_{3}} + {v_{1}^{3} v_{2} + v_{1} v_{2}^{3} + v_{1}^{3} v_{3} + v_{2}^{3} v_{3} + v_{1} v_{3}^{3} + v_{2} v_{3}^{3} \over v_{4}}\right) \ \cdot 
\eeq

\noindent Third, we use (\ref{eqn5}), (\ref{eqn2}), and (\ref{step1}) to obtain
\beq
\label{eqn7}
\left(v_{1}^{2}\right.&+&\left.v_{2}^{2} + v_{3}^{2} + v_{4}^{2}\right)\left(v_{1}v_{2}+v_{1} v_{3} + v_{1} v_{4} + v_{2} v_{3} + v_{2} v_{4} + v_{3} v_{4}\right)-\left(v_{1}^{2} v_{2} v_{3} + v_{1} v_{2}^{2} v_{3} + v_{1} v_{2} v_{3}^{2}\right.\nonumber \\
&+&\left.v_{1}^{2} v_{2} v_{4} + v_{1} v_{2}^{2} v_{4} + v_{1}^{2} v_{3} v_{4} + v_{2}^{2} v_{3} v_{4} + v_{1} v_{3}^{2} v_{4} +v_{2} v_{3}^{2} v_{4} + v_{1} v_{2} v_{4}^{2} + v_{1} v_{3} v_{4}^{2} + v_{2} v_{3} v_{4}^{2}\right)\nonumber \\
&=&v_{1}^{3} v_{2} + v_{1} v_{2}^{3} + v_{1}^{3} v_{3} + v_{2}^{3} v_{3} + v_{1} v_{3}^{3} + v_{2} v_{3}^{3} + v_{1}^{3} v_{4} + v_{2}^{3} v_{4} + v_{3}^{3} v_{4} + v_{1} v_{4}^{3} + v_{2} v_{4}^{3} + v_{3} v_{4}^{3}\nonumber \\
&=&(8c^{2}-2s_{1})(-4c^{2}+s_{1})-s_{2}^{2}=-2 (4 c^{2} - s_{1})^{2} - s_{2}^{2}\ .
\eeq

\noindent For each of the four terms in (\ref{s_{2}^3termintermediate}), we use (\ref{eqn7}) to simplify it.  For example, the first term in (\ref{s_{2}^3termintermediate}) simplifies as follows:
\beq
{v_{3} v_{4}^{3} + v_{2}^{3} v_{3} + v_{2} v_{3}^{3} + v_{2}^{3} v_{4} + v_{3}^{3} v_{4} + v_{2} v_{4}^{3} \over v_{1}}&=&
{-2 (4 c^{2} - s_{1})^{2} - s_{2}^{2} - v_{1}^{3} v_{2} - v_{1} v_{2}^{3} - v_{1}^{3} v_{3} - v_{1} v_{3}^{3} - v_{1}^{3} v_{4} - v_{1} v_{4}^{3} \over v_{1}}\nonumber \\
&=&{-2 (4 c^{2} - s_{1})^{2} - s_{2}^{2} \over v_{1}} - v_{1}^{2} v_{2} - v_{2}^{3} - v_{1}^{2} v_{3} - v_{3}^{3} - v_{1}^{2} v_{4} - v_{4}^{3}\nonumber \ .
\eeq

\noindent Likewise with the remaining terms in (\ref{s_{2}^3termintermediate}), so that the $s_{2}^{3}$-term reduces to
\beq
\label{s_{2}^3termalmost}
4c\,s_{2}^{5}\,\left(\left(-2 (4 c^{2}-s_{1})^{2} - s_{2}^{2}\right)\left({1\over v_{1}} + {1\over v_{2}} + {1\over v_{3}} + {1 \over v_{4}}\right)-v_{1}^{2} v_{2} - v_{1}^{2} v_{3} - v_{1}^{2} v_{4} - v_{1} v_{2}^{2} -v_{2}^{2} v_{3}\right.\nonumber \\
\Bigl.-v_{2}^{2} v_{4} - v_{1} v_{3}^{2} - v_{2} v_{3}^{2} - v_{3}^{2} v_{4} - v_{1} v_{4}^{2} - v_{2} v_{4}^{2} - v_{3} v_{4}^{2}-3\left(v_{1}^{3}+v_{2}^{3}+v_{2}^{3}+v_{2}^{3}\right)\Bigr)\ .
\eeq

\noindent Fourth, we use (\ref{eqn5}), (\ref{eqn1}), and (\ref{eqn6}) as follows:
\beq
\label{cubes}
\left(v_{1}^{2}\right.&+&\left.v_{2}^{2} + v_{3}^{2} + v_{4}^{2}\right)(v_{1} + v_{2} + v_{3} + v_{4})\nonumber \\
&-&\left(v_{1}^{2} v_{2} + v_{1} v_{2}^{2} + v_{1}^{2} v_{3} + v_{2}^{2} v_{3} + v_{1} v_{3}^{2} + v_{2} v_{3}^{2} + v_{1}^{2} v_{4} + v_{2}^{2} v_{4}+v_{3}^{2} v_{4} + v_{1} v_{4}^{2} + v_{2} v_{4}^{2} + v_{3} v_{4}^{2}\right)\nonumber \\
&=&v_{1}^{3}+v_{2}^{3} + v_{3}^{3} + v_{4}^{3}=(8c^{2}-2s_{1})\,0-6cs_{2}=-6cs_{2}\ .
\eeq

\noindent Fifth, we use (\ref{eqn3}) and (\ref{eqn4}) to obtain
\beq
\label{oneover}
{1\over v_{1}} + {1\over v_{2}} + {1\over v_{3}} + {1 \over v_{4}}={v_{2} v_{3} v_{4} + v_{1} v_{3} v_{4} + v_{1} v_{2} v_{4} + v_{1} v_{2} v_{3} \over v_{1} v_{2} v_{3} v_{4}}={8c \over s_{2}}\ \cdot
 \eeq

\noindent Finally, we insert (\ref{cubes}) and (\ref{oneover}) back into (\ref{s_{2}^3termalmost}) to obtain the simplified form of the $s_{2}^{3}$-term:
\beq
\label{finals_{2}^3term}
4c\,s_{2}^{5}\,\left(\left(-2 (4 c^{2}-s_{1})^{2} - s_{2}^{2}\right)\left({8c \over s_{2}}\right)-(6cs_{2})-3(-6cs_{2})\right)={\bf -1024 c^{6} s_{2}^{4} + 512 c^{4} s_{1} s_{2}^{4} - 64 c^{2} s_{1}^{2} s_{2}^{4} + 16 c^{2} s_{2}^{6}}\ .
\eeq

\vskip 24pt
\noindent {\bf The $s_{2}^{2}$-term:} Let us begin with the portion of this term with no $c$-dependence, namely,
\beq
\label{s_{2}^2nocterm}
16\,s_{2}^{2}\,\left(v_{1}^{4} v_{2}^{4} v_{3}^{2}\right.&+&\left.v_{1}^{4} v_{2}^{2} v_{3}^{4} + v_{1}^{2} v_{2}^{4} v_{3}^{4} + v_{1}^{4} v_{2}^{4} v_{4}^{2} + v_{1}^{4} v_{3}^{4} v_{4}^{2} + v_{2}^{4} v_{3}^{4} v_{4}^{2} + v_{1}^{4} v_{2}^{2} v_{4}^{4} + v_{1}^{2} v_{2}^{4} v_{4}^{4}\right.\nonumber \\
&+&\left.v_{1}^{4} v_{3}^{2} v_{4}^{4} +v_{2}^{4} v_{3}^{2} v_{4}^{4} + v_{1}^{2} v_{3}^{4} v_{4}^{4} + v_{2}^{2} v_{3}^{4} v_{4}^{4}\right) \ .
\eeq

Analogous to the $s_{2}^{3}$-term above, we begin by multiplying through twice by $1 = {\left({-s_{2}^{2} / 4}\right) \over v_{1} v_{2} v_{3} v_{4}}$, and then grouping together terms with a common denominator, to obtain
\beq
s_{2}^{6}\ \left({v_{2}^{2} v_{3}^{2} + v_{2}^{2} v_{4}^{2} + v_{3}^{2} v_{4}^{2} \over v_{1}^{2}} +  
{v_{1}^{2} v_{3}^{2} + v_{1}^{2} v_{4}^{2} + v_{3}^{2} v_{4}^{2} \over v_{2}^{2}} + 
{v_{1}^{2} v_{4}^{2} + v_{2}^{2} v_{4}^{2} + v_{1}^{2} v_{2}^{2} \over v_{3}^{2}} +
{v_{1}^{2} v_{2}^{2} + v_{1}^{2} v_{3}^{2} + v_{2}^{2} v_{3}^{2} \over v_{4}^{2}}\right)\nonumber \ \cdot
\eeq

\noindent Next, we use (\ref{step2}) on each of the four terms.  For example, the first term simplifies as follows:
\beq
 {v_{2}^{2} v_{3}^{2} + v_{2}^{2} v_{4}^{2} + v_{3}^{2} v_{4}^{2} \over v_{1}^{2}}&=&{-v_{1}^{2} v_{2}^{2} - v_{1}^{2} v_{3}^{2} - v_{1}^{2} v_{4}^{2} + (-4c^{2}+s_{1})^{2}-s_{2}^{2}/2\over v_{1}^{2}}\nonumber \\
 &=&-v_{2}^{2} - v_{3}^{2} - v_{4}^{2} + {(-4c^{2}+s_{1})^{2}-s_{2}^{2}/2 \over v_{1}^{2}}\nonumber \ .
\eeq

\noindent Likewise with the remaining terms, so that (\ref{s_{2}^2nocterm}) reduces to
\beq
s_{2}^{6}\ \left(-3\left(v_{1}^{2} + v_{2}^{2} + v_{3}^{2} + v_{4}^{2}\right) + \left((-4 c^{2} + s_{1})^{2} - {s_{2}^{2} \over 2}\right) \left({1\over v_{1}^{2}} + {1\over v_{2}^{2}} + {1\over v_{3}^{2}} + {1 \over v_{4}^{2}}\right)\right)\nonumber \ .
\eeq

\noindent Next, we use (\ref{step3}) and (\ref{eqn4}) to obtain
\beq
{1\over v_{1}^{2}} + {1\over v_{2}^{2}} + {1\over v_{3}^{2}} + {1 \over v_{4}^{2}}={v_{1}^{2} v_{2}^{2} v_{3}^{2} + v_{1}^{2} v_{2}^{2} v_{4}^{2} + v_{1}^{2} v_{3}^{2} v_{4}^{2} + v_{2}^{2} v_{3}^{2} v_{4}^{2} \over v_{1}^{2} v_{2}^{2} v_{3}^{2} v_{4}^{2}}={8(4c^{2}+s_{1}) \over s_{2}^{2}}\nonumber \ ,
\eeq

\noindent Using this and (\ref{eqn5}), we see that (\ref{s_{2}^2nocterm}) simplifies to
\beq
\label{finalnocanswer}
s_{2}^{6}\ \left(-3\left(8c^{2}-2s_{1}\right)+\left((-4 c^{2} + s_{1})^{2} - {s_{2}^{2} \over 2}\right)\left({8(4c^{2}+s_{1}) \over s_{2}^{2}}\right)\right)\nonumber \\
=512c^{6}s_{2}^{4}-128c^{4}s_{1}s_{2}^{4}-32c^{2}s_{1}^{2}s_{2}^{4}+8s_{1}^{3}s_{2}^{4}-40c^{2}s_{2}^{6}+2s_{1}s_{2}^6.
\eeq

\noindent There remains the portion of the $s_{2}^{2}$-term with a factor of $c^{2}$, namely
\beq
\label{s_{2}^2cterm}
64c^{2}\,s_{2}^{2}\,\left(v_{1}^{4} v_{2}^{2} v_{3} v_{4}\right.&+&\left.v_{1}^{2} v_{2}^{4} v_{3} v_{4} + v_{1}^{4} v_{2} v_{3}^{2} v_{4} + v_{1} v_{2}^{4} v_{3}^{2} v_{4} + v_{1}^{2} v_{2} v_{3}^{4} v_{4} +v_{1} v_{2}^{2} v_{3}^{4} v_{4}\right.\nonumber \\
&+&\left.v_{1}^{4} v_{2} v_{3} v_{4}^{2} + v_{1} v_{2}^{4} v_{3} v_{4}^{2} + v_{1} v_{2} v_{3}^{4} v_{4}^{2} + v_{1}^{2} v_{2} v_{3} v_{4}^{4} + v_{1} v_{2}^{2} v_{3} v_{4}^{4} + v_{1} v_{2} v_{3}^{2} v_{4}^{4}\right)\ ,
\eeq

\noindent which factors as
\beq
64c^{2}\,s_{2}^{2}\,v_{1} v_{2} v_{3} v_{4}\left(v_{1}^{3} v_{2}+v_{1} v_{2}^{3} + v_{1}^{3} v_{3} + v_{2}^{3} v_{3} + v_{1} v_{3}^{3} + v_{2} v_{3}^{3} + v_{1}^{3} v_{4} + v_{2}^{3} v_{4} + v_{3}^{3} v_{4} + v_{1} v_{4}^{3} + v_{2} v_{4}^{3} + v_{3} v_{4}^{3}\right)\nonumber \\
=-16 c^{2}\,s_{2}^{4}\,\left(v_{1}^{3} v_{2} + v_{1} v_{2}^{3} + v_{1}^{3} v_{3} + v_{2}^{3} v_{3} + v_{1} v_{3}^{3} + v_{2} v_{3}^{3} + v_{1}^{3} v_{4} + v_{2}^{3} v_{4} + v_{3}^{3} v_{4} + v_{1} v_{4}^{3} + v_{2} v_{4}^{3} + v_{3} v_{4}^{3}\right)\nonumber \ .
\eeq

\noindent To further simplify this expression, we use (\ref{eqn5}), (\ref{eqn2}), and (\ref{step1}):
\beq
\left(v_{1}^{2} + v_{2}^{2}\right.&+&\left.v_{3}^{2} + v_{4}^{2}\right)(v_{1}v_{2}+v_{1} v_{3} + v_{1} v_{4} + v_{2} v_{3} + v_{2} v_{4} + v_{3} v_{4})-\left(v_{1}^{2} v_{2} v_{3} + v_{1} v_{2}^{2} v_{3} + v_{1} v_{2} v_{3}^{2} + v_{1}^{2} v_{2} v_{4}\right.\nonumber \\
&+&\left.v_{1} v_{2}^{2} v_{4} + v_{1}^{2} v_{3} v_{4} + v_{2}^{2} v_{3} v_{4} + v_{1} v_{3}^{2} v_{4} +v_{2} v_{3}^{2} v_{4} + v_{1} v_{2} v_{4}^{2} + v_{1} v_{3} v_{4}^{2} + v_{2} v_{3} v_{4}^{2}\right)\nonumber \\
&=&v_{1}^{3} v_{2} + v_{1} v_{2}^{3} + v_{1}^{3} v_{3} + v_{2}^{3} v_{3} + v_{1} v_{3}^{3} + v_{2} v_{3}^{3} + v_{1}^{3} v_{4} + v_{2}^{3} v_{4} + v_{3}^{3} v_{4} + v_{1} v_{4}^{3} + v_{2} v_{4}^{3} + v_{3} v_{4}^{3}\nonumber \\
&=&(8c^{2}-2s_{1})(s_{1} - 4 c^{2})-s_{2}^{2}=-32c^{4}+16c^{2}s_{1}-2s_{1}^{2}-s_{2}^{2}\nonumber \ .
\eeq

\noindent Thus (\ref{s_{2}^2cterm}) simplifies to
\beq
\label{finalcanswer}
-16c^{2}s_{2}^{4}(-32c^{4}+16c^{2}s_{1}-2s_{1}^{2}-s_{2}^{2})=512c^{6}s_{2}^{4}-256c^{4}s_{1}s_{2}^{4}+32c^{2}s_{1}^{2}s_{2}^{4}+16c^{2}s_{2}^{6}\ .
\eeq

\noindent Finally, we add (\ref{finalnocanswer}) and (\ref{finalcanswer}) to obtain the simplified form of the $s_{2}^{2}$-term:
\beq
\label{finals_{2}^2term}
{\bf 1024 c^{6} s_{2}^{4} - 384 c^{4} s_{1} s_{2}^{4} + 8 s_{1}^{3} s_{2}^{4} - 24 c^{2} s_{2}^{6} + 2 s_{1} s_{2}^{6}}\ .
 \eeq

\vskip 24pt
\noindent {\bf The $s_{2}$-term.} First, we factor it as 
\beq
-64 c &s_{2}& v_{1} v_{2} v_{3} v_{4}\left(v_{1}^{3} v_{2}^{3} v_{3}+v_{1}^{3} v_{2} v_{3}^{3} + v_{1} v_{2}^{3} v_{3}^{3} + v_{1}^{3} v_{2}^{3} v_{4} + v_{1}^{3} v_{3}^{3} v_{4} + v_{2}^{3} v_{3}^{3} v_{4} + v_{1}^{3} v_{2} v_{4}^{3} + v_{1} v_{2}^{3} v_{4}^{3}\right.\nonumber \\
&&~~~~~~~~~~~~\left.+v_{1}^{3} v_{3} v_{4}^{3} + v_{2}^{3} v_{3} v_{4}^{3} + v_{1} v_{3}^{3} v_{4}^{3} + v_{2} v_{3}^{3} v_{4}^{3}\right)\nonumber \\
&=&16c\,s_{2}^{3}\,\left(v_{1}^{3} v_{2}^{3} v_{3}+v_{1}^{3} v_{2} v_{3}^{3} + v_{1} v_{2}^{3} v_{3}^{3} + v_{1}^{3} v_{2}^{3} v_{4} + v_{1}^{3} v_{3}^{3} v_{4} + v_{2}^{3} v_{3}^{3} v_{4} + v_{1}^{3} v_{2} v_{4}^{3} + v_{1} v_{2}^{3} v_{4}^{3}\right.\nonumber \\
&&~~~~~~~~~~~~\left.+v_{1}^{3} v_{3} v_{4}^{3} + v_{2}^{3} v_{3} v_{4}^{3} + v_{1} v_{3}^{3} v_{4}^{3} + v_{2} v_{3}^{3} v_{4}^{3}\right)\nonumber \ .
\eeq

\noindent Second, we multiply our equation by $1 = {\left({-s_{2}^{2} / 4}\right) \over v_{1} v_{2} v_{3} v_{4}}$, group terms together with the same denominator, and use (\ref{step2}) to obtain
\beq
4 c\,s_{2}^{5}\,\Bigl(v_{1}^{2} v_{2}&+&v_{1} v_{2}^{2} + v_{1}^{2} v_{3} + v_{2}^{2} v_{3} + v_{1} v_{3}^{2}+v_{2} v_{3}^{2} + v_{1}^{2} v_{4} + v_{2}^{2} v_{4} + v_{3}^{2} v_{4} + v_{1} v_{4}^{2} + v_{2} v_{4}^{2} + v_{3} v_{4}^{2}\nonumber \\
&-&\left((-4 c^{2} + s_{1})^{2} - {s_{2}^{2}\over 2}\right) \left({1\over v_{1}} + {1\over v_{2}} + {1\over v_{3}} + {1 \over v_{4}}\right)\Bigr)\nonumber \ .
\eeq

\noindent Finally, with the aid of (\ref{eqn6}) and (\ref{oneover}), the $s_{2}$-term simplifies to
\beq
\label{finals_{2}^1term}
4c\,s_{2}^{5}\,\left(6c s_{2} - \left((-4 c^{2} + s_{1})^{2} - {s_{2}^{2}\over 2}\right)\left({8c \over s_{2}}\right)\right)={\bf-512c^{6} s_{2}^{4}+256c^{4} s_{1} s_{2}^{4}-32c^{2} s_{1}^{2} s_{2}^{4}+40c^{2} s_{2}^{6}}\ .
\eeq

\vskip 24pt
\noindent {\bf The term with no $s_{2}$-dependence.} We factor this term as
\beq
64v_{1}^{2} v_{2}^{2} v_{3}^{2} v_{4}^{2}\left(v_{1}^{2} v_{2}^{2} v_{3}^{2} + v_{1}^{2} v_{2}^{2} v_{4}^{2} + v_{1}^{2} v_{3}^{2} v_{4}^{2} + v_{2}^{2} v_{3}^{2} v_{4}^{2}\right)\nonumber
\eeq

\noindent and then use (\ref{eqn4}) and (\ref{step3}) to obtain
\beq
\label{finals_{2}^0term}
64\left(-{s_{2}^{2}\over 4}\right)^{2}\left(2c^{2}s_{2}^2+{s_{1}s_{2}^2 \over 2}\right)={\bf 8c^{2}s_{2}^{6}+2s_{1}s_{2}^{6}}\ .
\eeq 

\vskip 24pt
\noindent We are done: for the simplified forms of (\ref{s_{2}^6term})--(\ref{s_{2}^0term}), namely, equations (\ref{finals_{2}^6term}), (\ref{finals_{2}^5term}), (\ref{finals_{2}^4term}), (\ref{finals_{2}^3term}), (\ref{finals_{2}^2term}), (\ref{finals_{2}^1term}), and (\ref{finals_{2}^0term}), sum to zero:
\beq
(\ref{finals_{2}^6term})&+&(\ref{finals_{2}^5term})+(\ref{finals_{2}^4term})+(\ref{finals_{2}^3term})+(\ref{finals_{2}^2term})+(\ref{finals_{2}^1term})+(\ref{finals_{2}^0term})\nonumber \\
&=&\left(8c^2s_{2}^{6}-2s_{1}s_{2}^{6}\right)+\left(-24c^{2}s_{2}^{6}\right)+\left(512 c^{6} s_{2}^{4} - 384 c^{4} s_{1} s_{2}^{4} + 96 c^{2} s_{1}^{2} s_{2}^{4} - 8 s_{1}^{3} s_{2}^{4} - 24 c^{2} s_{2}^{6} - 2 s_{1} s_{2}^{6}\right)\nonumber \\
&+&\left(-1024 c^{6} s_{2}^{4} + 512 c^{4} s_{1} s_{2}^{4} - 64 c^{2} s_{1}^{2} s_{2}^{4} + 16 c^{2} s_{2}^{6}\right)+\left(1024 c^{6} s_{2}^{4} - 384 c^{4} s_{1} s_{2}^{4} + 8 s_{1}^{3} s_{2}^{4} - 24 c^{2} s_{2}^{6} + 2 s_{1} s_{2}^{6}\right)\nonumber \\
&+&\left(-512c^{6} s_{2}^{4}+256c^{4} s_{1} s_{2}^{4}-32c^{2} s_{1}^{2} s_{2}^{4}+40c^{2} s_{2}^{6}\right)+\left(8c^{2}s_{2}^{6}+2s_{1}s_{2}^{6}\right)=0\nonumber \ .
\eeq

\noindent This completes the proof for the quantitative form of the lensing map in the neighborhood of an elliptic umbilic.\ $\square$

\subsection{Hyperbolic Umbilic}
\label{Appendix:hu}

The derivation of the quantitative form of the lensing map in the neighborhood of a hyperbolic umbilic critical point, can be found in \cite[Chap. 6]{Sch-EF}.  The resulting map is
\beq
\label{lenseqn3}
s_{1}&=&u^{2}+2cv\ ,\nonumber \\
s_{2}&=&v^{2}+2cu\ ,
\eeq  

\begin{figure}[ht]
\begin{center}
\includegraphics[scale=.5]{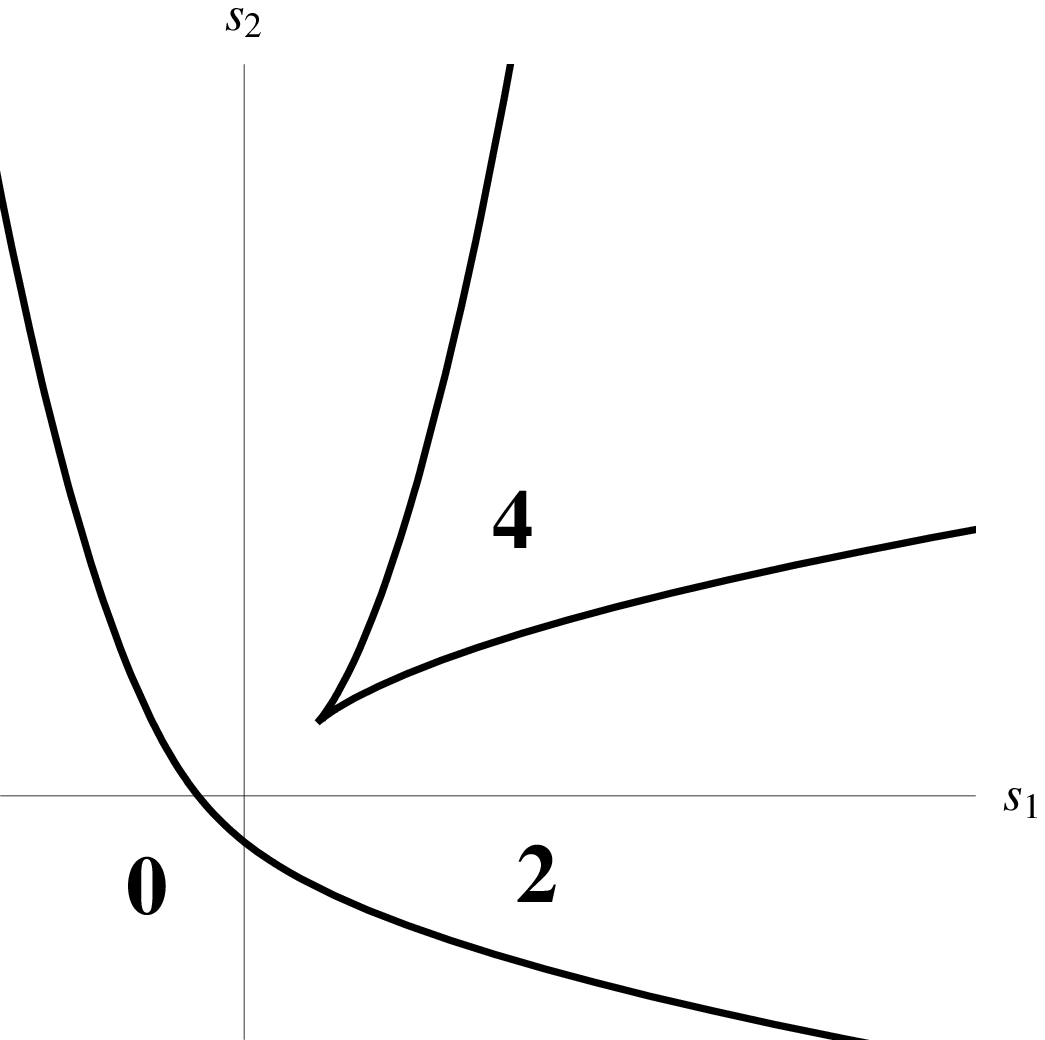}
\end{center}
\caption{Caustic curve for hyperbolic umbilic lensing
map $\blm_c$ in equation (\ref{lenseqn3}) or Table~\ref{table1}.  The number of lensed images for sources in their respective regions is indicated.}
\label{Figure4}
\end{figure}

\noindent and the corresponding magnification of an image $(u,v)$ is
\beq
\left({\rm det}({\rm Jac}\ {\bf s})\right)^{-1}(u,v)&=&{1 \over 4(uv-c^{2})}\nonumber \ \cdot
\eeq

\noindent The critical curves are hyperbolas given by
\beq
\label{hypcrit}
v&=&{c^{2} \over u}\ ,
\eeq

\noindent and the corresponding caustic curve is
\beq
\label{hypcaustic}
s_{1}&=&u^{2}+{2c^{3} \over u}\ ,\nonumber \\
s_{2}&=&2cu+{c^{4} \over u^{2}}\ \cdot
\eeq

\noindent The caustic curve is shown in Figure~\ref{Figure4}. The region {\lq\lq inside the beak\rq\rq} constitutes the four-image region.  We now show that for all sources inside this region, the total signed magnification is identically zero.

First, note from the second lens equation (\ref{lenseqn3}) that since $u_{i}(v_{i}) = {s_{2} - v_{i}^{2} \over 2 c}$, which is to say, since $v_{i}$ does not appear in the denominator, we do not need to restrict our analysis to the case where $v_{i} \not = 0$, as we did with the elliptic umbilic.  We will therefore let $(s_{1},s_{2})$ denote an arbitrary source in the four-image region.  The total signed magnification $\mu$ at $(s_{1},s_{2})$ is
\beq
\label{totalmag3}
\mu(u_{i},v_{i})={1 \over 4(u_{1}v_{1}-c^{2})} + {1 \over 4(u_{2}v_{2}-c^{2})} + {1 \over 4(u_{3}v_{3}-c^{2})} + {1 \over 4(u_{4}v_{4}-c^{2})}\ \cdot
\eeq 

\noindent We begin by eliminating $u$ from the lens equation (\ref{lenseqn3}) to obtain a (depressed) quartic in $v$:
\beq
v^{4}-2s_{2} v^{2}+8c^{3} v+(s_{2}^{2}-4c^{2}s_{1})=0\nonumber \ .
\eeq

\noindent Analogous to ({\ref{eqn1})--(\ref{eqn4}) and (\ref{eqn1a})--(\ref{eqn4a}), we obtain four equations involving only the $v_{i}$:
\beq
\label{eqn1b}
{\bf v_{1} + v_{2} + v_{3} + v_{4}}&=&{\bf 0}\ ,\\
\label{eqn2b}
{\bf v_{1}v_{2}+v_{1} v_{3} + v_{1} v_{4} + v_{2} v_{3} + v_{2} v_{4} + v_{3} v_{4}}&=&{\bf -2s_{2}}\ ,\\
\label{eqn3b}
{\bf v_{1} v_{2} v_{3} + v_{1} v_{2} v_{4} + v_{1} v_{3} v_{4} + v_{2} v_{3} v_{4}}&=&{\bf -8c^{3}}\ ,\\
\label{eqn4b}
{\bf v_{1} v_{2} v_{3} v_{4}}&=&{\bf -4c^{2}s_{1}+s_{2}^{2}}\ .
\eeq

\noindent We then insert $u_{i}(v_{i})$ into the total magnification (\ref{totalmag3}) and factor the numerator of the resulting expression in powers of $s_{2}$:
\beq
\label{1hyp}
c\,s_{2}^{3}\,(v_{1} v_{2} v_{3} + v_{1} v_{2} v_{4} + v_{1} v_{3} v_{4} + v_{2} v_{3} v_{4})\ ,
\eeq
\beq
\label{2hyp}
-c\,s_{2}^{2}\,\left(4 c^{3}\left(v_{1} v_{2}\right.\right.&+&\left.v_{1} v_{3} + v_{2} v_{3} + v_{1} v_{4} + v_{2} v_{4} + v_{3} v_{4}\right) + \left(v_{1}^{3} v_{2} v_{4} + v_{1} v_{2}^{3} v_{4} + v_{1}^{3} v_{3} v_{4} + v_{2}^{3} v_{3} v_{4}\right.\nonumber \\
&+&\left.\left.v_{1} v_{3}^{3} v_{4} + v_{2} v_{3}^{3} v_{4} + v_{1} v_{2} v_{4}^{3} + v_{1} v_{3} v_{4}^{3} + v_{2} v_{3} v_{4}^{3} + v_{1}^{3} v_{2} v_{3} + v_{1} v_{2}^{3} v_{3} + v_{1} v_{2} v_{3}^{3}\right)\right)\ ,
\eeq
\beq
\label{3hyp}
c\,s_{2}\,\left(12 c^{6}\left(v_{1}\right.\right.&+&\left.v_{2} + v_{3} + v_{4}\right) + 4 c^{3}\left(v_{1}^{3} v_{2} + v_{1} v_{2}^{3} + v_{1}^{3} v_{3} + v_{2}^{3} v_{3} + v_{1} v_{3}^{3} + v_{2} v_{3}^{3} + v_{1}^{3} v_{4} + v_{2}^{3} v_{4} + v_{3}^{3} v_{4}\right.\nonumber \\
&+&\left.v_{1} v_{4}^{3} + v_{2} v_{4}^{3} + v_{3} v_{4}^{3}\right) + \left(v_{1}^{3} v_{2}^{3} v_{3} +v_{1}^{3} v_{2} v_{3}^{3} + v_{1} v_{2}^{3} v_{3}^{3} + v_{1}^{3} v_{2}^{3} v_{4} +v_{1}^{3} v_{3}^{3} v_{4}\right.\nonumber \\
&+&\left.\left.v_{2}^{3} v_{3}^{3} v_{4}  + v_{1}^{3} v_{2} v_{4}^{3} + v_{1} v_{2}^{3} v_{4}^{3} + v_{1}^{3} v_{3} v_{4}^{3} + v_{2}^{3} v_{3} v_{4}^{3} + v_{1} v_{3}^{3} v_{4}^{3} + v_{2} v_{3}^{3} v_{4}^{3}\right)\right)\ ,
\eeq
\beq
\label{4hyp}
-32 c^{10}&-&12 c^{7}\left(v_{1}^{3} + v_{2}^{3} + v_{4}^{3} + v_{3}^{3}\right) - 4 c^{4}\left(v_{1}^{3} v_{3}^{3} + v_{2}^{3} v_{3}^{3} +v_{1}^{3} v_{2}^{3} + v_{1}^{3} v_{4}^{3} + v_{2}^{3} v_{4}^{3} + v_{3}^{3} v_{4}^{3}\right)\nonumber \\
&-&c\left(v_{1}^{3} v_{2}^{3} v_{4}^{3} + v_{1}^{3} v_{3}^{3} v_{4}^{3} + v_{2}^{3} v_{3}^{3} v_{4}^{3} + v_{1}^{3} v_{2}^{3} v_{3}^{3}\right)\ .
\eeq

\vskip 12pt

\noindent {\bf The $s_{2}^{3}$-term.} Using (\ref{eqn3b}), this term simplifies to
\beq
\label{hyps_{2}3term}
cs_{2}^{3}(-2s_{2})={\bf -8 c^{4} s_{2}^{3}}\ .
\eeq

\vskip 12pt

\noindent {\bf The $s_{2}^{2}$-term.} Analogous to (\ref{eqn5}), we have $v_{1}^{2} + v_{2}^{2} + v_{3}^{2} + v_{4}^{2}=4s_{2}$, which, when combined with (\ref{eqn1b}), (\ref{eqn3b}), and (\ref{eqn4b}), yields
\beq
\left(v_{1}^{2}\right.&+&\left.v_{2}^{2} + v_{3}^{2} + v_{4}^{2}\right)\left(v_{1} v_{2} v_{3} + v_{1} v_{2} v_{4} + v_{1} v_{3} v_{4} + v_{2} v_{3} v_{4}\right)-(v_{1} v_{2} v_{3} v_{4}) (v_{1} + v_{2} + v_{3} + v_{4})\nonumber \\
&=&v_{1}^{3} v_{2} v_{4} + v_{1} v_{2}^{3} v_{4} + v_{1}^{3} v_{3} v_{4}+v_{2}^{3} v_{3} v_{4} + v_{1} v_{3}^{3} v_{4} + v_{2} v_{3}^{3} v_{4} + v_{1} v_{2} v_{4}^{3} + v_{1} v_{3} v_{4}^{3} + v_{2} v_{3} v_{4}^{3}\nonumber \\
&+&v_{1}^{3} v_{2} v_{3} + v_{1} v_{2}^{3} v_{3} + v_{1} v_{2} v_{3}^{3}=(4s_{2})(-8c^{3})-(-4c^{2}s_{1})\,0=-32c^{3}s_{2}\nonumber \ .
\eeq

\noindent The $s_{2}^{2}$-term thus simplifies to
\beq
\label{hyps_{2}2term}
-c\ s_{2}^{2}\left(4c^{3}\left(-2s_{2}\right)-32c^{3}s_{2}\right)={\bf 40c^{4}s_{2}^{3}}\ .
\eeq

\vskip 12pt

\noindent {\bf The $s_{2}$-term.} We proceed in steps.  First, analogous to (\ref{step1}), we have
\beq
v_{1}^{2} v_{2} v_{3}&+&v_{1} v_{2}^{2} v_{3} + v_{1} v_{2} v_{3}^{2} + v_{1}^{2} v_{2} v_{4} + v_{1} v_{2}^{2} v_{4} + v_{1}^{2} v_{3} v_{4} + v_{2}^{2} v_{3} v_{4} + v_{1} v_{3}^{2} v_{4}\nonumber \\
&+&v_{2} v_{3}^{2} v_{4} + v_{1} v_{2} v_{4}^{2} + v_{1} v_{3} v_{4}^{2} + v_{2} v_{3} v_{4}^{2}\nonumber \\
&=&-4(-4c^{2}s_{1}+s_{2}^{2})=16c^{2}s_{1}-4s_{2}^{2}\nonumber \ ,
\eeq

\noindent which, as with (\ref{eqn7}) above, is used to obtain 
\beq
\label{X}
\left(v_{1}^{2}\right.&+&\left.v_{2}^{2} + v_{3}^{2} + v_{4}^{2}\right)(v_{1}v_{2}+v_{1} v_{3} + v_{1} v_{4} + v_{2} v_{3} + v_{2} v_{4} + v_{3} v_{4})-\left(v_{1}^{2} v_{2} v_{3} + v_{1} v_{2}^{2} v_{3} + v_{1} v_{2} v_{3}^{2} + v_{1}^{2} v_{2} v_{4}\right.\nonumber \\
&+&\left.v_{1} v_{2}^{2} v_{4} + v_{1}^{2} v_{3} v_{4} + v_{2}^{2} v_{3} v_{4} + v_{1} v_{3}^{2} v_{4} +v_{2} v_{3}^{2} v_{4} + v_{1} v_{2} v_{4}^{2} + v_{1} v_{3} v_{4}^{2} + v_{2} v_{3} v_{4}^{2}\right)\nonumber \\
&=&v_{1}^{3} v_{2} + v_{1} v_{2}^{3} + v_{1}^{3} v_{3} + v_{2}^{3} v_{3} + v_{1} v_{3}^{3} + v_{2} v_{3}^{3} + v_{1}^{3} v_{4} + v_{2}^{3} v_{4} + v_{3}^{3} v_{4} +v_{1} v_{4}^{3} + v_{2} v_{4}^{3} + v_{3} v_{4}^{3}\nonumber \\
&=&(4s_{2})(-2s_{2})-(-4s_{2}^{2}+16c^{2}s_{1})=-4s_{2}^{2}-16c^{2}s_{1}\ .
\eeq

\noindent Second, analogous to (\ref{eqn6}) and (\ref{step2}), we have
\beq
\label{Y}
v_{1}^{2} v_{2}+v_{1} v_{2}^{2} + v_{1}^2 v_{3} + v_{2}^2 v_{3} + v_{1} v_{3}^2 + v_{2} v_{3}^2 + v_{1}^2 v_{4} + v_{2}^2 v_{4} + v_{3}^2 v_{4} + v_{1} v_{4}^2 + v_{2} v_{4}^2 + v_{3} v_{4}^2=-3(-8c^{3})=24c^{3}
\eeq
\noindent and
\beq
v_{1}^{2} v_{2}^{2} + v_{1}^{2} v_{3}^{2} + v_{2}^{2} v_{3}^{2} + v_{1}^{2} v_{4}^{2} + v_{2}^{2} v_{4}^{2} + v_{3}^{2} v_{4}^{2} = 4 s_{2}^{2} + 2 (-4 c^{2} s_{1} + s_{2}^{2})=6s_{2}^{2}-8c^{2}s_{1}\nonumber \ .
\eeq
    
\noindent Third, we use (\ref{eqn4b}) and (\ref{Y}) to obtain
\beq
v_{1} v_{2} v_{3} v_{4}\left(v_{1}^{2} v_{2}\right.&+&\left.v_{1} v_{2}^{2} + v_{1}^2 v_{3} + v_{2}^2 v_{3} + v_{1} v_{3}^2 + v_{2} v_{3}^2 + v_{1}^2 v_{4} + v_{2}^2 v_{4} + v_{3}^2 v_{4} + v_{1} v_{4}^2 + v_{2} v_{4}^2 + v_{3} v_{4}^2\right)\nonumber \\     
&=&v_{1}^3 v_{2}^2 v_{3} v_{4} + v_{1}^2 v_{2}^3 v_{3} v_{4} + v_{1}^3 v_{2} v_{3}^2 v_{4} + v_{1} v_{2}^3 v_{3}^2 v_{4} + v_{1}^2 v_{2} v_{3}^3 v_{4} + v_{1} v_{2}^2 v_{3}^3 v_{4}\nonumber \\
&+&v_{1}^3 v_{2} v_{3} v_{4}^2 + v_{1} v_{2}^3 v_{3} v_{4}^2 + v_{1} v_{2} v_{3}^3 v_{4}^2 + v_{1}^2 v_{2} v_{3} v_{4}^3 + v_{1} v_{2}^2 v_{3} v_{4}^3 + v_{1} v_{2} v_{3}^2 v_{4}^3\nonumber \\
&=&(s_{2}^{2}-4 c^{2} s_{1})24 c^3\nonumber \ .
\eeq

\noindent Fourth, we combine these three equations to obtain
\beq
\label{Z}
\left(v_{1}^2 v_{2}^2\right.&+&\left.v_{1}^2 v_{3}^2 + v_{2}^2 v_{3}^2 + v_{1}^2 v_{4}^2 + v_{2}^2 v_{4}^2 + v_{3}^2 v_{4}^2\right)\left(v_{1} v_{2} v_{3} + v_{1} v_{2} v_{4} + v_{1} v_{3} v_{4} + v_{2} v_{3} v_{4}\right) - \left(v_{1}^3 v_{2}^2 v_{3} v_{4} + v_{1}^2 v_{2}^3 v_{3} v_{4} + v_{1}^3 v_{2} v_{3}^2 v_{4}\right.\nonumber \\
&+&\left.v_{1} v_{2}^3 v_{3}^2 v_{4} + v_{1}^2 v_{2} v_{3}^3 v_{4} + v_{1} v_{2}^2 v_{3}^3 v_{4} + v_{1}^3 v_{2} v_{3} v_{4}^2 + v_{1} v_{2}^3 v_{3} v_{4}^2 + v_{1} v_{2} v_{3}^3 v_{4}^2 + v_{1}^2 v_{2} v_{3} v_{4}^3 + v_{1} v_{2}^2 v_{3} v_{4}^3 + v_{1} v_{2} v_{3}^2 v_{4}^3\right)\nonumber \\ 
&=&v_{1}^3 v_{2}^3 v_{3} + v_{1}^3 v_{2} v_{3}^3 + v_{1} v_{2}^3 v_{3}^3 + v_{1}^3 v_{2}^3 v_{4} + v_{1}^3 v_{3}^3 v_{4} + v_{2}^3 v_{3}^3 v_{4} + v_{1}^3 v_{2} v_{4}^3 + v_{1} v_{2}^3 v_{4}^3 + v_{1}^3 v_{3} v_{4}^3 + v_{2}^3 v_{3} v_{4}^3 + v_{1} v_{3}^3 v_{4}^3 + v_{2} v_{3}^3 v_{4}^3\nonumber \\
&=&(6s_{2}^{2}-8c^{2}s_{1})(-8c^{3}) - (s_{2}^{2}-4 c^{2} s_{1})24 c^{3} =160 c^{5} s_{1} - 72 c^{3} s_{2}^{2}\ .
\eeq

\noindent Finally, we use (\ref{eqn4b}), (\ref{X}), and (\ref{Z}) to simplify the $s_{2}$-term:
\beq
\label{hyps_{2}term}
c\,s_{2}\,\left(12c^{6}(0)+4c^{3}\left(-4s_{2}^{2}-16c^{2}s_{1}\right)+\left(160 c^{5} s_{1} - 72 c^{3} s_{2}^{2}\right)\right)={\bf 96 c^{6} s_{1} s_{2} - 88 c^{4} s_{2}^{3}}\ .
\eeq

\vskip 12pt

\noindent {\bf The term with no $s_{2}$-dependence.} Again, we proceed in steps.  First, analogous to (\ref{step3}) and (\ref{cubes}), we have
\beq
v_{1}^2 v_{2}^2 v_{3}^2 + v_{1}^2 v_{2}^2 v_{4}^2 + v_{1}^2 v_{3}^2 v_{4}^2 + v_{2}^2 v_{3}^2 v_{4}^2=(-8 c^3)^2 - 2 (s_{2}^2 - 4 c^2 s_{1}) (-2 s_{2})=64 c^6 + 4 s_{2} (-4 c^2 s_{1} + s_{2}^2)\nonumber
\eeq

\noindent and
\beq
\label{cube}
v_{1}^{3}+v_{2}^{3} + v_{3}^{3} + v_{4}^{3}=-24c^{3}\ .
\eeq

\noindent Second, we have
\beq
\left(v_{1}^2 v_{2} v_{3}\right.&+&\left.v_{1} v_{2}^2 v_{3} + v_{1} v_{2} v_{3}^2 + v_{1}^2 v_{2} v_{4} + v_{1} v_{2}^2 v_{4} + v_{1}^2 v_{3} v_{4} + v_{2}^2 v_{3} v_{4} + v_{1} v_{3}^2 v_{4} + v_{2} v_{3}^2 v_{4} + v_{1} v_{2} v_{4}^2 + v_{1} v_{3} v_{4}^2\right.\nonumber \\
&+&\left.v_{2} v_{3} v_{4}^2\right)(v_{1} v_{2} + v_{1} v_{3} + v_{2} v_{3} + v_{1} v_{4} + v_{2} v_{4} + v_{3} v_{4}) - 3\left(v_{1}^2 v_{2}^2 v_{3}^2 + v_{1}^2 v_{2}^2 v_{4}^2 + v_{1}^2 v_{3}^2 v_{4}^2 + v_{2}^2 v_{3}^2 v_{4}^2\right)\nonumber \\
&-&3v_{1} v_{2} v_{3} v_{4}\left(v_{1}^2 + v_{2}^2 + v_{3}^2 + v_{4}^2\right) - 3v_{1} v_{2} v_{3} v_{4} (v_{1} v_{2} + v_{1} v_{3} + v_{2} v_{3}+v_{1} v_{4} + v_{2} v_{4} + v_{3} v_{4})\nonumber \\
&=&v_{1}^3 v_{2}^2 v_{3} + v_{1}^2 v_{2}^3 v_{3} + v_{1}^3 v_{2} v_{3}^2 + v_{1} v_{2}^3 v_{3}^2 + v_{1}^2 v_{2} v_{3}^3 + v_{1} v_{2}^2 v_{3}^3 + v_{1}^3 v_{2}^2 v_{4} + v_{1}^2 v_{2}^3 v_{4} +v_{1}^2 v_{2}^2 v_{3} v_{4}\nonumber \\
&+&v_{1}^3 v_{3}^2 v_{4} + v_{1}^2 v_{2} v_{3}^2 v_{4} + v_{1} v_{2}^2 v_{3}^2 v_{4} + v_{2}^3 v_{3}^2 v_{4} + v_{1}^2 v_{3}^3 v_{4} + v_{2}^2 v_{3}^3 v_{4} + v_{1}^3 v_{2} v_{4}^2 + v_{1} v_{2}^3 v_{4}^2\nonumber \\
&+&v_{1}^3 v_{3} v_{4}^2 + v_{1}^2 v_{2} v_{3} v_{4}^2 + v_{1} v_{2}^2 v_{3} v_{4}^2 + v_{2}^3 v_{3} v_{4}^2 + v_{1} v_{2} v_{3}^2 v_{4}^2 + v_{1} v_{3}^3 v_{4}^2 + v_{2} v_{3}^3 v_{4}^2\nonumber \\
&+&v_{1}^2 v_{2} v_{4}^3 + v_{1} v_{2}^2 v_{4}^3 + v_{1}^2 v_{3} v_{4}^3 + v_{2}^2 v_{3} v_{4}^3 + v_{1} v_{3}^2 v_{4}^3 + v_{2} v_{3}^2 v_{4}^3\nonumber \\
&=&(-4s_{2}^{2}+16c^{2}s_{1})(-2s_{2})-3\left(64 c^6 + 4 s_{2} (-4 c^2 s_{1} + s_{2}^2)\right)-3(s_{2}^{2}-4c^{2}s_{1})(4s_{2})-3(s_{2}^{2}-4c^{2}s_{1})(-2s_{2})\nonumber \\
&=&-192 c^6 + 40 c^2 s_{1} s_{2} - 10 s_{2}^{3}\nonumber \ ,
\eeq

\noindent which we use to obtain
\beq
\label{cubecube}
\left(v_{1}^2 v_{2}^2\right.&+&\left.v_{1}^2 v_{3}^2 + v_{2}^2 v_{3}^2 + v_{1}^2 v_{4}^2 + v_{2}^2 v_{4}^2 + v_{3}^2 v_{4}^2\right) (v_{1} v_{2} + v_{1} v_{3} + v_{2} v_{3} + v_{1} v_{4} + v_{2} v_{4} + v_{3} v_{4})\nonumber\\
&-&\left(v_{1}^3 v_{2}^2 v_{3} + v_{1}^2 v_{2}^3 v_{3} + v_{1}^3 v_{2} v_{3}^2 + v_{1} v_{2}^3 v_{3}^2 + v_{1}^2 v_{2} v_{3}^3 + v_{1} v_{2}^2 v_{3}^3 + v_{1}^3 v_{2}^2 v_{4} + v_{1}^2 v_{2}^3 v_{4} + v_{1}^2 v_{2}^2 v_{3} v_{4}\right.\nonumber \\
&+&v_{1}^3 v_{3}^2 v_{4} + v_{1}^2 v_{2} v_{3}^2 v_{4} + v_{1} v_{2}^2 v_{3}^2 v_{4} + v_{2}^3 v_{3}^2 v_{4} + v_{1}^2 v_{3}^3 v_{4} + v_{2}^2 v_{3}^3 v_{4} + v_{1}^3 v_{2} v_{4}^2 + v_{1} v_{2}^3 v_{4}^2\nonumber \\
&+&v_{1}^3 v_{3} v_{4}^2 + v_{1}^2 v_{2} v_{3} v_{4}^2 + v_{1} v_{2}^2 v_{3} v_{4}^2 + v_{2}^3 v_{3} v_{4}^2 + v_{1} v_{2} v_{3}^2 v_{4}^2 + v_{1} v_{3}^3 v_{4}^2 + v_{2} v_{3}^3 v_{4}^2\nonumber \\
&+&\left.v_{1}^2 v_{2} v_{4}^3 + v_{1} v_{2}^2 v_{4}^3 + v_{1}^2 v_{3} v_{4}^3 + v_{2}^2 v_{3} v_{4}^3 + v_{1} v_{3}^2 v_{4}^3 + v_{2} v_{3}^2 v_{4}^3\right)\nonumber \\
&=&v_{1}^3 v_{2}^3 + v_{1}^3 v_{3}^3 + v_{2}^3 v_{3}^3 + v_{1}^3 v_{4}^3 + v_{2}^3 v_{4}^3 + v_{3}^3 v_{4}^3\nonumber \\
&=&\left(4 s_{2}^2 + 2 \left(-4 c^2 s_{1} + s_{2}^2\right)\right) (-2 s_{2}) - (-192 c^6 + 40 c^2 s_{1} s_{2} - 10 s_{2}^3)\nonumber \\
&=&192 c^6 - 24 c^2 s_{1} s_{2} - 2 s_{2}^3\ .
\eeq

\noindent Third, we have
\beq
&&(v_{1} v_{2} v_{3}+v_{1} v_{2} v_{4} + v_{1} v_{3} v_{4} + v_{2} v_{3} v_{4}) (v_{1} v_{2} + v_{1} v_{3} + v_{1} v_{4} + v_{2} v_{3} + v_{2} v_{4} + v_{3} v_{4}) - 3 (v_{1} + v_{2} + v_{3} + v_{4}) (v_{1} v_{2} v_{3} v_{4})\nonumber \\
&=&v_{1}^2 v_{2}^2 v_{3} + v_{1}^2 v_{2} v_{3}^2 + v_{1} v_{2}^2 v_{3}^2 + v_{1}^2 v_{2}^2 v_{4} + v_{1}^2 v_{3}^2 v_{4} + v_{2}^2 v_{3}^2 v_{4} + v_{1}^2 v_{2} v_{4}^2 + v_{1} v_{2}^2 v_{4}^2 + v_{1}^2 v_{3} v_{4}^2 + v_{2}^2 v_{3} v_{4}^2 + v_{1} v_{3}^2 v_{4}^2 + v_{2} v_{3}^2 v_{4}^2\nonumber \\
&=&(-8 c^3) (-2 s_{2}) =16 c^3 s_{2}\nonumber \ ,
\eeq

\noindent which we use to obtain
\beq
\label{cubecubecube}
\left(v_{1}^2 v_{2}^2 v_{3}^2\right.&+&\left.v_{1}^2 v_{2}^2 v_{4}^2 + v_{1}^2 v_{3}^2 v_{4}^2 + v_{2}^2 v_{3}^2 v_{4}^2\right)(v_{1} v_{2} v_{3} + v_{1} v_{2} v_{4} + v_{1} v_{3} v_{4} + v_{2} v_{3} v_{4}) - v_{1} v_{2} v_{3} v_{4}\left(v_{1}^2 v_{2}^2 v_{3} + v_{1}^2 v_{2} v_{3}^2 + v_{1} v_{2}^2 v_{3}^2\right.\nonumber \\
&+&\left.v_{1}^2 v_{2}^2 v_{4} + v_{1}^2 v_{3}^2 v_{4} + v_{2}^2 v_{3}^2 v_{4} + v_{1}^2 v_{2} v_{4}^2 + v_{1} v_{2}^2 v_{4}^2 + v_{1}^2 v_{3} v_{4}^2 + v_{2}^2 v_{3} v_{4}^2 + v_{1} v_{3}^2 v_{4}^2 + v_{2} v_{3}^2 v_{4}^2\right)\nonumber \\
&=&v_{1}^3 v_{2}^3 v_{3}^3 + v_{1}^3 v_{2}^3 v_{4}^3 + v_{1}^3 v_{3}^3 v_{4}^3 + v_{2}^3 v_{3}^3 v_{4}^3\nonumber \\
&=&(64 c^6 + 4 s_{2}\left(-4 c^2 s_{1} + s_{2}^2)\right) (-8 c^3) - (-4 c^2 s_{1} + s_{2}^2) (16 c^3 s_{2})\nonumber \\
&=&-512 c^9 + 192 c^5 s_{1} s_{2} - 48 c^3 s_{2}^3\ .
\eeq

\noindent Finally, we use (\ref{cube}), (\ref{cubecube}), and (\ref{cubecubecube}) to simplify the term with no $s_{2}$-dependence:
\beq
\label{hyps_{2}0term}
&-&32c^{10}-12c^7(-24^3)-4c^4\left(192 c^6 - 24 c^2 s_{1} s_{2} - 2 s_{2}^3\right)-c\left(-512 c^9 + 192 c^5 s_{1} s_{2} - 48 c^3 s_{2}^3\right)\nonumber \\
&=&{\bf -96 c^6 s_{1} s_{2} + 56 c^4 s_{2}^3}\ .
\eeq

\vskip 12pt

\noindent We can now verify that the simplified forms of (\ref{1hyp})--(\ref{4hyp}), namely, equations (\ref{hyps_{2}3term}), (\ref{hyps_{2}2term}), (\ref{hyps_{2}term}), and (\ref{hyps_{2}0term}), sum to zero: $$(\ref{hyps_{2}3term})+(\ref{hyps_{2}2term})+(\ref{hyps_{2}term})+(\ref{hyps_{2}0term}) = \left( -8 c^{4} s_{2}^{3}\right)+\left(40c^{4}s_{2}^{3}\right)+\left(96 c^{6} s_{1} s_{2} - 88 c^{4} s_{2}^{3}\right)+\left(-96 c^6 s_{1} s_{2} + 56 c^4 s_{2}^3\right)= 0\ .$$

\noindent This completes the proof for the quantitative form of  the lensing map in the neighborhood of a hyperbolic umbilic.\ $\square$

\section{Proof of the Main Theorem for Generic Mappings}
\label{Appendix:ProofGeneric}

\subsection{Elliptic Umbilic}

\begin{figure}[ht]
\begin{center}
\includegraphics[scale=.5]{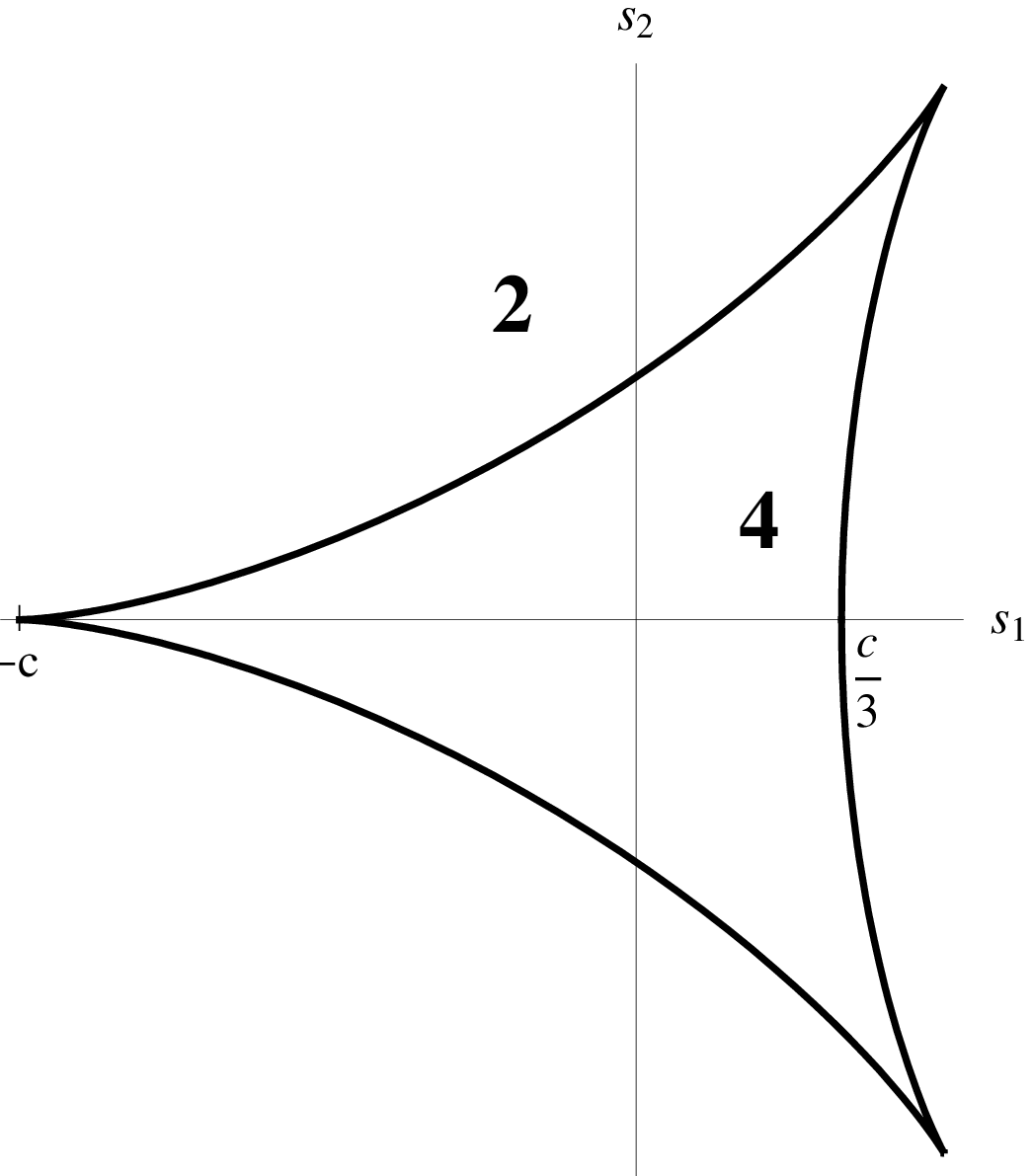}
\end{center}
\caption{Caustic curve for the generic elliptic umbilic
mapping $\bbf_c$ in equation (\ref{lenseqn2}) or Table~\ref{table1}.  The number of lensed images for sources in their respective regions is indicated.}
\label{Figure5}
\end{figure}

The derivation of the generic form of a one-parameter family of maps between planes in the neighborhood of an elliptic umbilic critical point, can be found in \cite[Chap. 8]{Majthay}.  The resulting map is
\beq
\label{lenseqn2}
s_{1}&=&3v^{2}-3u^{2}-2cu\ ,\nonumber \\
s_{2}&=&6uv-2cv\ ,
\eeq  

\noindent and the corresponding magnification of an image $(u,v)$ is
\beq
\label{mag2}
\left({\rm det}({\rm Jac}\ {\bf s})\right)^{-1}(u,v)&=&{1 \over 4c^{2} - 36(u^{2}+v^{2})}\ \cdot
\eeq

\noindent (Note that our notation differs from that of \cite{Majthay}.)  The caustic curve is shown in Figure~\ref{Figure5}.  Although these equations are noticeably different from their lensing map analogues (\ref{lenseqn}) and (\ref{mag}), it is nonetheless true that the total signed magnification for quadruply-imaged sources is zero in this the generic case as well.  The proof, in fact, is virtually identical to the lensing map case in Appendix~\ref{Appendix:eu}, as we now show.

As before, we begin by considering the special case of a source in the four-image region with $s_{2}=0$.  In this case the generic lens equation (\ref{lenseqn2}) is solvable.  The lensed images are $\left({1\over 3}\left(-c\pm \sqrt{c^{2}-3s_{1}}\,\right),0\right), \left({c \over 3},\pm {\sqrt{c^{2}+s_{1} \over 3}}\right)$, all of which are real because $-c < s_{1} < {c \over 3}$ inside the caustic curve (see Figure~\ref{Figure5}).  The total signed magnification, obtained by inserting each of these four images into (\ref{mag2}) and summing over, will be zero.  Once again, therefore, we will restrict ourselves to sources $(s_{1},s_{2})$ inside the caustic curve with $s_{2} \not = 0$.  Note from the second lens equation (\ref{lenseqn2}) that $s_{2} \not = 0$ forces the $v$-coordinate of each lensed image to be nonzero: $v_{i} \not = 0$.

\vskip 12pt

Let $(s_{1},s_{2} \not = 0)$ denote the position of an arbitrary source lying off the $s_{1}$-axis inside the caustic curve.  Let $(u_{i},v_{i})$ once again denote the corresponding lensed images.
\noindent The total signed magnification $\mu$ at $(s_{1},s_{2})$ is
\beq
\label{totalmag2}
\mu(u_{i},v_{i})={1 \over 4c^{2} - 36(u_{1}^{2}+v_{1}^{2})} + {1 \over 4c^{2} - 36(u_{2}^{2}+v_{2}^{2})} + {1 \over 4c^{2} - 36(u_{3}^{2}+v_{3}^{2})} + {1 \over 4c^{2} - 36(u_{4}^{2}+v_{4}^{2})}\ \cdot
\eeq 

\noindent We begin by eliminating $u$ from the lens equation (\ref{lenseqn2}) to obtain a (depressed) quartic in $v$:
\beq
\label{vquartic2}
v^{4} - {1 \over 3}(s_{1}+c^{2})v^{2} - {2 \over 9}cs_{2}v - {s_{2}^{2}\over 36}=0\ .
\eeq

\noindent Knowing that this quartic must factor as
\beq
\label{roots_{2}}
(v - v_{1})(v - v_{2})(v - v_{3})(v - v_{4})=0 \ ,
\eeq 

\noindent we expand (\ref{roots_{2}}) and equate its coefficients to those of (\ref{vquartic2}).  As a result we obtain four equations involving the $v_{i}$:
\beq
\label{eqn1a}
{\bf v_{1} + v_{2} + v_{3} + v_{4}}&=&{\bf 0}\ ,\\
\label{eqn2a}
{\bf v_{1}v_{2}+v_{1} v_{3} + v_{1} v_{4} + v_{2} v_{3} + v_{2} v_{4} + v_{3} v_{4}}&=&{\bf - {1 \over 3}(s_{1}+c^{2})}\ ,\\
\label{eqn3a}
{\bf v_{1} v_{2} v_{3} + v_{1} v_{2} v_{4} + v_{1} v_{3} v_{4} + v_{2} v_{3} v_{4}}&=&{\bf {2 \over 9}cs_{2}}\ ,\\
\label{eqn4a}
{\bf v_{1} v_{2} v_{3} v_{4}}&=&{\bf -{s_{2}^{2}\over 36}}\ \cdot
\eeq

\noindent Next, we use the second equation in (\ref{lenseqn2}) to express each $u_{i}$ in terms of $v_{i}$, bearing in mind that all $v_{i} \not = 0$,
\beq
\label{uintermsofv_{2}}
u_{i}(v_{i}) = {s_{2} + 2 c v_{i} \over 6 v_{i}}\ \cdot
\eeq

\noindent Once again, our procedure will be to insert (\ref{uintermsofv_{2}}) into the total magnification (\ref{totalmag2}), thereby obtaining an expression involving only the $v_{i}$, $\mu = \mu(v_{i})$.  When we do so, and factor according to powers of $s_{2}$, we obtain
\beq
\label{s_{2}^6termgen}
-s_{2}^{6}\,\left(v_{1}^{2} + v_{2}^{2} + v_{3}^{2} + v_{4}^{2}\right)\nonumber \ ,
\eeq
\beq
\label{s_{2}^5termgen}
-4\,c\,s_{2}^{5}\,\left(v_{1}^{2} v_{2} + v_{1} v_{2}^{2} + v_{1}^{2} v_{3} + v_{2}^{2} v_{3} + v_{1} v_{3}^{2} + v_{2} v_{3}^{2} + v_{1}^{2} v_{4} + v_{2}^{2} v_{4} + v_{3}^{2} v_{4} + v_{1} v_{4}^{2} + v_{2} v_{4}^{2} + v_{3} v_{4}^{2}\right)\nonumber \ ,
\eeq
\beq
\label{s_{2}^4termgen}
-4\,s_{2}^{4}\,\left(9\left(v_{1}^{4} v_{2}^{2}\right.\right.&+&\left.\left.v_{1}^{2} v_{2}^{4} + v_{1}^{4} v_{3}^{2} + v_{2}^{4} v_{3}^{2} + v_{1}^{2} v_{3}^{4} + v_{2}^{2} v_{3}^{4} + v_{1}^{4} v_{4}^{2} + v_{2}^{4} v_{4}^{2} + v_{3}^{4} v_{4}^{2} + v_{1}^{2} v_{4}^{4} + v_{2}^{2} v_{4}^{4} + v_{3}^{2} v_{4}^{4}\right.\right)\nonumber \\
&+&4 c^{2}\left(v_{1}^{2} v_{2} v_{3} + v_{1} v_{2}^{2} v_{3} + v_{1} v_{2} v_{3}^{2} + v_{1}^{2} v_{2} v_{4} + v_{1} v_{2}^{2} v_{4} + v_{1}^{2} v_{3} v_{4} + v_{2}^{2} v_{3} v_{4}\right.\nonumber \\
&+&\left.\left.v_{1} v_{3}^{2} v_{4} +v_{2} v_{3}^{2} v_{4} + v_{1} v_{2} v_{4}^{2} + v_{1} v_{3} v_{4}^{2} + v_{2} v_{3} v_{4}^{2}\right)\right)\nonumber \ ,
\eeq
\beq
\label{s_{2}^3termgen}
-16\,c\,s_{2}^{3}\,\left(9\left(v_{1}^{4} v_{2}^{2} v_{3}\right.\right.&+&\left.\left.v_{1}^{2} v_{2}^{4} v_{3} + v_{1}^{4} v_{2} v_{3}^{2} + v_{1} v_{2}^{4} v_{3}^{2} + v_{1}^{2} v_{2} v_{3}^{4} + v_{1} v_{2}^{2} v_{3}^{4} + v_{1}^{4} v_{2}^{2} v_{4} + v_{1}^{2} v_{2}^{4} v_{4} + v_{1}^{4} v_{3}^{2} v_{4} + v_{2}^{4} v_{3}^{2} v_{4}\right.\right.\nonumber \\
&+&v_{1}^{2} v_{3}^{4} v_{4} + v_{2}^{2} v_{3}^{4} v_{4} + v_{1}^{4} v_{2} v_{4}^{2} + v_{1} v_{2}^{4} v_{4}^{2} + v_{1}^{4} v_{3} v_{4}^{2} + v_{2}^{4} v_{3} v_{4}^{2} + v_{1} v_{3}^{4} v_{4}^{2} + v_{2} v_{3}^{4} v_{4}^{2}\nonumber \\
&+&\left.v_{1}^{2} v_{2} v_{4}^{4} + v_{1} v_{2}^{2} v_{4}^{4} + v_{1}^{2} v_{3} v_{4}^{4} + v_{2}^{2} v_{3} v_{4}^{4} + v_{1} v_{3}^{2} v_{4}^{4} + v_{2} v_{3}^{2} v_{4}^{4}\right)\nonumber \\
&-&4c^{2} \left.\left(v_{1}^{2} v_{2} v_{3} v_{4} + 4 v_{1} v_{2}^{2} v_{3} v_{4} + 4 v_{1} v_{2} v_{3}^{2} v_{4} + 4 v_{1} v_{2} v_{3} v_{4}^{2}\right)\right)\nonumber \ ,
\eeq
\beq
\label{s_{2}^2termgen}
-144\,s_{2}^{2}\,\left(9\left(v_{1}^{4} v_{2}^{4} v_{3}^{2}\right.\right.&+&\left.\left.v_{1}^{4} v_{2}^{2} v_{3}^{4} + v_{1}^{2} v_{2}^{4} v_{3}^{4} + v_{1}^{4} v_{2}^{4} v_{4}^{2} + v_{1}^{4} v_{3}^{4} v_{4}^{2} + v_{2}^{4} v_{3}^{4} v_{4}^{2} + v_{1}^{4} v_{2}^{2} v_{4}^{4} + v_{1}^{2} v_{2}^{4} v_{4}^{4} + v_{1}^{4} v_{3}^{2} v_{4}^{4} + v_{2}^{4} v_{3}^{2} v_{4}^{4}\right.\right.\nonumber \\
&+&\left.v_{1}^{2} v_{3}^{4} v_{4}^{4} + v_{2}^{2} v_{3}^{4} v_{4}^{4}\right) + 4c^{2} \left(v_{1}^{4} v_{2}^{2} v_{3} v_{4} + v_{1}^{2} v_{2}^{4} v_{3} v_{4} + v_{1}^{4} v_{2} v_{3}^{2} v_{4} + v_{1} v_{2}^{4} v_{3}^{2} v_{4} + v_{1}^{2} v_{2} v_{3}^{4} v_{4}\right.\nonumber \\
&+&\left.\left.v_{1} v_{2}^{2} v_{3}^{4} v_{4} + v_{1}^{4} v_{2} v_{3} v_{4}^{2} + v_{1} v_{2}^{4} v_{3} v_{4}^{2} + v_{1} v_{2} v_{3}^{4} v_{4}^{2} + v_{1}^{2} v_{2} v_{3} v_{4}^{4} + v_{1} v_{2}^{2} v_{3} v_{4}^{4} + v_{1} v_{2} v_{3}^{2} v_{4}^{4}\right)\right)\nonumber \ ,
\eeq
\beq
\label{s_{2}^1termgen}
-5184\,c\,s_{2}\,\left(v_{1}^{4} v_{2}^{4} v_{3}^{2} v_{4}\right.&+&v_{1}^{4} v_{2}^{2} v_{3}^{4} v_{4} + v_{1}^{2} v_{2}^{4} v_{3}^{4} v_{4} + v_{1}^{4} v_{2}^{4} v_{3} v_{4}^{2} + v_{1}^{4} v_{2} v_{3}^{4} v_{4}^{2} + v_{1} v_{2}^{4} v_{3}^{4} v_{4}^{2} + v_{1}^{4} v_{2}^{2} v_{3} v_{4}^{4}\nonumber \\
&+&\left.v_{1}^{2} v_{2}^{4} v_{3} v_{4}^{4} + v_{1}^{4} v_{2} v_{3}^{2} v_{4}^{4} + v_{1} v_{2}^{4} v_{3}^{2} v_{4}^{4} + v_{1}^{2} v_{2} v_{3}^{4} v_{4}^{4} + v_{1} v_{2}^{2} v_{3}^{4} v_{4}^{4}\right)\nonumber \ ,
\eeq
\beq
\label{s_{2}^0termgen}
-46656 \left(v_{1}^{4} v_{2}^{4} v_{3}^{4} v_{4}^{2} + v_{1}^{4} v_{2}^{4} v_{3}^{2} v_{4}^{4} + v_{1}^{4} v_{2}^{2} v_{3}^{4} v_{4}^{4} + v_{1}^{2} v_{2}^{4} v_{3}^{4} v_{4}^{4}\right)\nonumber \ .
\eeq

\noindent Perusal of these expressions shows that aside from constant factors, the form of each of the expressions in parentheses is {\it identical} to its counterpart in the lensing map case in Appendix~\ref{Appendix:eu}.  This means that the procedures we employed above, (\ref{eqn5})--(\ref{finals_{2}^0term}), carry through without modification.  Of course, we expect different final answers, since the right-hand sides of (\ref{eqn2a})--(\ref{eqn4a}) differ from those of (\ref{eqn2})--(\ref{eqn4}).  With that said, we can forgo the labor and only state the final results:

\vskip 12pt

{\bf The $s_{2}^{6}$-term:} $-{2 \over 3} c^{2} s_{2}^{6} - {2 \over 3} s_{1} s_{2}^{6}$\ ,

\vskip 12pt

{\bf The $s_{2}^{5}$-term:} ${8 \over 3} c^{2} s_{2}^{6}$\ ,

\vskip 12pt

{\bf The $s_{2}^{4}$-term:} $-{8 \over 3} c^{6} s_{2}^{4} - 8 c^{4} s_{1} s_{2}^{4} - 8 c^{2} s_{1}^{2} s_{2}^{4} - {8 \over 3} s_{1}^{3} s_{2}^{4} + {26 \over 9} c^{2} s_{2}^{6} - {2 \over 9} s_{1} s_{2}^{6}$\ ,

\vskip 12pt

{\bf The $s_{2}^{3}$-term:} ${64 \over 9} c^{6} s_{2}^{4} + {128 \over 9} c^{4} s_{1} s_{2}^{4} + {65 \over 9} c^{2} s_{1}^{2} s_{2}^{4} - {16 \over 9} c^{2} s_{2}^{6}$\ ,

\vskip 12pt

{\bf The $s_{2}^{2}$-term:} $-8 c^{6} s_{2}^{4} - {40 \over 3} c^{4} s_{1} s_{2}^{4} - {8 \over 3} c^{2} s_{1}^{2} s_{2}^{4} + {8 \over 3} s_{1}^{3} s_{2}^{4} + {22 \over 9} c^{2} s_{2}^{6} + {2 \over 3} s_{1} s_{2}^{6}$\ ,

\vskip 12pt

{\bf The $s_{2}$-term:} ${32 \over 9} c^{6} s_{2}^{4} + {64 \over 9} c^{4} s_{1} s_{2}^{4} + {32 \over 9} c^{2} s_{1}^{2} s_{2}^{4} - {40 \over 9} c^{2} s_{2}^{6}$\ ,

\vskip 12pt

{\bf The term with no $s_{2}$-dependence:} $-{10 \over 9} c^{2} s_{2}^{6} + {2 \over 3} s_{1} s_{2}^{6}$\ .

\vskip 12pt

\noindent We can now verify that these terms sum to zero.  This completes the proof for the generic form of a one-parameter family of maps between planes in the neighborhood of an elliptic umbilic.\ $\square$

\subsection{Hyperbolic Umbilic}

The derivation of the generic form of a one-parameter family of maps between planes in the neighborhood of a hyperbolic umbilic critical point, can be found in \cite[Chap. 8]{Majthay}.  The resulting map is
\beq
\label{lenseqn4}
s_{1}&=&-3u^{2}-cv\ ,\nonumber \\
s_{2}&=&-3v^{2}-cu\ ,
\eeq

\noindent and the corresponding magnification of an image $(u,v)$ is
\beq
\label{mag4}
\left({\rm det}({\rm Jac}\ {\bf s})\right)^{-1}(u,v)&=&{1 \over -c^{2} + 36uv}\ \cdot
\eeq

\noindent The caustic curve is shown in Figure~\ref{Figure6}.  We will show that the total signed magnification for quadruply-imaged sources is zero in this the generic case as well.  What is more, just as the proof for the generic elliptic umbilic was virtually identical to its lensing map analogue in Appendix~\ref{Appendix:eu}, so will be the case here.

First, note from the second lens equation (\ref{lenseqn4}) that since $u_{i}(v_{i}) = {-s_{2} - 3v_{i}^{2} \over c}$, which is to say, since $v_{i}$ does not appear in the denominator, we do not need to restrict our analysis to the case where $v_{i} \not = 0$.  Once again, therefore, we will let $(s_{1},s_{2})$ denote an arbitrary source in the four-image region.  The total signed magnification $\mu$ at $(s_{1},s_{2})$ is
\beq
\label{totalmag4}
\mu(u_{i},v_{i})={1 \over -c^{2} + 36u_{1}v_{1}} + {1 \over -c^{2} + 36u_{2}v_{2}} + {1 \over -c^{2} + 36u_{3}v_{3}} + {1 \over -c^{2} + 36u_{4}v_{4}}\ \cdot
\eeq 

\noindent This time the (depressed) quartic in $v$ is
\beq
v^{4}+{2\over 3}s_{2} v^{2}+{c^{3}\over 27} v+{3s_{2}^{2}+c^{2}s_{1}\over 27}=0\nonumber
\eeq

\noindent and the corresponding four equations involving only the $v_{i}$ are
\beq
\label{eqn1bgen}
{\bf v_{1} + v_{2} + v_{3} + v_{4}}&=&{\bf 0}\ ,\\
\label{eqn2bgen}
{\bf v_{1}v_{2}+v_{1} v_{3} + v_{1} v_{4} + v_{2} v_{3} + v_{2} v_{4} + v_{3} v_{4}}&=&{\bf {2 \over 3}s_{2}}\ ,\\
\label{eqn3bgen}
{\bf v_{1} v_{2} v_{3} + v_{1} v_{2} v_{4} + v_{1} v_{3} v_{4} + v_{2} v_{3} v_{4}}&=&{\bf -{c^{3} \over 27}}\ ,\\
\label{eqn4bgen}
{\bf v_{1} v_{2} v_{3} v_{4}}&=&{\bf {c^{2}s_{1}+3s_{2}^{2}\over 27}}\ \cdot
\eeq

\begin{figure}[ht]
\begin{center}
\includegraphics[scale=.5]{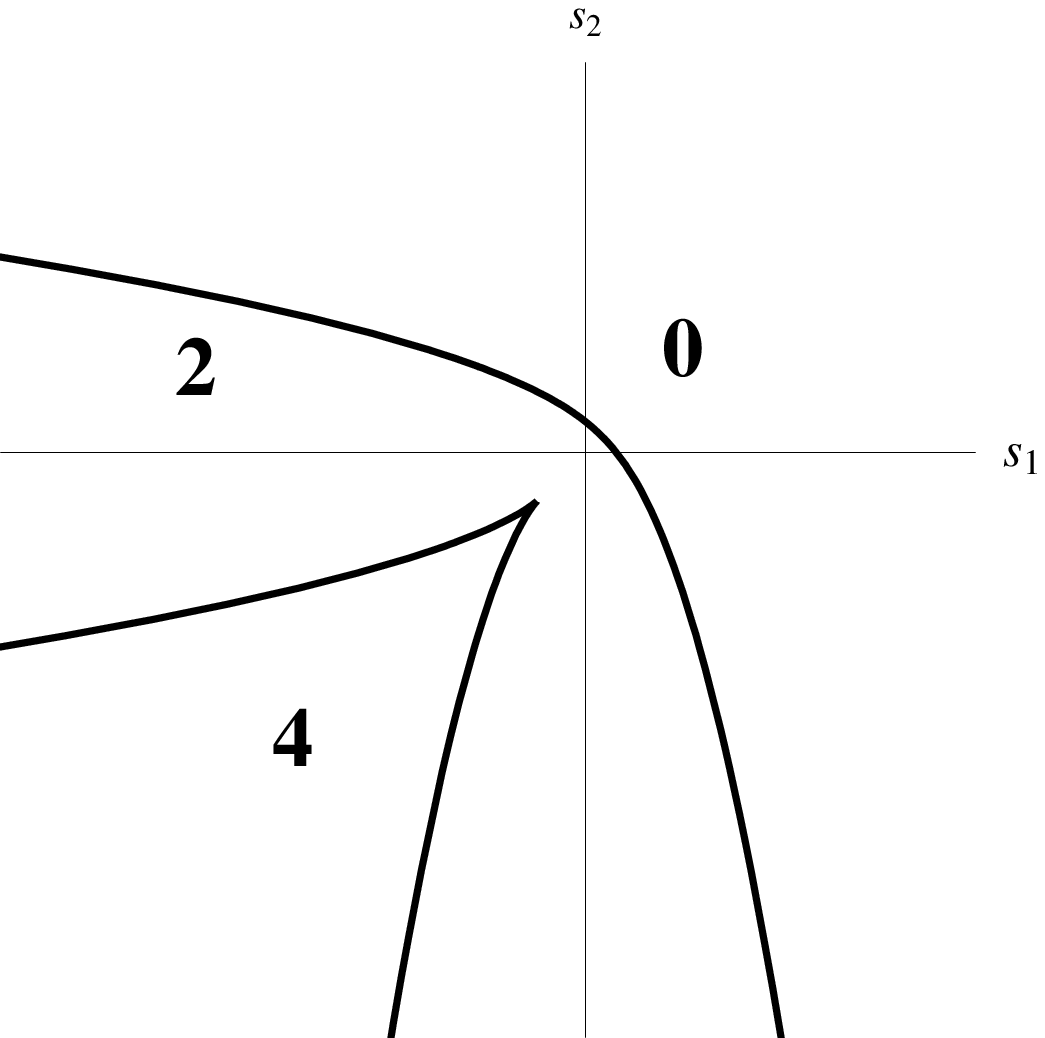}
\end{center}
\caption{Caustic curve for the generic hyperbolic umbilic mapping
$\bbf_c$ in equation (\ref{lenseqn4}) or Table~\ref{table1}.  The number of lensed images for sources in their respective regions is indicated.}
\label{Figure6}
\end{figure}

\noindent We now insert $u_{i}(v_{i})$ into the total signed magnification (\ref{totalmag4}) and factor the numerator of the resulting expression in powers of $s_{2}$:
\beq
\label{1hypgen}
-46656\,c\,s_{2}^{3}\,(v_{1} v_{2} v_{3} + v_{1} v_{2} v_{4} + v_{1} v_{3} v_{4} + v_{2} v_{3} v_{4})\ ,
\eeq
\beq
\label{2hypgen}
-c\,s_{2}^{2}\,\left(2592\,c^{3}\left(v_{1} v_{2}\right.\right.&+&\left.v_{1} v_{3} + v_{2} v_{3} + v_{1} v_{4} + v_{2} v_{4} + v_{3} v_{4}\right) + 139968\left(v_{1}^{3} v_{2} v_{4} + v_{1} v_{2}^{3} v_{4} + v_{1}^{3} v_{3} v_{4} + v_{2}^{3} v_{3} v_{4}\right.\nonumber \\
&+&\left.\left.v_{1} v_{3}^{3} v_{4} + v_{2} v_{3}^{3} v_{4} + v_{1} v_{2} v_{4}^{3} + v_{1} v_{3} v_{4}^{3} + v_{2} v_{3} v_{4}^{3} + v_{1}^{3} v_{2} v_{3} + v_{1} v_{2}^{3} v_{3} + v_{1} v_{2} v_{3}^{3}\right)\right)\ ,
\eeq
\beq
\label{3hypgen}
c\,s_{2}\,\left(108\,c^{6}\left(v_{1}\right.\right.&+&\left.v_{2} + v_{3} + v_{4}\right) + 7776\,c^{3}\left(v_{1}^{3} v_{2} + v_{1} v_{2}^{3} + v_{1}^{3} v_{3} + v_{2}^{3} v_{3} + v_{1} v_{3}^{3} + v_{2} v_{3}^{3} + v_{1}^{3} v_{4} + v_{2}^{3} v_{4} + v_{3}^{3} v_{4}\right.\nonumber \\
&+&\left.v_{1} v_{4}^{3} + v_{2} v_{4}^{3} + v_{3} v_{4}^{3}\right) +419904\left(v_{1}^{3} v_{2}^{3} v_{3} +v_{1}^{3} v_{2} v_{3}^{3} + v_{1} v_{2}^{3} v_{3}^{3} + v_{1}^{3} v_{2}^{3} v_{4} +v_{1}^{3} v_{3}^{3} v_{4}\right.\nonumber \\
&+&\left.\left.v_{2}^{3} v_{3}^{3} v_{4}  + v_{1}^{3} v_{2} v_{4}^{3} + v_{1} v_{2}^{3} v_{4}^{3} + v_{1}^{3} v_{3} v_{4}^{3} + v_{2}^{3} v_{3} v_{4}^{3} + v_{1} v_{3}^{3} v_{4}^{3} + v_{2} v_{3}^{3} v_{4}^{3}\right)\right)\ ,
\eeq
\beq
\label{4hypgen}
-4\,c^{10}&-&324\,c^{7}\left(v_{1}^{3} + v_{2}^{3} + v_{4}^{3} + v_{3}^{3}\right) - 23328\,c^{4}\left(v_{1}^{3} v_{3}^{3} + v_{2}^{3} v_{3}^{3} +v_{1}^{3} v_{2}^{3} + v_{1}^{3} v_{4}^{3} + v_{2}^{3} v_{4}^{3} + v_{3}^{3} v_{4}^{3}\right)\nonumber \\
&-&1259710\,c\left(v_{1}^{3} v_{2}^{3} v_{4}^{3} + v_{1}^{3} v_{3}^{3} v_{4}^{3} + v_{2}^{3} v_{3}^{3} v_{4}^{3} + v_{1}^{3} v_{2}^{3} v_{3}^{3}\right)\ .
\eeq

\noindent Perusal of these expressions shows that aside from constant factors, the form of each expression in parentheses is once again {\it identical} to its counterpart in the lensing map case in Appendix~\ref{Appendix:hu}.  This means that the procedures we employed above, (\ref{hyps_{2}3term})--(\ref{hyps_{2}0term}), carry through without modification.  Of course, as with the elliptic umbilic, we expect different final answers for each term, since the right-hand sides of (\ref{eqn2bgen})--(\ref{eqn4bgen}) differ from those of (\ref{eqn2b})--(\ref{eqn4b}).  With that said, we once again forgo the labor and only state the final results:

\vskip 12pt

{\bf The $s_{2}^{3}$-term:} $1728 c^{4} s_{2}^3$\ ,

\vskip 12pt

{\bf The $s_{2}^{2}$-term:} $-8640 c^{4} s_{2}^3$\ ,

\vskip 12pt

{\bf The $s_{2}$-term:} $1728 c^{6} s_{1} s_{2} + 19008 c^{4} s_{2}^{3}$\ ,

\vskip 12pt

{\bf The term with no $s_{2}$-dependence:} $-1728 c^{6} s_{1} s_{2} - 12096 c^{4} s_{2}^{3}$\ .

\vskip 12pt

\noindent We can now verify these terms sum to zero.\ $\square$

\subsection{Swallowtail}

The generic form of a one-parameter family of maps between planes in the neighborhood of a swallowtail critical point can be found in Golubitsky \& Guillemin 1973 \cite[p. 176]{Gol-G}.  The resulting map is
\beq
\label{lenseqn5}
s_{1}&=&u v+c u^2+u^4\ ,\nonumber \\
s_{2}&=&v\ ,
\eeq

\begin{figure}[ht]
\begin{center}
\includegraphics[scale=.5]{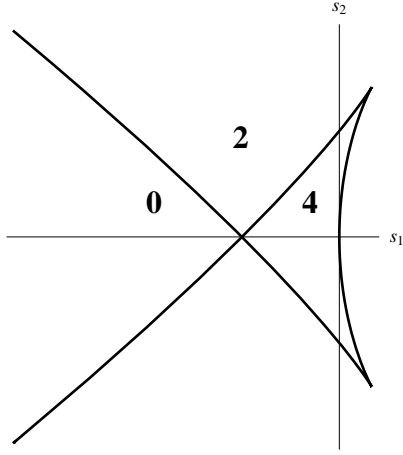}
\end{center}
\caption{Caustic curve for the generic swallowtail mapping 
$\bbf_c$ in equation (\ref{lenseqn5}) or Table~\ref{table1}.  The number of lensed images for sources in their respective regions is indicated.  Note that in order to produce the tail, we must have $c < 0$ for the given
form of the swallowtail in (\ref{lenseqn5}).}
\label{Figure7}
\end{figure}

\noindent and the corresponding magnification of an image $(u,v)$ is
\beq
\label{mag5}
\left({\rm det}({\rm Jac}\ {\bf s})\right)^{-1}(u,v)&=&{1 \over 2cu+4u^3+v}\ \cdot
\eeq

\noindent The caustic curve is shown in Figure~\ref{Figure7}.  The {\lq\lq tail\rq\rq} constitutes the four-image region.  We will show that the total signed magnification for all sources inside this region is identically zero.

Let $(s_{1},s_{2})$ denote an arbitrary source in the four-image region, and $(u_{i},v_{i})$ the corresponding lensed images.  The total signed magnification $\mu$ at $(s_{1},s_{2})$ is
\beq
\label{totalmag5}
\mu(u_{i})={1 \over 2 c u_{1} + 4 u_{1}^3 + s_{2}} + {1 \over 2 c u_{2} + 4 u_{2}^3 + s_{2}} + {1 \over 2 c u_{3} + 4 u_{3}^3 + s_{2}} + {1 \over 2 c u_{4} + 4 u_{4}^3 + s_{2}}\ ,
\eeq 
\noindent where we have used the second lens equation (\ref{lenseqn5}) to substitute $v_{i} = s_{2}$ into the magnification of each lensed image.  This time the (depressed) quartic is in $u$,
\beq
u^{4}+c u^{2}+s_{2} u-s_{1}=0\nonumber \ ,
\eeq

\noindent and the corresponding four equations involving only the $u_{i}$ are
\beq
\label{eqn1sw}
{\bf u_{1} + u_{2} + u_{3} + u_{4}}&=&{\bf 0}\ ,\\
\label{eqn2sw}
{\bf u_{1} u_{2}+u_{1} u_{3} + u_{1} u_{4} + u_{2} u_{3} + u_{2} u_{4} + u_{3} u_{4}}&=&{\bf c}\ ,\\
\label{eqn3sw}
{\bf u_{1} u_{2} u_{3} + u_{1} u_{2} u_{4} + u_{1} u_{3} u_{4} + u_{2} u_{3} u_{4}}&=&{\bf -s_{2}}\ ,\\
\label{eqn4sw}
{\bf u_{1} u_{2} v_{3} v_{4}}&=&{\bf -s_{1}}\ .
\eeq

\noindent In what is now becoming a familiar story, we write (\ref{totalmag5}) over a common denominator and then factor the resulting numerator in powers of $s_{2}$:
\beq
\label{1sw}
-4\,s_{2}^{3}\nonumber \ ,
\eeq
\beq
\label{2sw}
s_{2}^2\,\left(6c\left(u_{1} + u_{2} + u_{3} + u_{4}\right) + 12\left(u_{1}^3 + u_{2}^3 + u_{3}^3 + u_{4}^3\right)\right)\nonumber \ ,
\eeq
\beq
\label{3sw}
s_{2}\left(8c^2\left(u_{1} u_{2}+u_{1} u_{3} + u_{2} u_{3} + u_{1} u_{4} + u_{2} u_{4} + u_{3} u_{4}\right) + 16c\left(u_{1}^3 u_{2} + u_{1} u_{2}^3 + u_{1}^3 u_{3} + u_{2}^3 u_{3} + u_{1} u_{3}^3 + u_{2} u_{3}^3\right.\right.\nonumber \\
+\left.\left.u_{1}^3 u_{4} + u_{2}^3 u_{4} + u_{3}^3 u_{4} + u_{1} u_{4}^3 + u_{2} u_{4}^3 + u_{3} u_{4}^3\right)+32\left( u_{1}^3 u_{2}^3 + u_{1}^3 u_{3}^3 + u_{2}^3 u_{3}^3 + u_{1}^3 u_{4}^3 + u_{2}^3 u_{4}^3 + u_{3}^3 u_{4}^3\right)\right)\nonumber \ ,
\eeq
\beq
\label{4sw}
8c^3 \left(u_{1} u_{2} u_{3}\right.&+&\left.u_{1} u_{2} u_{4} + u_{1} u_{3} u_{4} + u_{2} u_{3} u_{4}\right) + 16c^2\left(u_{1}^3 u_{2} u_{3} + u_{1} u_{2}^3 u_{3} + u_{1} u_{2} u_{3}^3 + u_{1}^3 u_{2} u_{4} + u_{1} u_{2}^3 u_{4}\right.\nonumber \\
&+&\left.u_{1}^3 u_{3} u_{4} + u_{2}^3 u_{3} u_{4} + u_{1} u_{3}^3 u_{4} + u_{2} u_{3}^3 u_{4} + u_{1} u_{2} u_{4}^3 + u_{1} u_{3} u_{4}^3 + u_{2} u_{3} u_{4}^3\right)+32c\left(u_{1}^3 u_{2}^3 u_{3} \right.\nonumber \\
&+&\left.u_{1}^3 u_{2} u_{3}^3 + u_{1} u_{2}^3 u_{3}^3 + u_{1}^3 u_{2}^3 u_{4} + u_{1}^3 u_{3}^3 u_{4} + u_{2}^3 u_{3}^3 u_{4} + u_{1}^3 u_{2} u_{4}^3+u_{1} u_{2}^3 u_{4}^3 + u_{1}^3 u_{3} u_{4}^3\right.\nonumber \\
&+&\left.u_{2}^3 u_{3} u_{4}^3 + u_{1} u_{3}^3 u_{4}^3 + u_{2} u_{3}^3 u_{4}^3\right) + 64\left(u_{1}^3 u_{2}^3 u_{3}^3 + u_{1}^3 u_{2}^3 u_{4}^3+u_{1}^3 u_{3}^3 u_{4}^3 + u_{2}^3 u_{3}^3 u_{4}^3\right)\nonumber \ .
\eeq

\noindent Perusal of these expressions shows that every polynomial in parentheses has already been calculated in the case of the hyperbolic umbilic.  Once again, we expect different final answers for each term, since the right-hand sides of (\ref{eqn2sw})--(\ref{eqn4sw}) differ from those of (\ref{eqn2b})--(\ref{eqn4b}).  With that said, we once again forgo the labor and only state the final results:

\vskip 12pt

{\bf The $s_{2}^{3}$-term:} $4s_{2}^3$\ ,

\vskip 12pt

{\bf The $s_{2}^{2}$-term:} $-36s_{2}^3$\ ,

\vskip 12pt

{\bf The $s_{2}$-term:} $8 c^3 s_{2} + 32 c s_{1} s_{2} + 96 s_{2}^3$\ ,

\vskip 12pt

{\bf The term with no $s_{2}$-dependence:} $-8 c^3 s_{2} - 32 c s_{1} s_{2} - 64 s_{2}^3$\ .

\vskip 12pt

\noindent We can now verify that these terms sum to zero.\ $\square$

\newpage



\begin{thebibliography}{}

\bibitem{Petters}
A. O. Petters, H. Levine, and J. Wambsganss, {\it Singularity Theory
and Gravitational Lensing} (Birkh{\"a}user, 2001).

\bibitem{Witt-Mao}
H. J. Witt, S. Mao,
{\it Astrophys. J. Lett.} (1995), {\bf 447}, 105.

\bibitem{Rhie}
S. H. Rhie,
{\it Astrophys. J.} (1997), {\bf 484}, 67.

\bibitem{Dalal}
N. Dalal,
{\it Astrophys. J.} (1998), {\bf 509}, 13.

\bibitem{Witt-Mao2}
H. J. Witt, S. Mao,
{\it Mon. Not. Roy. Astron. Soc.} (2000), {\bf 311}, 689.

\bibitem{Dalal-Rabin}
N. Dalal, J. M. Rabin,
{\it J. Math. Phys.} (2001), {\bf 42}, 1818.

\bibitem{Hunter-Evans}
C. Hunter, N. W. Evans,
{\it Astrophys. J.} (2001), {\bf 554}, 1227.

\bibitem{Werner}
M. Werner,
{\it J. Math. Phys.} (2007), {\bf 48}, 052501.

\bibitem{Blandford90}
R. D. Blandford,
{\it Q. Jl. Roy. Astron. Soc.} (1990), {\bf 31}, 305.

\bibitem{Sch-Weiss92}
P. Schneider, A. Weiss,
{\it Astron. Astrophys.} (1992), {\bf 260}, 1.

\bibitem{Petters93}
A. O. Petters,
{\it J. Math. Phys.} (1993), {\bf 33}, 3555

\bibitem{Sch-EF}
P. Schneider, J. Ehlers,  E. Falco, {\it Gravitational Lenses}
(Springer, 1992).

\bibitem{KGP-cusps}
C. Keeton, S. Gaudi, and A. O. Petters,
{\it Astrophys. J.} (2003), {\bf 598}, 138.

\bibitem{KGP-folds}
C. Keeton, S. Gaudi, and A. O. Petters,
{\it Astrophys. J.} (2005), {\bf 635}, 35.

\bibitem{Blan-Nar}
R. Blandford, R. Narayan,
{\it Astrophys. J.} (1986), {\bf 310}, 568.

\bibitem{Zakharov}
A. Zakharov,
{\it Astron. Astrophys.} (1995), {\bf 293}, 1.

\bibitem{Arnold86}
V. I. Arnold,
{\it J. Sov. Math.} (1986), {\bf 32}, 229.

\bibitem{Majthay}
A. Majthay, {\it Foundations of Catastrophe Theory} (Pitman, 1985).

\bibitem{C-Hayes}
D. Castrigiano, S. Hayes, {\it Catastrophe Theory} (Westview, 2004).

\bibitem{Gol-G}
M. Golubitsky, V. Guillemin, {\it Stable Mappings and Their Singularities} (Springer, 1973).

\bibitem{Arnold73}
V. I. Arnold,
{\it Func. Anal. Appl.} (1973), {\bf 6}, 254.

\bibitem{AGV1}
V. I. Arnold, S. M. Gusein-Zade, and A.N. Varchenko,
{\it Singularities of Differentiable Maps, vol. I} (Birkh{\"a}user, 1985).

\bibitem{Mao-Sch}
S. Mao, P. Schneider,
{\it Mon. Not. Roy. Astron. Soc.} (1998), {\bf 295}, 587.

\bibitem{M-M}
R. B. Metcalf, P. Madau,
{\it Astrophys. J.} (2001), {\bf 563}, 9.

\bibitem{Chiba}
M. Chiba,
{\it Astrophys. J.} (2002), {\bf 565}, 17.

\bibitem{Keeton}
C. Keeton,
astro-ph/0102340.

\end{thebibliography}
\end{document}